\documentclass[acmtog,authorversion]{acmart}
\acmSubmissionID{969}

\usepackage{pgfplots}
\usetikzlibrary{math, calc, positioning}
\usepackage{ifthen}

\pgfplotsset{compat=1.18}

\usepackage{multirow}

\usepackage[ruled]{algorithm2e} % For algorithms

\SetAlFnt{\small}
\SetAlCapFnt{\small}
\SetAlCapNameFnt{\small}
\SetAlCapHSkip{0pt}

\newcommand{\change}[1]{{#1}}

\acmJournal{TOG}
\copyrightyear{2025}
\acmYear{2025}
\acmConference[SIGGRAPH Conference Papers '25]{Special Interest Group on Computer Graphics and Interactive Techniques Conference Conference Papers }{August 10--14, 2025}{Vancouver, BC, Canada}
\acmBooktitle{Special Interest Group on Computer Graphics and Interactive Techniques Conference Conference Papers (SIGGRAPH Conference Papers '25), August 10--14, 2025, Vancouver, BC, Canada}
\acmDOI{10.1145/3721238.3730721}
\acmISBN{979-8-4007-1540-2/2025/08}

\begin{document}
% Title portion
\title{A Fluorescent Material Model for Non-Spectral Editing \& Rendering}

% DO NOT ENTER AUTHOR INFORMATION FOR ANONYMOUS TECHNICAL PAPER SUBMISSIONS TO SIGGRAPH 2019!
\author{Laurent Belcour}
\orcid{0000-0002-1982-0717}
\affiliation{%
 \institution{Intel Corporation}
 \country{France}}
\email{laurent.belcour@gmail.com}

\author{Alban Fichet}
\orcid{0000-0002-2227-0891}
\affiliation{%
 \institution{Intel Corporation}
 \country{France}}

\author{Pascal Barla}
\orcid{0000-0003-2844-6656}
\affiliation{%
 \institution{Inria Bordeaux}
 \country{France}}

\renewcommand\shortauthors{Belcour, L. et al}

\begin{teaserfigure}
    \center
    \scalebox{0.9}{
    \begin{tikzpicture}
        \begin{scope}
        \begin{scope}
            \node[anchor=south west,draw=black,thick,inner sep=0pt] (teaser) { \includegraphics[width=0.5\textwidth]{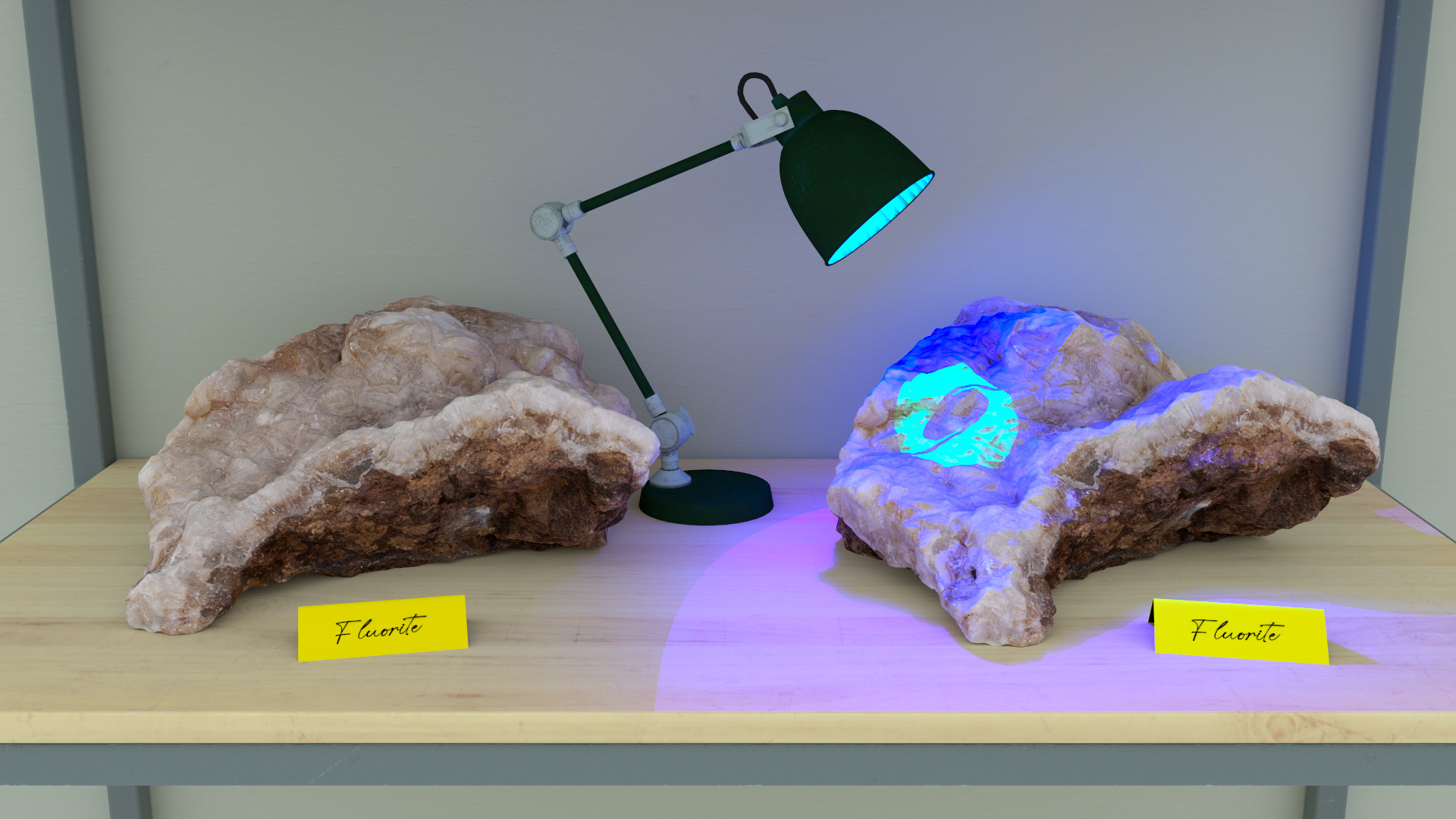} };
        \end{scope}
        %
        % Left insets
        \draw[thick, color=orange, fill] (2.0cm, 1.1cm)  circle (1pt) -- (-1.0cm, 1.1cm);
        \begin{scope}[xshift=-3.5cm,yshift=8pt]
            \node[anchor=south west, inner sep=0pt] (spec_P) at (0,0.099) { \includegraphics[height=2cm]{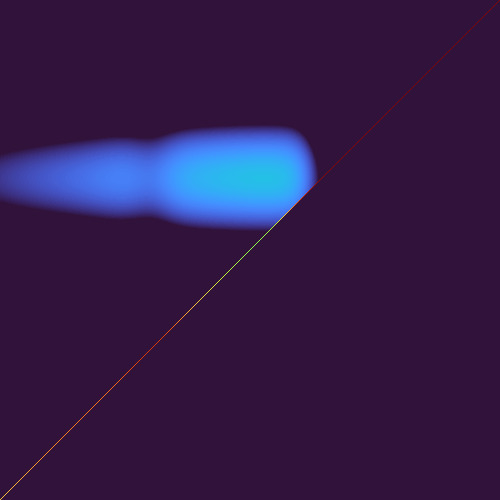} };
            \draw (0.0,0.1) rectangle ++(2cm,2cm);
            \node at (1cm,-5pt) {\footnotesize Our model};
            \draw[thick, color=olive, ->] (0,1cm)  -- (-0.2cm,1cm);
            \draw[thick, color=olive, fill] (0,1cm)  circle (1pt);
        \end{scope}
        \begin{scope}[xshift=-2.2cm,yshift=2.96cm]
            \node[anchor=south west, inner sep=0pt] (spec_P) at (0,0.099) { \includegraphics[height=2cm]{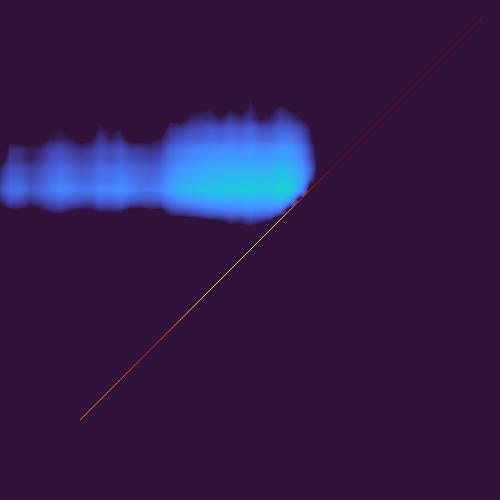} };
            \draw (0.0,0.1) rectangle ++(2cm,2cm);
            \node at (1cm,-5pt) {\footnotesize Measured reradiation};
            % \draw[thick, color=orange, ->] (0,1cm)  -- (-0.5cm,1cm) -- (-0.5cm,-1.8cm) -- (-0.1cm, -1.8cm);
            \draw[thick, color=orange, ->] (0,1cm)  -- (-0.5cm,1cm) -- (-0.5cm,-0.55cm);
            \draw[thick, color=orange, fill] (0, 1.0cm)  circle (1pt);
            \draw[thick, color=orange, ->] (0.7,-1.85cm)  -- ++(0.3cm,0cm);
            \draw[thick, color=orange, fill] (0.7,-1.85cm)  circle (1pt);
            \draw[thick, color=olive, ->] (0,1.2cm)  -- (-1.4cm,1.2cm);
            \draw[thick, color=olive, fill] (0,1.2cm)  circle (1pt);
        \end{scope}
        \begin{scope}[xshift=-0.6cm, yshift=1.1cm]
            \node[anchor=north east, inner sep=0pt] (red_P) at (0.5, 0.5) { \includegraphics[height=1cm]{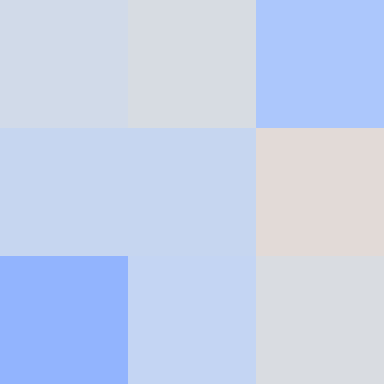} };
            \foreach \v in {1,2} {
            \draw[thin] ($(red_P.north west)+(\v/3, 0)$) -- ($(red_P.south west)+(\v/3, 0)$);
            \draw[thin] ($(red_P.north west)-(0, \v/3)$) -- ($(red_P.north east)-(0, \v/3)$);
            }
            \draw (red_P.south west) rectangle (red_P.north east);
            \node at (0,-0.7) (ana) { {\tiny Analytical} };
            \node[below=-2pt of ana.south,inner sep=0pt] { {\tiny reduction} };
        \end{scope}
        % Inset rendering A (measurement)
        \begin{scope}[xshift=-5.7cm,yshift=3.06cm]
            \node[anchor=south west, inner sep=0pt] (specA) at (0.0, 0.0) { \includegraphics[height=2cm]{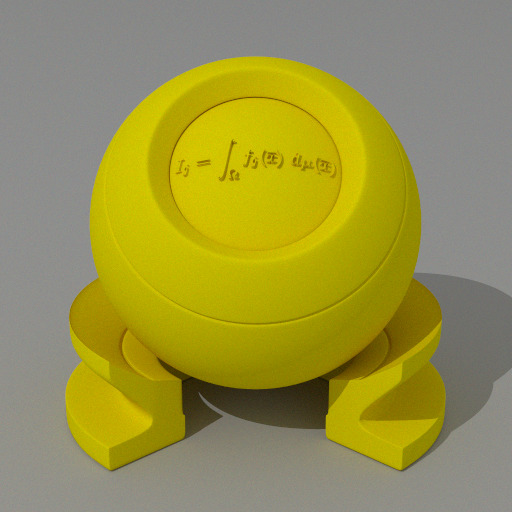} };
            \begin{scope}
                \clip(0,0) rectangle (1,2);
                \node[anchor=south west, inner sep=0pt] (reducedA) at (0.0, 0.0) { \includegraphics[height=2cm]{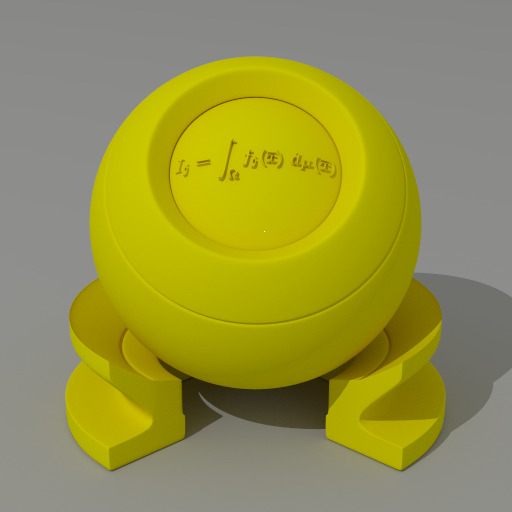} };
                \draw[color=black]  (1,0) -- (1,2);
            \end{scope}
            \node[above=-1pt of specA.south east, anchor=south east, minimum width=1cm] {\tiny \change{Spectral}};
            \node[above=0pt of specA.south west, anchor=south west, minimum width=1cm] {\tiny \change{XYZU}};
            \draw (0.0,0.0) rectangle ++(2cm,2cm);
        \end{scope}
        % Inset rendering B (model)
        \begin{scope}[xshift=-5.7cm,yshift=0.38cm]
            \node[anchor=south west, inner sep=0pt] (specB) at (0.0, 0.0) { \includegraphics[height=2cm]{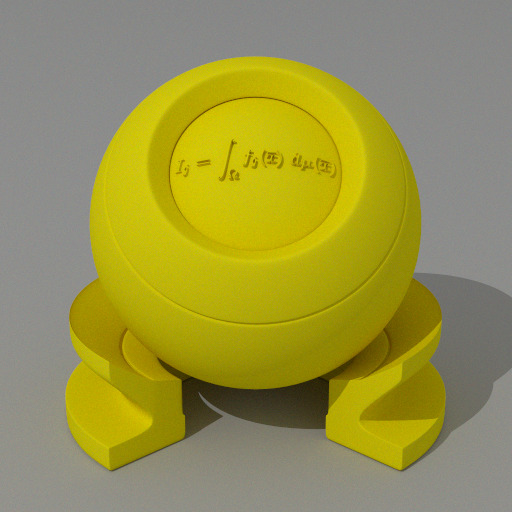} };
            \begin{scope}
                \clip(0,0) rectangle (1,2);
                \node[anchor=south west, inner sep=0pt] (reducedB) at (0.0, 0.0) { \includegraphics[height=2cm]{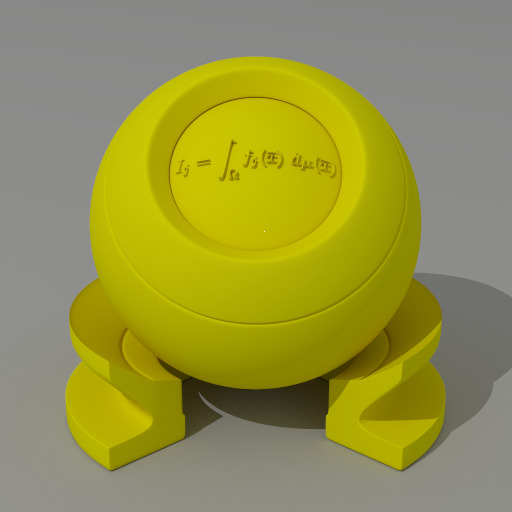} };
                \draw[color=black]  (1,0) -- (1,2);
            \end{scope}
            \node[above=-1pt of specB.south east, anchor=south east, minimum width=1cm] {\tiny \change{Spectral}};
            \node[above=0pt of specB.south west, anchor=south west, minimum width=1cm] {\tiny \change{XYZU}};
            \draw (0.0,0.0) rectangle ++(2cm,2cm);
            \node[below=2pt of specB] {\footnotesize \change{Renderings}};
        \end{scope}
        %
        %
        % Right insets
        \draw[thick, color=red, fill] (3.5,2) circle (1pt);
        \draw[thick, color=red] (3.5,2) circle (1pt) -- (3.5,0.7) -- (10,0.7);
        \draw[thick, color=red, fill] (6.2,1.5)  circle (1pt) -- (6.2,0.7);
        \begin{scope}[xshift=0.512\textwidth,yshift=8pt]
            \node[anchor=south west, inner sep=0pt] (spec_P) at (0,0.099) {
            \includegraphics[height=2cm]{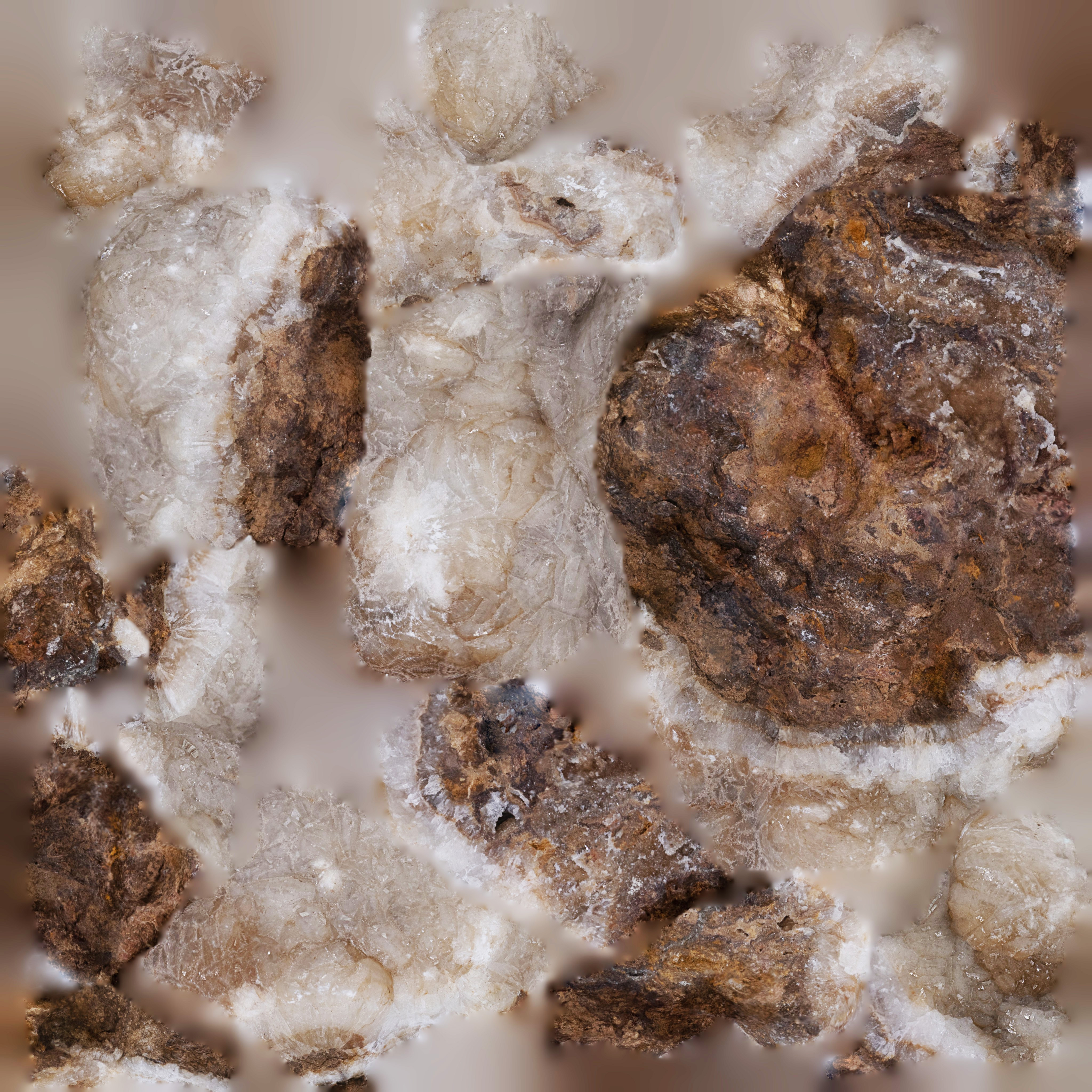} };
            \draw (0.0,0.1) rectangle ++(2cm,2cm);
            \node[anchor=south west, inner sep=0pt] (spec_P) at (1,0.0) {
            \includegraphics[height=2cm]{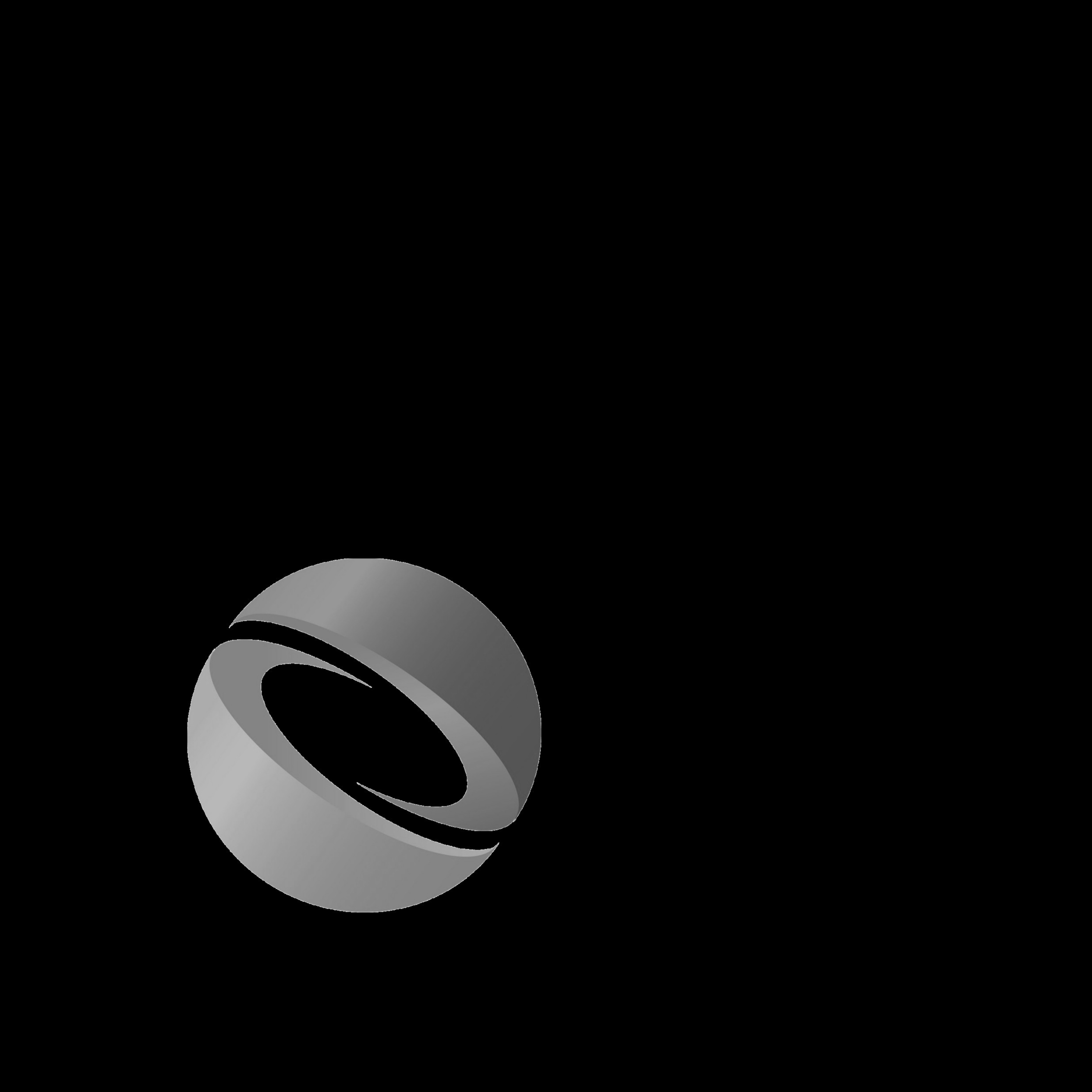} };
            \draw (1,0.0) rectangle ++(2cm,2cm);
            \node at (1.5cm,-5pt) {\footnotesize Albedo \& Fluorescence strength};
        \end{scope}
        \draw[thick, color=blue, fill] (5.2,4) circle (1pt);
        \draw[thick, color=blue] (5.2,4) -- ++(1.3,0);
        \begin{scope}[xshift=0.36\textwidth,yshift=3.4cm]
            \begin{axis}[width=4.0cm, height=3.0cm, xmin=300, xmax=750, ymin=0, x tick label style={font=\tiny}, y tick label style={font=\tiny},grid, yshift=0.1cm,axis background/.style={fill=white, opacity=0.6}]
            \addplot[mark=none, color=blue, very thick] table [
                x index=0,
                y index=1,
                col sep=comma,
                ] {./Figures/Teaser/450nm.csv};
            \end{axis}
            % \draw (0.0,0.1) rectangle ++(3cm,2cm);
            \node at (1.25cm,-8pt) {\tiny Light spectrum};
        \end{scope}
        \begin{scope}[xshift=0.6\textwidth,yshift=2.96cm]
            \node[anchor=south west, inner sep=0pt] (spec_P) at (0,0.099) { \includegraphics[height=2cm]{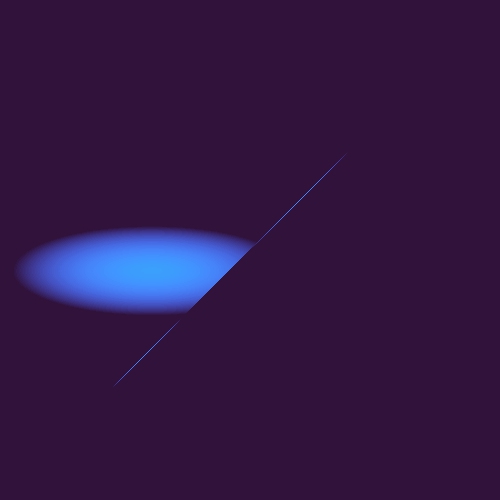} };
            \draw (0.0cm,0.1) rectangle ++(2cm,2cm);
            \node at (1.0cm,-5pt) {\footnotesize Our model};
        \end{scope}
        \begin{scope}[xshift=0.545\textwidth, yshift=4.0cm]
            \node[anchor=north east, inner sep=0pt] (red_P) at (0.5, 0.5) { \includegraphics[height=1cm]{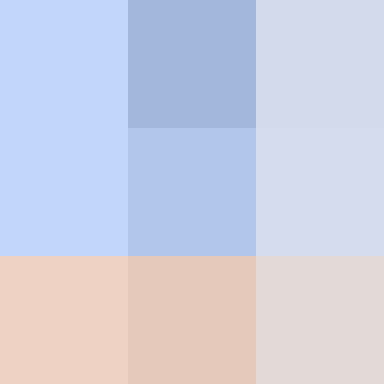} };
            \foreach \v in {1,2} {
            \draw[thin] ($(red_P.north west)+(\v/3, 0)$) -- ($(red_P.south west)+(\v/3, 0)$);
            \draw[thin] ($(red_P.north west)-(0, \v/3)$) -- ($(red_P.north east)-(0, \v/3)$);
            }
            \draw (red_P.south west) rectangle (red_P.north east);
            \node at (0,-0.7) (ana) { {\tiny Analytical} };
            \node[below=-2pt of ana.south,inner sep=0pt] { {\tiny reduction} };
        \end{scope}
        \draw[thick, color=red] (10.70,0.60) rectangle ++(0.1, 0.1);
        % \draw[thick, color=red, fill] (12.0,1) circle (1pt);
        \draw[thick, color=red, ->] (10.80,0.65) -- ++(2.5,0) -- ++(0,3.3) -- ++(-0.4,0);
        \draw[thick, color=red, fill] (10.75,4) circle (1pt);
        \draw[thick, color=red, ->] (10.75,4) -- ++(-0.3,0);
        \node[white, anchor=south west, xshift=3pt] at (teaser.south west) {\textbf{XYZ Rendering}};
        \end{scope}
    \end{tikzpicture}}
    \vspace{-10pt}
    \caption{
    We introduce a material model for the rendering and editing of fluorescence.
	Our approach relies on a Gaussian approximation of the fluorescence component of the bi-spectral reradiation, which reproduces measured data accurately thanks to an energy conserving decomposition.
	Our model allows for analytical spectral integration, making it compatible with non-spectral and real-time rendering, and eases the editing of fluorescent material textures.
    }
    \label{fig:teaser}
\end{teaserfigure}
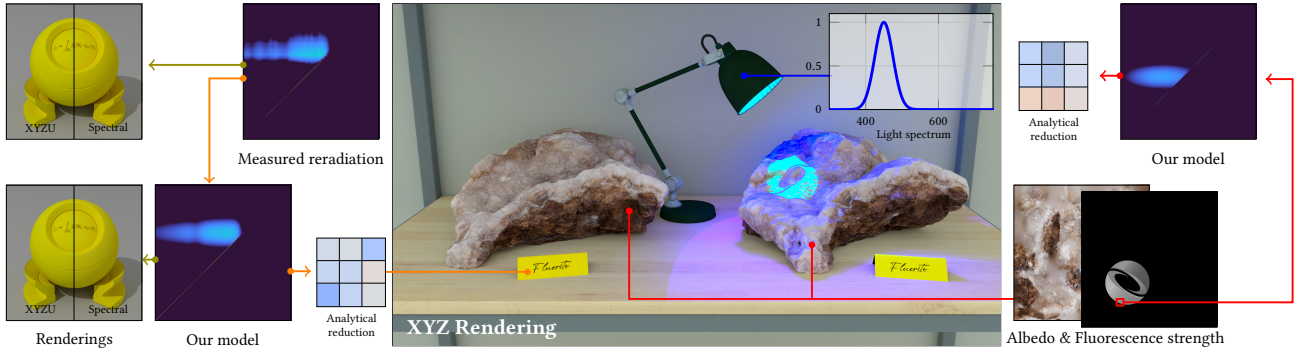

\begin{abstract}
Fluorescent materials are characterized by a spectral reradiation toward longer wavelengths.
Recent work~\cite{Fichet2024} has shown that the rendering of fluorescence in a non-spectral engine is possible through the use of appropriate reduced reradiation matrices.
But the approach has limited expressivity, as it requires the storage of one reduced matrix per fluorescent material, and only works with measured fluorescent assets.

In this work, we introduce an analytical approach to the editing and rendering of fluorescence in a non-spectral engine.
It is based on a decomposition of the reduced reradiation matrix, and an analytically-integrable Gaussian-based model of the fluorescent component.
The model reproduces the appearance of fluorescent materials accurately, especially with the addition of a UV basis.
Most importantly, it grants variations of fluorescent material parameters in real-time, either for the editing of fluorescent materials, or for the dynamic spatial variation of fluorescence properties across object surfaces.
A simplified one-Gaussian fluorescence model even allows for the artist-friendly creation of plausible fluorescent materials from scratch, requiring only a few reflectance colors as input.
\end{abstract}




% \keywords{Material Models, Fluorescence, Appearance Editing, Physically-based Rendering.}

\maketitle

\section{Introduction \& related work}
\label{sec:intro}

\paragraph{Context}
\change{Fluorescence is widespread in plants~\cite{Zenchyzen2024}, animals~\cite{Travouillon2023} and minerals~\cite{Robbins1983}.}
\change{It} is visually characterized by an apparent change of reflectance color depending on the environment lighting.
This may result in a change of hue (e.g., for a same material when going from daylight to indoor lighting), and/or in an increase of reflected intensity compared to the surrounding (e.g., for materials that reflect non-visible ultraviolet light in the human visible range).
Formally, fluorescence occurs when light of an incoming wavelength $\lambda_i$ is reradiated at another outgoing wavelength $\lambda_o$~\cite{Glassner95}.
This is characterized by spectral reradiation matrices $\mathcal{P}(\lambda_i, \lambda_o)$.
Such matrices may be assumed triangular (i.e., $\mathcal{P}(\lambda_i, \lambda_o) = 0 \ \forall \lambda_o < \lambda_i$) since in the vast majority of cases, reradiation only occurs toward smaller energies, or larger wavelengths.
In diffuse fluorescent materials, $\mathcal{P}$ replaces and augments the conventional spectral albedo, which then corresponds to values along its diagonal: $\rho(\lambda) = \mathcal{P}(\lambda, \lambda)$.

\vspace{-10pt}
\paragraph{Related work}
In Computer Graphics and Optics, most previous methods (e.g.,\change{~\cite{Jung18,Wilkie2001,Wilkie2006,Hua23,Iser2023}}) render fluorescence using a single measured reradiation matrix $\mathcal{P}$ per material, assuming fluorescence and scattering properties are uncorrelated.
There are two notable exceptions in the literature.
The work of Hullin et al.~\shortcite{Hullin10} projects measured bispectral BSDFs (Bidirectional Scattering Distribution Functions) on an angularly-dependent linear combination of a few reradiation matrices.
Unfortunately, measured data for that model have not been made available.
The method of Benamira et al.~\shortcite{Benamira23} suggests a fully analytical bispectral BSDF inspired from Quantum Dots (nano-particles distributed on a microsurface).
Such a structural assumption is very specific and unlikely to correspond to common fluorescent materials; in particular, the angularly-varying reradiation matrices produced by their model differ significantly from the reradiation matrices measured in diffuse fluorescent materials.
In the absence of reliable angularly-varying references for common materials, we chose to focus in this paper on \emph{diffuse} fluorescence.

\paragraph{Problem}
All of the aforementioned methods require a spectral rendering engine.
The rendering of fluorescence in non-spectral engines was first suggested by Hulin et al.~\shortcite{Hullin10}.
It has been recently improved by Fichet et al.~\shortcite{Fichet2024}, who construct a reduced reradiation matrix $\mathrm{P}$ from the spectral reradiation matrix $\mathcal{P}$ using:
\begin{equation}
\label{eqn:reduced-rerad}
  \mathrm{P} = S^{\top} \mathcal{P} \tilde{S},
\end{equation}
where $S$ is the $N \times K$ matrix of sensitivity functions ($K$ functions discretized over $N$ wavelengths), and $\tilde{S} = S(S^{\top}S)^{-1}$ is the $N \times K$ matrix of \emph{dual} sensitivity functions (i.e., verifying $S^{\top} \tilde{S}=I_K$, the $K \times K$ identity matrix).
An outgoing color $\mathbf{c}_o$ is then approximated by a matrix-vector product with an incoming color $\mathbf{c}_i$:
\begin{equation}
\label{eqn:reduced-out-col}
   \mathbf{c}_o \approx \mathrm{P} \times \mathbf{c}_i.
\end{equation}
The method yields satisfying approximations when rendering measured fluorescent materials using the CIE XYZ Color Matching Functions as bases for $S$ (i.e., using $K=3$), and is even more accurate with the addition of an UV basis function (i.e., using $K=4$).

However, the method also presents two shortcomings: 1) for each new fluorescent material in a 3D scene, one must compute and store an additional reduced reradiation matrix $\mathrm{P}$; 2) the method takes as input \emph{measured} spectral reradiation matrices $\mathcal{P}$, for which there exist only a few databases~\cite{Gonzalez00,Iser2023}.
As a result, the approach is limited in terms of expressivity: it does not provide an artist-friendly fluorescence creation or editing solution; and spatially-varying fluorescence is restricted to interpolations of measured materials, with no independent control over reflectance and fluorescence components.

\paragraph{Contributions}
In this paper, we address these two limitations through the introduction of an analytical fluorescent material model tailored to non-spectral rendering.
In Section~\ref{sec:decomp} we first explain how we decompose a reduced reradiation matrix into fluorescent and non-fluorescent components with explicit constraints on energy conservation.
The fluorescent component is then modeled in Section~\ref{sec:model} as a sum of Gaussians, similarly to previous work~\cite{Iser2023}.
Our main contribution then lies in an analytical method to integrate this model over Gaussian-based sensitivity functions, granting on-the-fly editing of fluorescent material properties \change{similarly to previous work on iridescence (e.g.,~\cite{belcour2017,Fourneau2024})}.
Lastly, we show in Section~\ref{sec:edit} that a simplified, one-Gaussian version of the model allows for the \emph{creation} of spectral fluorescent materials with artist-friendly controls.

\section{Reduced reradiation matrix decomposition}
\label{sec:decomp}

Before we introduce a fluorescence model that grants analytic, on-the-fly integration in the next section, we first discuss the decomposition of spectral and reduced reradiation matrices.

\paragraph{Spectral decomposition}
Spectral reradiation may be additively decomposed into reflectance and fluorescent components:
\begin{equation}
\label{eqn:spec-decomp}
  \mathcal{P}(\lambda_i, \lambda_o) = \mathcal{R}(\lambda_i, \lambda_o) + \mathcal{F}(\lambda_i, \lambda_o),
\end{equation}
where $\mathcal{R}(\lambda_i, \lambda_o) = \delta(\lambda_o-\lambda_i) \rho(\lambda_o)$ is a diagonal spectral reflectance matrix with $\delta$ the Kronecker delta, and $\mathcal{F}(\lambda_i, \lambda_o)$ holds the remaining off-diagonal (fluorescent) spectral reradiation.

In contrast to methods that rely on measured fluorescence data that are naturally energy-conserving, our model has to remain physically-plausible when fluorescence properties are modified by artists.
The energy conservation constraint is defined with respect to spectral reradiation by:
\begin{equation}
   \label{eqn:energy-conservation}
   \int_0^{\infty} \mathcal{P}(\lambda_i, \lambda_o) d\lambda_o \le 1 \ \mbox{for all} \ \lambda_i.
\end{equation}

Enforcing such a constraint is problematic in our context as it depends both on the reflectance $\mathcal{R}$ and fluorescence  $\mathcal{F}$ in Equation~\ref{eqn:spec-decomp}.
In order to grant independent control over fluorescence while conserving energy (assuming $\rho(\lambda_i)\le1 \, \forall \lambda_i$), we introduce a normalized fluorescent term $\mathcal{\bar{F}}$ verifying:%, defined with respect to $\mathcal{F}$ by:
\begin{equation}
    \label{eqn:normalized-fluo}
    \mathcal{F}(\lambda_i, \lambda_o) = \mathcal{\bar{F}}(\lambda_i, \lambda_o) (1-\mathcal{R}(\lambda_i,\lambda_o)).
\end{equation}

\begin{figure}[t]
    \centering
    \resizebox{\linewidth}{!}{\input{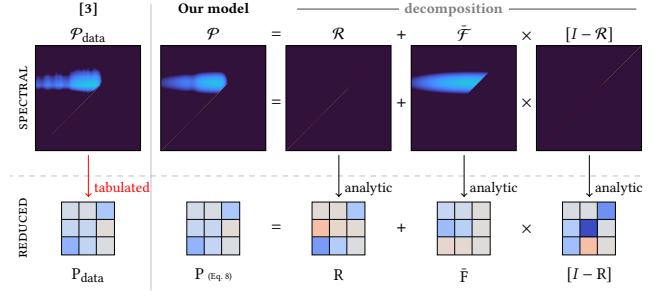}}
    \vspace{-17pt}
    \caption{\textbf{Model overview.} With our decomposition, our model faithfully reproduces a measured reradiation matrix (HERPIPIN) with as low as one Gaussian for the reradiation. Our matrix decomposition works both on full spectral reradiation matrices (Equation~\ref{eqn:normalized-fluo}) and with reduced matrices (Equation~\ref{eqn:reduced-decomp}). This ensures energy conservation during the edition without the need to recompute the reduction for all components.
    \label{fig:model-overview}
    \vspace{-10pt}
    }
\end{figure}

Using Equation~\ref{eqn:normalized-fluo} in Equation~\ref{eqn:spec-decomp} and performing the integration of Equation~\ref{eqn:energy-conservation} yields the following energy conservation constraint:
\begin{equation}
   \label{eqn:energy-conservation2}
   \int_0^{\infty} \mathcal{\bar{F}}(\lambda_i, \lambda_o) d\lambda_o \le 1 \ \mbox{for all} \ \lambda_i,
\end{equation}
which may be enforced regardless of reflectance, as desired.
The decomposition of the spectral reradiation in terms of reflectance and normalized fluorescence is illustrated in the top row of Figure~\ref{fig:model-overview}.
In spite of the simplicity of the normalized fluorescence component $\mathcal{\bar{F}}$, the measured reradiation is accurately reproduced, thanks to multiplication by the $[1-\mathcal{R}]$ term.

\paragraph{Reduced decomposition}
The reduced reradiation matrix is decomposed by applying Equation~\ref{eqn:reduced-rerad} to each component of Equation~\ref{eqn:spec-decomp}:
\begin{equation}
\label{eqn:reduced-decomp}
    \mathrm{P} = \mathrm{R} + \mathrm{F} = S^T \mathcal{R} \tilde{S} + S^T \mathcal{F} \tilde{S},
\end{equation}
with $\mathrm{R}$ and $\mathrm{F}$ denoting reduced reflectance and fluorescence matrices respectively.
The latter may be re-expressed in terms of the reduced normalized fluorescence matrix $\mathrm{\bar{F}}=S^{\top} \mathcal{\bar{F}}\tilde{S}$ similarly to Equation~\ref{eqn:normalized-fluo}:
\begin{equation}
    \label{eqn:reduced-normalized-fluo}
    \mathrm{F} = \mathrm{\bar{F}} \left[I-\mathrm{R}\right].
\end{equation}

\begin{proof}
We start from $\mathrm{F}=S^{\top} \mathcal{F} \tilde{S} = S^{\top} \mathcal{\bar{F}}\tilde{S} - S^{\top} \mathcal{\bar{F}} \mathcal{R} \tilde{S}$ using Equation~\ref{eqn:normalized-fluo}.
The first term corresponds to $\mathrm{\bar{F}}$.
We next observe that $\tilde{S}S^{\top} = (SS^{\top})^{-1} SS^{\top} = I$ from the definition of dual sensitivity functions.
We may thus write $S^{\top} \mathcal{\bar{F}} \mathcal{R} \tilde{S} = S^{\top} \mathcal{\bar{F}} \tilde{S} S^{\top} \mathcal{R} \tilde{S} = \mathrm{\bar{F}}\mathrm{R}$ for the second term.
As a result, we obtain $\mathrm{F} = \mathrm{\bar{F}} - \mathrm{\bar{F}}\mathrm{R} = \mathrm{\bar{F}} \left[I-\mathrm{R}\right]$.
\end{proof}

The reduced reradiation matrix of Equation~\ref{eqn:reduced-decomp} then becomes:
\begin{equation}
    \label{eqn:reduced-decomp2}
    \mathrm{P} = \mathrm{R} + \mathrm{\bar{F}} \left[I-\mathrm{R}\right],
\end{equation}
which is illustrated in the bottom row of Figure~\ref{fig:model-overview}, with positive (resp. negative) coefficients shown in blue (resp. red).

The outgoing color $\mathbf{c}_o$ may itself be decomposed by plugging the reduced reradiation matrix of Equation~\ref{eqn:reduced-decomp2} into Equation~\ref{eqn:reduced-out-col}:
\begin{equation}
    \label{eqn:reduced-out-col2}
    \mathbf{c}_o = \mathbf{c}_{o,r} + \mathbf{c}_{o,f} \approx \mathrm{R} \mathbf{c}_i + \mathrm{\bar{F}} \left[I - \mathrm{R}\right] \mathbf{c}_i,
\end{equation}
with $\mathbf{c}_{o,r}$ and $\mathbf{c}_{o,f}$ denoting reflected and fluorescent outgoing colors respectively.
The term $\left[I - \mathrm{R}\right] \mathbf{c}_i$ may then be interpreted as the incoming color available to fluorescent reradiation.

\section{Analytic fluorescence model}
\label{sec:model}

\begin{figure}[t]
    \centering
    \resizebox{\linewidth}{!}{\input{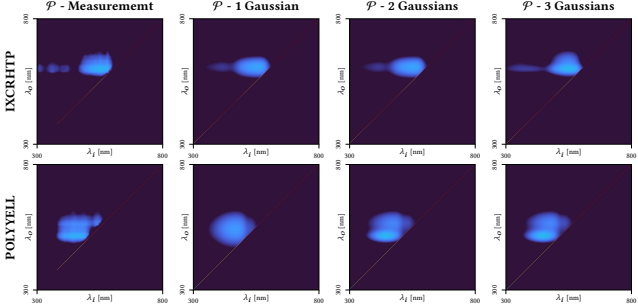}}
    \vspace{-23pt}
    \caption{\textbf{Fitting of measured materials with our model.}
    Our reradiation rescaling approach not only ensures energy conservation but also better reproduces the shape of the reradiative part for most of the measurements, even when using a single Gaussian.
    }
    \vspace{-15pt}
    \label{fig:fitting-measurements}
\end{figure}

Our focus in this paper is on the editing and rendering of fluorescence.
In the following, we show that a Gaussian-based model of the normalized spectral fluorescence  $\mathcal{\bar{F}}$ has two key advantages: a small number of Gaussians is sufficient to accurately reproduce the reflected colors of measured fluorescent materials under various illuminants (Section~\ref{sec:gauss-model}); and it may be analytically integrated over arbitrary Gaussian-based sensitivity functions to yield a reduced normalized fluorescence matrix $\mathrm{\bar{F}}$ for use in non-spectral rendering (Section~\ref{sec:gauss-int}).
For completeness, we provide Gaussian fits of XYZ sensitivity functions, and suggest an optional Gaussian basis for ultraviolet (UV) reradiation (Section~\ref{sec:gauss-bases}).

\subsection{Gaussian-based normalized fluorescence}
\label{sec:gauss-model}

% Our goal is now to find a spectral model for the fluorescent component that may be integrated on the fly to yield $\mathrm{\bar{F}}$.
Iser et al.~\shortcite{Iser2023} have shown that an axis-aligned Gaussian Mixture Model (GMM) of the fluorescent component $\mathcal{F}$ may accurately fit spectral measured data.
We take inspiration from their work, with a few key differences: we use a GMM model to fit the \emph{normalized} fluorescent component $\mathcal{\bar{F}}$; and we evaluate its accuracy in terms of \emph{reradiated colors} under various illuminants.

% \paragraph{Spectral fluorescence model}
Formally, we assume that the normalized spectral fluorescence $\mathcal{\bar{F}}$ may be approximated by a combination of $Q$ \emph{axis-aligned} 2D Gaussian functions:
% %
% \begin{equation}
% \label{eqn:fluo_coeff}
%     \mathcal{\bar{F}}(\lambda_i,\lambda_o) =  \sum_{q=1}^Q \alpha_q \mathcal{A}_q(\lambda_i) \mathcal{E}_q(\lambda_o)
%     \mathcal{H}(\lambda_o-\lambda_i),
% \end{equation}
% %
%
\begin{equation}
\label{eqn:fluo-model}
    \mathcal{\bar{F}}(\boldsymbol{\lambda}) \approx  \sum_{q=1}^Q \mathcal{G}_q(\boldsymbol{\lambda}; \alpha_q, \boldsymbol{\mu}_q, \Sigma_q)
    \mathcal{H}(\lambda_o-\lambda_i) \;\; \mbox{with}\ \boldsymbol{\lambda} = \begin{bmatrix}\lambda_i \\ \lambda_o \end{bmatrix},
\end{equation}
where the $\mathcal{G}_q$ are 2D Gaussians of intensity $\alpha_q$, mean $\pmb{\mu}_q$ and \emph{diagonal} co-variance matrix $\Sigma_q$, while $\mathcal{H}$ is a heaviside function used to enforce reradiation toward smaller energies (see Figure~\ref{fig:model-overview}). As shown in Figure~\ref{fig:fitting-measurements}, this decomposition reproduces details of the reradiation, even for a small $Q = 1$ number of Gaussians.

In order to evaluate the accuracy of this spectral fluorescence model, we study how colors reflected by measured materials under various spectral illuminants differ from the ground-truth when using our GMM approximation.
Formally, we compute:
\begin{equation}
    \label{eqn:spectral-out-col}
    \mathbf{c}_o = \iint \mathcal{P}(\lambda_i, \lambda_o) \mathcal{L}(\lambda_i) \mathbf{s}(\lambda_o) d\lambda_i d\lambda_o,
\end{equation}
where $\mathbf{s}=[\bar{x}, \bar{y}, \bar{z}]^{\top}$ is the vector of XYZ CMFs, $\mathcal{L}$ is a spectral illuminant, and $\mathcal{P} = \mathcal{R} + (1-\mathcal{R}) \mathcal{\bar{F}}$ with $\mathcal{\bar{F}}$ coming either from the measured material or from the approximation of Equation~\ref{eqn:fluo-model}.
Note that the spectral reflectance $\mathcal{R}$ remains unaltered.

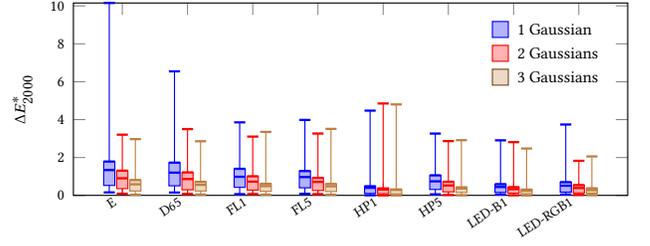
\begin{figure}[t]
    \centering
    \resizebox{\linewidth}{!}{\pgfplotsset{
    box plot/.style={
        /pgfplots/.cd,
        black,
        only marks,
        mark=-,
        mark size=\pgfkeysvalueof{/pgfplots/box plot width},
        /pgfplots/error bars/y dir=plus,
        /pgfplots/error bars/y explicit,
        /pgfplots/table/x index=\pgfkeysvalueof{/pgfplots/box plot x index},
    },
    box plot box/.style={
        /pgfplots/error bars/draw error bar/.code 2 args={%
            \draw  ##1 -- ++(\pgfkeysvalueof{/pgfplots/box plot width},0pt) |- ##2 -- ++(-\pgfkeysvalueof{/pgfplots/box plot width},0pt) |- ##1 -- cycle;
        },
        /pgfplots/table/.cd,
        y index=\pgfkeysvalueof{/pgfplots/box plot box top index},
        y error expr={
            \thisrowno{\pgfkeysvalueof{/pgfplots/box plot box bottom index}}
            - \thisrowno{\pgfkeysvalueof{/pgfplots/box plot box top index}}
        },
        /pgfplots/box plot
    },
    box plot top whisker/.style={
        /pgfplots/error bars/draw error bar/.code 2 args={%
            \pgfkeysgetvalue{/pgfplots/error bars/error mark}%
            {\pgfplotserrorbarsmark}%
            \pgfkeysgetvalue{/pgfplots/error bars/error mark options}%
            {\pgfplotserrorbarsmarkopts}%
            \path ##1 -- ##2;
        },
        /pgfplots/table/.cd,
        y index=\pgfkeysvalueof{/pgfplots/box plot whisker top index},
        y error expr={
            \thisrowno{\pgfkeysvalueof{/pgfplots/box plot box top index}}
            - \thisrowno{\pgfkeysvalueof{/pgfplots/box plot whisker top index}}
        },
        /pgfplots/box plot
    },
    box plot bottom whisker/.style={
        /pgfplots/error bars/draw error bar/.code 2 args={%
            \pgfkeysgetvalue{/pgfplots/error bars/error mark}%
            {\pgfplotserrorbarsmark}%
            \pgfkeysgetvalue{/pgfplots/error bars/error mark options}%
            {\pgfplotserrorbarsmarkopts}%
            \path ##1 -- ##2;
        },
        /pgfplots/table/.cd,
        y index=\pgfkeysvalueof{/pgfplots/box plot whisker bottom index},
        y error expr={
            \thisrowno{\pgfkeysvalueof{/pgfplots/box plot box bottom index}}
            - \thisrowno{\pgfkeysvalueof{/pgfplots/box plot whisker bottom index}}
        },
        /pgfplots/box plot
    },
    box plot median/.style={
        /pgfplots/box plot,
        /pgfplots/table/y index=\pgfkeysvalueof{/pgfplots/box plot median index}
    },
    box plot width/.initial=1em,
    box plot x index/.initial=0,
    box plot median index/.initial=1,
    box plot box top index/.initial=2,
    box plot box bottom index/.initial=3,
    box plot whisker top index/.initial=4,
    box plot whisker bottom index/.initial=5,
}

\newcommand{\boxplot}[2][]{
    \addplot [box plot median,#1] table[col sep=comma] {#2};
    \addplot [forget plot, box plot box,#1] table[col sep=comma] {#2};
    \addplot [forget plot, box plot top whisker,#1] table[col sep=comma] {#2};
    \addplot [forget plot, box plot bottom whisker,#1] table[col sep=comma] {#2};
}

\begin{tikzpicture}
    \begin{axis} [
        name=moustache,
        box plot width=0.7mm,
        enlargelimits=false,
        width=\linewidth,
        height=4cm,
        xmin=0.25,
        xmax=8.75,
        xtick=data,
        xtick={1,...,8},
        xticklabels from table={Figures/DE/data/illu.csv}{Illuminant},
        x tick label style={rotate=33,anchor=east,font=\tiny},
        y tick label style={font=\tiny},
        ylabel={\tiny{$\Delta E_{2000}^*$}},
        ylabel near ticks,
        ]
        \boxplot [
            thick,
            blue,
            fill=blue!30!white,
            forget plot,
            box plot whisker bottom index=2,
            box plot box bottom index=3,
            box plot median index=1,
            box plot box top index=5,
            box plot whisker top index=6
        ] {Figures/DE/data/with_diag/de_patches_1-norm.csv}

        \boxplot [
            thick,
            red,
            fill=red!30!white,
            forget plot,
            box plot whisker bottom index=2,
            box plot box bottom index=3,
            box plot median index=1,
            box plot box top index=5,
            box plot whisker top index=6
        ] {Figures/DE/data/with_diag/de_patches_2-norm.csv}

        \boxplot [
            thick,
            brown,
            fill=brown!30!white,
            forget plot,
            box plot whisker bottom index=2,
            box plot box bottom index=3,
            box plot median index=1,
            box plot box top index=5,
            box plot whisker top index=6
        ] {Figures/DE/data/with_diag/de_patches_3-norm.csv}
    \end{axis}

    \node[xshift=-1.5cm, yshift=-0.15cm, anchor=north west] (key_g1) at (moustache.north east) {\scriptsize{1 Gaussian}};
    \node[xshift=-1.5cm, yshift=-0.45cm, anchor=north west] (key_g2) at (moustache.north east) {\scriptsize{2 Gaussians}};
    \node[xshift=-1.5cm, yshift=-0.75cm, anchor=north west] (key_g3) at (moustache.north east) {\scriptsize{3 Gaussians}};

    \draw[blue,fill=blue!30!white]            ($(key_g1.west) - (.1cm, 0) + (.1cm, .1cm)$) rectangle ($(key_g1.west) - (.1cm, 0) - (.1cm, .1cm)$);
    \draw[red,fill=red!30!white]              ($(key_g2.west) - (.1cm, 0) + (.1cm, .1cm)$) rectangle ($(key_g2.west) - (.1cm, 0) - (.1cm, .1cm)$);
    \draw[brown!60!black,fill=brown!30!white] ($(key_g3.west) - (.1cm, 0) + (.1cm, .1cm)$) rectangle ($(key_g3.west) - (.1cm, 0) - (.1cm, .1cm)$);
\end{tikzpicture}}
    \vspace{-23pt}
    \caption{\textbf{Accuracy of our model.}
    We compare the reradiation of standard illuminants on the measured reradiation matrices from Gonzalez and Fairchild~\shortcite{Gonzalez00} with the fitting of our model in the spectral domain. We use a different number of Gaussians for the reradiation and six Gaussians for the diagonal. The whiskers bottom and top represent resp. the minimum and the maximum, the box bottom and top represent resp. the first and third quartiles and the central line represent the average $\Delta E^*_{2000}$ distribution.
    %\alban{J'utilise un fit Gaussien sur la diag, cela peut ajouter de l'erreur (deux materiaux problématiques car refl > 1 clampée). Je devrais peut-être remplacer par la diagonale originale et le fit que par la RR.}
    \label{fig:accuracy-wrt-nb-gaussians}
    \vspace{-10pt}
    }
\end{figure}

As shown in Figure~\ref{fig:accuracy-wrt-nb-gaussians}, using $Q=1$ Gaussian function is enough to capture the effect of fluorescence on reflected colors of the measured materials we have tested.
We obtain almost unnoticeable differences ($ < 2$ in $\Delta E^*_{2000}$) between most of reflected colors obtained with the ground truth and the GMM approximations for a large number of illuminants.
We refer the reader to Supplemental materials for a thorough evaluation on 143 measured materials from \cite{Gonzalez00}.

% through the method of Fichet et al.~\shortcite{Fichet2024} to the ground truth colors, using two variants: either directly using the reduced reradiation matrix $\mathrm{P}$ in Equation~\ref{eqn:out-col} (i.e., numerical reduction); or using the reduced normalized fluorescence matrix $\mathrm{\bar{F}}$ in Equation~\ref{eqn:out-col-R} whose coefficients are computed through Equation~\ref{eqn:reduced-fluo-coeff} with the GMM of Equation~\ref{eqn:fluo-model} (i.e., Gaussian reduction).
% In the latter case, the reflectance matrix is computed from measured data (i.e., $\mathrm{R} = S^{\top} \mathcal{R} \tilde{S}$).
% These experiments show that Gaussian reduction using $Q=2$ performs nearly as good as numerical reduction on measured materials containing a single fluophore.

\subsection{Gaussian-based spectral integration}
\label{sec:gauss-int}

For non-spectral rendering, we rely on the reduced reradiation approach of Fichet et al.~\shortcite{Fichet2024}, where the outgoing color $\mathbf{c}_o$ is approximated by the matrix-vector product of Equation~\ref{eqn:reduced-out-col}.
In our case the reduced reradiation matrix $\mathrm{P}$ is further decomposed according to Equation~\ref{eqn:reduced-decomp2}, yielding Equation~\ref{eqn:reduced-out-col2} for $\mathbf{c}_o$.
Since our main objective is the control of fluorescence, we next assume that the $\mathrm{R}$ matrix is known, and focus on the $\mathrm{\bar{F}}$ matrix.
Its coefficients are given by:
\begin{equation}
    \label{eqn:reduced-fluo-coeff}
    \mathrm{\bar{F}}_{jk} =
    \iint
    \mathcal{\bar{F}}(\lambda_i,\lambda_o) \tilde{s}_j(\lambda_i) s_k(\lambda_o) d\lambda_i d\lambda_o,
\end{equation}
where $s_k$ ($k \in [1,K]$) and $\tilde{s}_j$ ($j \in [1,K]$) are \emph{arbitrary} sensitivity and dual sensitivity functions respectively.
When these functions are discretized, they correspond to the columns of $S$ and $\tilde{S}$ respectively.

We now show that Equation~\ref{eqn:reduced-fluo-coeff} may be integrated analytically provided sensitivity functions are given in Gaussian form.
Specific examples of such Gaussian bases will be given later, in Section~\ref{sec:gauss-bases}.

\paragraph{Gaussian sensitivity functions}
A first issue is that dual sensitivity functions are usually not Gaussian-shaped.
Fortunately, since $\tilde{S} S^{\top} = I$ by design, we have $\tilde{S} S^{\top} S = S$ and thus $\tilde{S}= S (S^{\top}S)^{-1}$. Hence, dual sensitivity functions can be expressed as weighted sums of the sensitivity functions.
We then obtain:
\begin{equation}
    \label{eqn:corrected-fluo}
    \mathrm{\bar{F}}
    % = S^{\top} \mathcal{\bar{F}} \tilde{S}
    = S^{\top} \mathcal{\bar{F}} S (S^{\top}S)^{-1}.
\end{equation}
Intuitively, this means that \textit{for Gaussian sensitivity functions, dual sensitivity functions are defined as a weighted sum of Gaussians}.

% where $C=(S^{\top}S)^{-1}$ is a $K \times K$ correction matrix.

% \begin{proof}
%     Since $\tilde{S} S^{\top} = I$ by design, we have $\tilde{S} S^{\top} S = S$ and thus $\tilde{S}= S (S^{\top}S)^{-1} = SC$.
%     Pluging it in $\mathrm{\bar{F}} = S^{\top} \mathcal{\bar{F}}\tilde{S}$ yields Equation~\ref{eqn:corrected-fluo}.
% \end{proof}

% Indeed, since $\tilde{S} S^{\top} = I$ by construction, we have $\tilde{S} S^{\top} S = S$ and thus $\tilde{S}= S (S^{\top}S)^{-1}$; we may then rewrite the normalized reduced fluorescence matrix as $\mathrm{\bar{F}} = S^{\top} \mathcal{\bar{F}} S C$ where $C=(S^{\top}S)^{-1}$ is a $K \times K$ correction matrix that is applied to attenuated incoming color coefficients.
%
A second issue is that a single Gaussian may not be sufficient to approximate a sensitivity function.
Let's assume that $M \ge K$ Gaussians are required in the general case.
We then rewrite the matrix of sensitivity functions as $S \approx GT_G$, with $G$ a $N \times M$ matrix whose columns are discretized 1D Gaussian functions, and $T_G$ a $M \times K$ transfer matrix that linearly combines 1D Gaussians contributing to a same sensitivity function.

The reduced normalized fluorescence matrix may now be written:
\begin{equation}
    \mathrm{\bar{F}}
    % = T^{\top}G^{\top} \mathcal{\bar{F}} G T (S^{\top}S)^{-1}
    = T_G^{\top} \mathrm{\bar{F}^{\circ}} T_G C ,
\end{equation}
where we have introduced
$C=(S^{\top}S)^{-1}$, a $K \times K$ matrix that is precomputed once for a given choice of sensitivity functions; and
$\mathrm{\bar{F}^{\circ}} = G^{\top} \mathcal{\bar{F}} G$, a $M \times M$ matrix whose coefficients are given by:
\begin{equation}
\label{eqn:gaussian-fluo-coeff}
    \mathrm{\bar{F}^{\circ}}_{jk} =
    % \int_{\lambda_{\min}}^{\lambda_{\max}}
    % \int_{\lambda_{\min}}^{\lambda_{\max}}
    \iint
    \mathcal{\bar{F}}(\lambda_i,\lambda_o) g_j(\lambda_i) g_k(\lambda_o) d\lambda_i d\lambda_o,
\end{equation}
where $g_m$ ($m \in [1,M]$) are 1D Gaussian functions, which when discretized correpond to the columns of $G$.
Our next goal is to find a method to integrate Equation~\ref{eqn:gaussian-fluo-coeff} analytically.

\paragraph{Analytical integration}
For simplicity and without loss of generality, we will now consider a single 2D Gaussian ($Q=1$) in Equation~\ref{eqn:fluo-model}, dropping the $q$ subscript: $\mathcal{\bar{F}}(\lambda_i,\lambda_o) = \mathcal{G}(\pmb{\lambda}; \alpha, \pmb{\mu}, \Sigma) \mathcal{H}(\lambda_o-\lambda_i)$.

Equation~\ref{eqn:gaussian-fluo-coeff} may then be rewritten as:
\begin{equation}
\label{eqn:gaussian-fluo-coeff2}
    \mathrm{\bar{F}^{\circ}}_{jk} =
    % \int_{\lambda_{\min}}^{\lambda_{\max}}
    % \int_{\lambda_{\min}}^{\lambda_{\max}}
    \int
    \mathcal{G}_{jk}(\pmb{\lambda}; \alpha_{jk}, \pmb{\mu}_{jk}, \Sigma_{jk}) \, \mathcal{H}(\lambda_o-\lambda_i) \, d\pmb{\lambda},
\end{equation}
with $\mathcal{G}_{jk}$ a 2D Gaussian function defined by:
\begin{equation}
    \mathcal{G}_{jk}(\pmb{\lambda}; \alpha_{jk}, \pmb{\mu}_{jk}, \Sigma_{jk}) = \mathcal{G}(\pmb{\lambda}; \alpha, \pmb{\mu}, \Sigma) \, g_j(\lambda_i) \, g_k(\lambda_o),
\end{equation}
where Gaussian product rules are applied to the $\lambda_i$ and $\lambda_o$ dimensions separately to yield $\alpha_{jk}$, $\pmb{\mu}_{jk}=[\mu_j, \mu_k]^{\top}$ and $\Sigma_{jk}=\mbox{diag}(\sigma^2_{j},\sigma^2_{k})$.

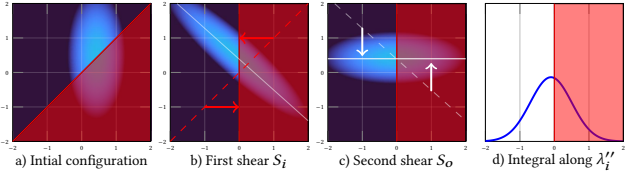
\begin{figure}[t]
    % \hspace{-10pt}
    \resizebox{\linewidth}{!}{
    \begin{tikzpicture}[font=\scriptsize]
        \newlength{\x}
        \setlength{\x}{0.23\linewidth}
        \newlength{\dx}
        \setlength{\dx}{8pt}

        % \begin{scope}
        %     \node[anchor=south west, inner sep=0pt] (img) { \includegraphics[width=\matW cm]{./Figures/AltDerivation/gaussian_01.jpg}};
        %     \begin{axis} [
        %         width=\matW cm,
        %         height=\matW cm,
        %         enlargelimits=false,
        %         ymin=-2, ymax=2,
        %         xmin=-2, xmax=2,
        %         grid style={line width=.1pt, draw=gray!10, opacity=0.2},
        %         tick label style = {font=\tiny, scale=0.5},
        %         grid
        %         ]
        %     \end{axis}
        %     \draw[fill=red, fill opacity=0.5] (img.south west) -- (img.north east) -- (img.south east) -- cycle;
        %     \draw[color=red] (img.south west) -- (img.north east);
        %     % \draw[color=red] (0,0) rectangle (\x,\x);
        %     % \draw[color=white, opacity=0.5] (0,0.5\x) -- +(\x,0);

        %     \node at (0.5\x,-8pt) { a) Intial configuration };
        % \end{scope}

        % Initial configuration
        \begin{scope}
            \node[anchor=south west, inner sep=0pt] (img) { \includegraphics[width=\x]{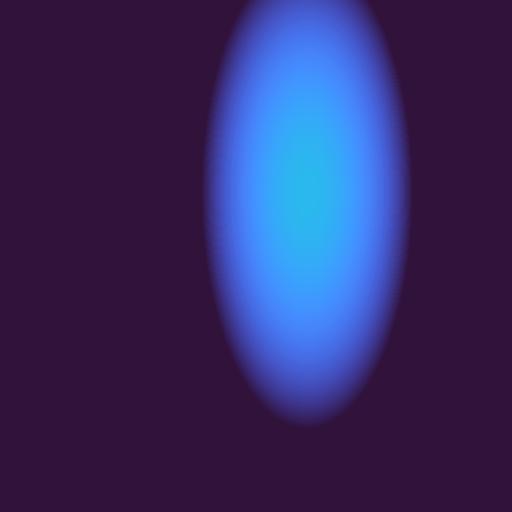}};
            \begin{axis} [
                width=3.55cm,
                height=3.55cm,
                enlargelimits=false,
                ymin=-2, ymax=2,
                xmin=-2, xmax=2,
                grid style={line width=.1pt, draw=gray!10, opacity=0.2},
                tick label style = {font=\tiny, scale=0.5},
                grid
                ]
            \end{axis}
            \draw[fill=red, fill opacity=0.5] (0,0) -- (\x,\x) -- (\x,0) -- (0,0);
            \draw[color=red] (0,0) -- (\x,\x);
            % \draw[color=red] (0,0) rectangle (\x,\x);
            % \draw[color=white, opacity=0.5] (0,0.5\x) -- +(\x,0);

            \node at (0.5\x,-8pt) { a) Intial configuration };
        \end{scope}

        % First shear
        \begin{scope}[xshift=\x+\dx]
            \node[anchor=south west, inner sep=0pt] { \includegraphics[width=\x]{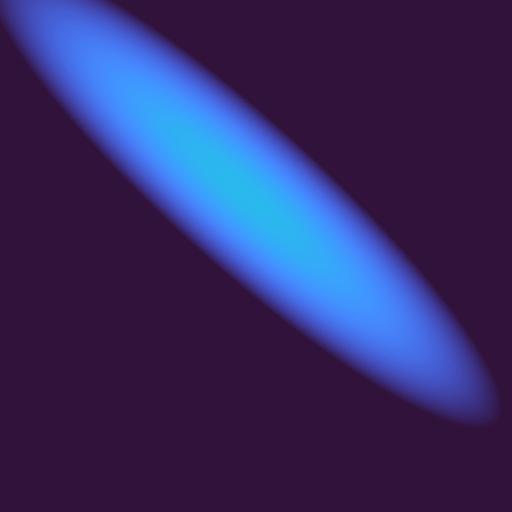}};
            \begin{axis} [
                width=3.55cm,
                height=3.55cm,
                enlargelimits=false,
                ymin=-2, ymax=2,
                xmin=-2, xmax=2,
                grid style={line width=.1pt, draw=gray!10, opacity=0.2},
                tick label style = {font=\tiny, scale=0.5},
                grid
                ]
            \end{axis}
            \draw[fill=red, fill opacity=0.5] (0.5\x,0.0) -- (0.5\x,\x) -- (\x,\x) -- (\x,0) -- (0.5\x,0);
            \draw[color=red] (0.5\x,0.0) -- (0.5\x,\x);
            % \draw[color=red] (0,0) rectangle (\x,\x);
            % \draw[color=white, opacity=0.5] (0,0.5\x) -- +(\x,0);
            \draw[color=white, opacity=0.5] (0.05\x,\x) -- (\x,0.15\x);
            \draw[color=red, ->, thick] (0.75\x,0.75\x) -- +(-0.25\x,0);
            \draw[color=red, ->, thick] (0.25\x,0.25\x) -- +( 0.25\x,0);
            \draw[color=red, dashed] (0,0) -- (\x,\x);

            \node at (0.5\x,-8pt) { b) First shear $S_i$ };
        \end{scope}

        % Second shear
        \begin{scope}[xshift=2*(\x+\dx)]
            \node[anchor=south west, inner sep=0pt] { \includegraphics[width=\x]{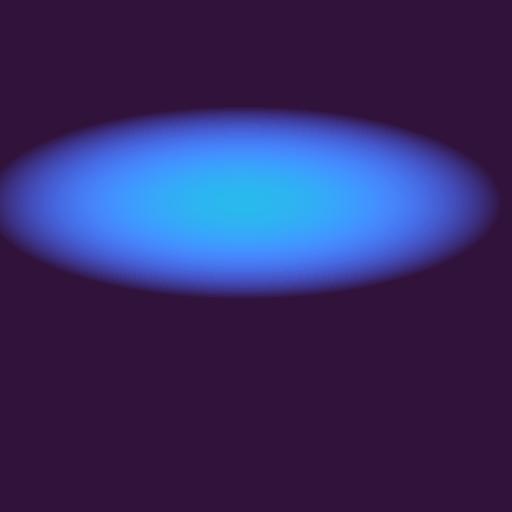}};
            \begin{axis} [
                width=3.55cm,
                height=3.55cm,
                enlargelimits=false,
                ymin=-2, ymax=2,
                xmin=-2, xmax=2,
                grid style={line width=.1pt, draw=gray!10, opacity=0.2},
                tick label style = {font=\tiny, scale=0.5},
                grid
                ]
            \end{axis}
            \draw[fill=red, fill opacity=0.5] (0.5\x,0.0) -- (0.5\x,\x) -- (\x,\x) -- (\x,0) -- (0.5\x,0);
            \draw[color=red] (0.5\x,0.0) -- (0.5\x,\x);
            % \draw[color=red] (0,0) rectangle (\x,\x);
            % \draw[color=white, opacity=0.5] (0,0.5\x-0.416\x) -- (\x,+0.9166\x);
            % \draw[color=white, opacity=0.5, dashed] (0,0.5\x) -- +(\x,0);
            % \draw[color=white, opacity=0.5, ->, thick] (0.25\x,0.5\x) -- +(0,-0.20\x);
            % \draw[color=white, opacity=0.5, ->, thick] (0.75\x,0.5\x) -- +(0,+0.20\x);
            \draw[color=white, opacity=0.5, dashed] (0.05\x,\x) -- (\x,0.15\x);
            \draw[color=white] (0,0.6\x) -- (\x,0.6\x);
            \draw[color=white,->, thick, line cap = round] (0.25\x,0.82\x) -- +(0,-0.20\x);
            \draw[color=white,->, thick, line cap = round] (0.75\x,0.37\x) -- +(0,+0.20\x);

            \node at (0.5\x,-8pt) { c) Second shear $S_o$};
        \end{scope}

        % 1D integral
        \begin{scope}[xshift=3*(\x+\dx)]
            % \node[anchor=south west, inner sep=0pt] { \includegraphics[width=\x]{./Figures/AltDerivation/gaussian_03.jpg}};
            \newcommand\gauss[3]{#3/(#2*sqrt(2*pi))*exp(-((x-#1)^2)/(2*#2^2))}
            \begin{axis} [
                width=3.55cm,
                height=3.55cm,
                enlargelimits=false,
                ymin=0, ymax=1,
                every axis plot post/.append style={thick,mark=none,domain=-2:2,samples=50,smooth},
                ytick=\empty,
                tick label style = {font=\tiny, scale=0.5},
                grid,
                ]
                \addplot {\gauss{-0.1}{0.6}{0.7}};
            \end{axis}
            \draw[fill=red, fill opacity=0.5] (0.5\x,0.0) -- (0.5\x,\x) -- (\x,\x) -- (\x,0) -- (0.5\x,0);
            \draw[color=red] (0.5\x,0.0) -- (0.5\x,\x);
            % \draw[color=red] (0,0) rectangle (\x,\x);
            % \draw[color=white, opacity=0.5] (0,0) (0,0.5\x) -- +(\x,0);

            \node at (0.5\x,-8pt) { d) Integral along $\lambda^{\prime\prime}_i$ };
        \end{scope}
    \end{tikzpicture}
    }
    \vspace{-0.7cm}
    \caption{
    \textbf{Analytical integration of Gaussian-based Fluorescence.}
    We show that we can analytically integrate our 2D model (a) by changing the shape of the integration domain (in red) with a serie of shears (b-c) that produce an axis aligned Gaussian and map the Heaviside to the half plane  $\lambda^{\prime\prime}_i > 0$. This later stage enables to integrate separately the dimensions (c). In this Figure, the 2D Gaussian is scaled for visualization purpose and will not spread in the negative domain in practical cases.
    \label{fig:analytical_integral}
    }
\end{figure}

The key idea of our analytical method is then illustrated in Figure~\ref{fig:analytical_integral}: instead of integrating the 2D Gaussian $\mathcal{G}_{jk}$ over a triangular domain as in Equation~\ref{eqn:gaussian-fluo-coeff2}, we perform two changes of variable through a pair of skew transforms.
The first skew $\mathcal{S}_i$ along $\lambda_i$ converts the diagonal boundary into a vertical boundary, which turns $\mathcal{G}_{jk}$ into an \emph{oblique} 2D Gaussian (Fig.~\ref{fig:analytical_integral}~(b)); the second skew $\mathcal{S}_o$ along $\lambda_o$ leaves the vertical boundary unchanged, but produces a novel \emph{axis-aligned} 2D Gaussian (Fig.~\ref{fig:analytical_integral}~(c)) that is analytically integrated (Fig.~\ref{fig:analytical_integral}~(d)).

The first change of variable $\pmb{\lambda}' = \mathcal{S}_i \pmb{\lambda}$ maps the $\lambda_o-\lambda_i = 0$ boundary to $\lambda^\prime_i = 0$, which is achieved by a horizontal shear:
\begin{equation}
    \label{eqn:first_shear_formula}
    \mathcal{S}_i =
    \begin{bmatrix}
        1 & -1 \\
        0 &  1
    \end{bmatrix}.
\end{equation}
Since shearing does not change the space measure, this yields:
\begin{eqnarray}
    \mathrm{\bar{F}^{\circ}}_{jk}
    & = &
    \int
    \mathcal{G}_{jk}(\mathcal{S}_i^{-1} \pmb{\lambda}'; \alpha_{jk}, \pmb{\mu}_{jk}, \Sigma_{jk}) \, \mathcal{H}(-\lambda_i') \, d\pmb{\lambda}',\\
    & = &
    \int
    \mathcal{G}_{jk}(\pmb{\lambda}'; \alpha_{jk}, \mathcal{S}_i \pmb{\mu}_{jk}, \mathcal{S}_i \Sigma_{jk} \mathcal{S}_i^{\top}) \, \mathcal{H}(-\lambda_i') \, d\pmb{\lambda}',
\end{eqnarray}
where we have used an equivalent 2D Gaussian with sheared parameters in the second line, with the covariance matrix given by:
\begin{equation}
\label{eqn:oblique-covariance}
    \mathcal{S}_i \Sigma_{jk} \mathcal{S}_i^{\top} =
    \begin{bmatrix}
         \sigma^2_{j} + \sigma^2_{k} & -\sigma^2_{k} \\
        -\sigma^2_{k} &  \sigma^2_{k} \\
    \end{bmatrix}.
\end{equation}

The goal of the second change of variable $\pmb{\lambda}'' = \mathcal{S}_o \pmb{\lambda}'$ is to turn the oblique 2D Gaussian into an axis-aligned 2D Gaussian.
$\mathcal{S}_o$ should thus transform the matrix of Equation~\ref{eqn:oblique-covariance} into a diagonal matrix:
\begin{align}
    \mathcal{S}_o =
    \begin{bmatrix}
         1 & 0 \\
        {\sigma^2_{k} \over \sigma^2_{j}+\sigma^2_{k}} &  1 \\
    \end{bmatrix}.
\end{align}
Applying the second change of variables then yields:
\begin{eqnarray}
    \label{eqn:gaussian-fluo-coeff3}
    \mathrm{\bar{F}^{\circ}}_{jk}
    & = &
    \int
    \mathcal{G}_{jk}(\pmb{\lambda}''; \alpha_{jk}, \hat{\pmb{\mu}}_{jk}, \hat{\Sigma}_{jk}) \, \mathcal{H}(-\lambda_i'') \, d\pmb{\lambda}'',
\end{eqnarray}
where the new covariance matrix $\hat{\Sigma}_{jk}$ is diagonal as expected:
\begin{align}
\label{eqn:final-covariance}
    \hat{\Sigma}_{jk} = \mathcal{S}_o \mathcal{S}_i \Sigma_{jk} \mathcal{S}_i^T \mathcal{S}_o^T =
    \begin{bmatrix}
         \sigma^2_{j} + \sigma^2_{k} & 0 \\
        0 &  {\sigma^2_{j} \sigma^2_{k} \over \sigma^2_{j} + \sigma^2_{k}}
    \end{bmatrix}
    \coloneq
    \begin{bmatrix}
        \hat{\sigma}^2_{j} & 0 \\
        0 &  \hat{\sigma}^2_{k}
    \end{bmatrix};
\end{align}
and the new mean vector $\hat{\pmb{\mu}}_{jk}$ is given by:
\begin{align}
\label{eqn:final-mean}
    \hat{\pmb{\mu}}_{jk} = \mathcal{S}_o \mathcal{S}_i \pmb{\mu}_{jk} =
    \begin{bmatrix}
        \mu_{j}-\mu_{k} \\
        {\sigma^2_{j} \mu_{k} + \sigma^2_{k} \mu_{j} \over \sigma^2_{j} + \sigma^2_{k}}
    \end{bmatrix}
    \coloneq
    \begin{bmatrix}
        \hat{\mu}_{j} \\
        \hat{\mu}_{k}
    \end{bmatrix}.
\end{align}

Equation~\ref{eqn:gaussian-fluo-coeff3} now has a simple analytical solution since the Heaviside term $\mathcal{H}$ only applies to the $\lambda_i''$ dimension and the integrated 2D Gaussian is axis-aligned.
Carrying integration, we obtain:
\begin{equation}
\label{eqn:gaussian-fluo-coeff-final1}
    \mathrm{\bar{F}^{\circ}}_{jk} = \alpha_{jk} \times
    \frac{1}{2}\sqrt{2 \pi}\hat{\sigma}_{j} \,
    \left[ 1 - \text{erf}\left( {{ {\hat{\mu}}_{j}} \over \sqrt{2} \hat{\sigma}_{j}} \right) \right]
    \times \sqrt{2 \pi} \hat{\sigma}_{k},
\end{equation}
where the second (resp. third) factor corresponds to bounded (resp. unbounded) integration along the $\lambda_i''$ (resp. $\lambda_o''$) dimension.

After a few simplifications using Equations~\ref{eqn:final-covariance} and~\ref{eqn:final-mean}, we obtain:
\begin{equation}
\label{eqn:gaussian-fluo-coeff-final2}
    \boxed{
        \mathrm{\bar{F}^{\circ}}_{jk} =
        \pi \, \alpha_{jk}
        \sigma_{j} \sigma_{k} \,  \left[ 1 - \text{erf}\left( { \mu_{j} - \mu_{k} \over  \sqrt{2 \left({\sigma^2_{j} + \sigma^2_{k} }\right)}} \right) \right]
        .
    }
\end{equation}

\subsection{Gaussian basis functions}
\label{sec:gauss-bases}

\begin{figure}[t]
    % \centering
    \resizebox{\linewidth}{!}{\pgfplotsset{
    legend image with text/.style={
        legend image code/.code={%
            \node[anchor=center] at (0.3cm,0cm) {#1};
        }
    },
}

\begin{tikzpicture}[font=\footnotesize]
    \begin{axis}[
        % at={(0, -2.5 cm)},
        % anchor=north west,
        legend pos=outer north east,
        legend columns=2,
        width=1.0\linewidth,
        height=0.55\linewidth,
        xmin=300,
        xmax=750,
        ymin=0,
        x tick label style={font=\tiny},
        y tick label style={font=\tiny},
        enlarge x limits={abs=0.6},
        xlabel={wavelength [$\mathrm{nm}$]},
        ylabel near ticks,
        % cycle list name=delist
        ]

    \addlegendimage{legend image with text=CIE XYZ}
    \addlegendentry{}

    \addlegendimage{legend image with text=Gaussian}
    \addlegendentry{}

    % --------------------------------------------------------------------------

    % add the data rows
    \addplot[mark=none, color=purple, thick, opacity=0.6] table [
        x index=0,
        y index=1,
        col sep=comma,
        ] {Figures/GaussianCMF/data/cmf_cie.csv};
    \addlegendentry{}

    \addplot[mark=none, color=purple, very thick, dashed] table [
        x index=0,
        y index=1,
        col sep=comma,
        ] {Figures/GaussianCMF/data/cmf_optim.csv};
    \addlegendentry{$\bar{x}$}

    \addplot[mark=none, color=teal, thick, opacity=0.6] table [
        x index=0,
        y index=2,
        col sep=comma,
        ] {Figures/GaussianCMF/data/cmf_cie.csv};
    \addlegendentry{}

    \addplot[mark=none, color=teal, very thick, dashed] table [
        x index=0,
        y index=2,
        col sep=comma,
        ] {Figures/GaussianCMF/data/cmf_optim.csv};
    \addlegendentry{$\bar{y}$}

    \addplot[mark=none, color=blue, thick, opacity=0.6] table [
        x index=0,
        y index=3,
        col sep=comma,
        ] {Figures/GaussianCMF/data/cmf_cie.csv};
    \addlegendentry{}

    \addplot[mark=none, color=blue, very thick, dashed] table [
        x index=0,
        y index=3,
        col sep=comma,
        ] {Figures/GaussianCMF/data/cmf_optim.csv};
    \addlegendentry{$\bar{z}$}

    % --------------------------------------------------------------------------

    \addplot[mark=none, color=white] {0};
    \addlegendentry{}

    \addplot[mark=none, color=violet, very thick, dashed] table [
        x index=0,
        y index=4,
        col sep=comma,
        ] {Figures/GaussianCMF/data/cmf_optim.csv};
    \addlegendentry{UV}

    \end{axis}
\end{tikzpicture}}
    \vspace{-20pt}
    \caption{\textbf{Gaussian basis functions.}
    Our analytical bispectral integration relies on the availability of Gaussian sensitivity functions.
    We fit the CIE XYZ 2006 $2^\circ$ $\bar{x}$ sensitivity function (in red) with a pair of Gaussians, while $\bar{y}$ and $\bar{z}$ only require a single Gaussian.
    In order to account for reradiation from non-visible to visible light, we provide an additional UV basis (in purple) whose parameters are chosen to best approximate measured materials.}
    \label{fig:gauss-bases}
\end{figure}
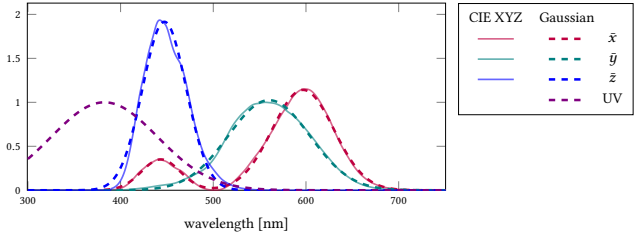

The analytical bispectral integration process described in the previous section relies on the availability of Gaussian basis functions.
As shown in Figure~\ref{fig:gauss-bases}, the CIE XYZ 2006 $2^\circ$ observer tabulated sensitivity functions are well approximated by Gaussian bases.
In particular, $\bar{x}$ requires a pair of Gaussians, while a single Gaussian is sufficient for both $\bar{y}$ or $\bar{z}$.
Gaussian basis parameters are provided in Table~\ref{tbl:gauss-params}.

\begin{table}[h!]
    \caption{\label{tbl:gauss-params} \textbf{Gaussian basis parameters.} We fit the CIE XYZ 2006 2$\deg$ using Gaussians ($\bar{x}, \bar{y}, \bar{z}$) and report the magnitude, mean and standard deviation here. We further include a Gaussian sensitivity function for the UV band.}
    \vspace{-7pt}
    \begin{tabular}{|c|c|c|c|}
        \hline
        & intensity & mean (nm) & std (nm)\\
        \hline
        \multirow{2}{*}{$\bar{x}$} & 0.35087 & 443.412226 & 20.838149 \\
        & 1.141263 & 596.813847 & 33.276659 \\
        \hline
        $\bar{y}$ & 1.024335 & 560.186336 & 43.898132 \\
        \hline
        $\bar{z}$ & 1.915863 & 447.268188 & 23.542626 \\
        \hline
        UV & 1 & 382.535501 & 57.432550 \\
        \hline
    \end{tabular}
    \vspace{-7pt}
\end{table}

\paragraph{Additional UV basis.}
We use an additional basis in the UV range to account for reradiation from non-visible to visible light, similarly to Fichet et al.~\shortcite{Fichet2024}.
Contrary to them, we use a single Gaussian function for this additional channel.
The Gaussian parameters of this extra band (also provided in Table~\ref{tbl:gauss-params}) are optimized over a set of measured materials~\cite{Gonzalez00}, as detailed in Supplemental material.
In a nutshell, we searched for the Gaussian parameters that minimize the $\Delta E^*_{2000}$ difference between the color obtained with the full spectral reradiation (e.g., using Equation~\ref{eqn:spectral-out-col}) and the color obtained with a reduced $4 \times 4$ XYZU reradiation matrix (i.e., using Equation~\ref{eqn:reduced-out-col}), for several illuminants.

\paragraph{Transfer matrices.}
Recall that the matrix of sensitivity functions is given by $S = G T_G$.
When working with XYZU colors, our approach requires $5$ Gaussian basis functions.
Hence $G$ is the $N \times 5$ matrix of discretized 1D Gaussians, and $T_G$ is a $5 \times 4$ matrix given by:
\begin{equation}
    T_G =  \left[
    \begin{array}{cccc}
         	1 & 0 & 0 & 0\\
         	1 & 0 & 0 & 0\\
         	0 & 1 & 0 & 0\\
         	0 & 0 & 1 & 0\\
	0 & 0 & 0 & 1
    \end{array}
    \right].
\end{equation}

When using a different color space for light transport than the one of the sensor -- say CIE XYZ 2006 $2^\circ$, we also have to project the outgoing color $\mathbf{c}_o$ back to the sensor color space.
As already mentioned in the work of Fichet et al.~\shortcite{Fichet2024}, this is done through $T  \mathbf{c}_{o}$, where $T=S_{XYZ}^{\top}\ \tilde{S}$ is a final $3 \times K$ transfer matrix.
When working with XYZU colors ($K=4$), it simply amounts to:
\begin{equation}
    T =  \left[
    \begin{array}{cccc}
         	1 & 0 & 0 & 0\\
         	0 & 1 & 0 & 0\\
         	0 & 0 & 1 & 0
    \end{array}
    \right].
\end{equation}

%%
%\begin{equation}
%   \mathbf{c} = T \ \mathbf{c}_{o}= S_{XYZ}^{\top}\ \tilde{S} \ \mathbf{c}_{o} ,
%\end{equation}
%%
%where $\tilde{S}_{G}$ is the dual basis of Gaussian based CMFs (with or without the additional UV band).
%\pascal{pas fan des notations ci-dessus, faudrait trouver mieux. L'ordre des multiplications devrait être inversé aussi, non ?}

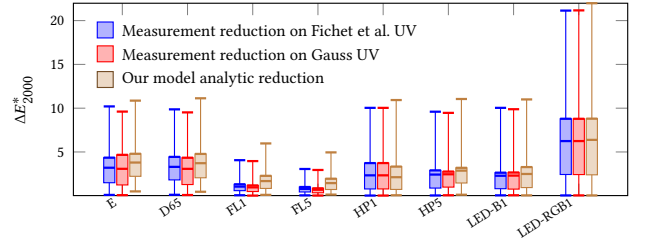
\begin{figure}[t]
    % \centering
    \resizebox{\linewidth}{!}{\pgfplotsset{
    box plot/.style={
        /pgfplots/.cd,
        black,
        only marks,
        mark=-,
        mark size=\pgfkeysvalueof{/pgfplots/box plot width},
        /pgfplots/error bars/y dir=plus,
        /pgfplots/error bars/y explicit,
        /pgfplots/table/x index=\pgfkeysvalueof{/pgfplots/box plot x index},
    },
    box plot box/.style={
        /pgfplots/error bars/draw error bar/.code 2 args={%
            \draw  ##1 -- ++(\pgfkeysvalueof{/pgfplots/box plot width},0pt) |- ##2 -- ++(-\pgfkeysvalueof{/pgfplots/box plot width},0pt) |- ##1 -- cycle;
        },
        /pgfplots/table/.cd,
        y index=\pgfkeysvalueof{/pgfplots/box plot box top index},
        y error expr={
            \thisrowno{\pgfkeysvalueof{/pgfplots/box plot box bottom index}}
            - \thisrowno{\pgfkeysvalueof{/pgfplots/box plot box top index}}
        },
        /pgfplots/box plot
    },
    box plot top whisker/.style={
        /pgfplots/error bars/draw error bar/.code 2 args={%
            \pgfkeysgetvalue{/pgfplots/error bars/error mark}%
            {\pgfplotserrorbarsmark}%
            \pgfkeysgetvalue{/pgfplots/error bars/error mark options}%
            {\pgfplotserrorbarsmarkopts}%
            \path ##1 -- ##2;
        },
        /pgfplots/table/.cd,
        y index=\pgfkeysvalueof{/pgfplots/box plot whisker top index},
        y error expr={
            \thisrowno{\pgfkeysvalueof{/pgfplots/box plot box top index}}
            - \thisrowno{\pgfkeysvalueof{/pgfplots/box plot whisker top index}}
        },
        /pgfplots/box plot
    },
    box plot bottom whisker/.style={
        /pgfplots/error bars/draw error bar/.code 2 args={%
            \pgfkeysgetvalue{/pgfplots/error bars/error mark}%
            {\pgfplotserrorbarsmark}%
            \pgfkeysgetvalue{/pgfplots/error bars/error mark options}%
            {\pgfplotserrorbarsmarkopts}%
            \path ##1 -- ##2;
        },
        /pgfplots/table/.cd,
        y index=\pgfkeysvalueof{/pgfplots/box plot whisker bottom index},
        y error expr={
            \thisrowno{\pgfkeysvalueof{/pgfplots/box plot box bottom index}}
            - \thisrowno{\pgfkeysvalueof{/pgfplots/box plot whisker bottom index}}
        },
        /pgfplots/box plot
    },
    box plot median/.style={
        /pgfplots/box plot,
        /pgfplots/table/y index=\pgfkeysvalueof{/pgfplots/box plot median index}
    },
    box plot width/.initial=1em,
    box plot x index/.initial=0,
    box plot median index/.initial=1,
    box plot box top index/.initial=2,
    box plot box bottom index/.initial=3,
    box plot whisker top index/.initial=4,
    box plot whisker bottom index/.initial=5,
}

\newcommand{\boxplot}[2][]{
    \addplot [box plot median,#1] table[col sep=comma] {#2};
    \addplot [forget plot, box plot box,#1] table[col sep=comma] {#2};
    \addplot [forget plot, box plot top whisker,#1] table[col sep=comma] {#2};
    \addplot [forget plot, box plot bottom whisker,#1] table[col sep=comma] {#2};
}

\begin{tikzpicture}
    \begin{axis} [
        name=moustache,
        box plot width=0.7mm,
        enlargelimits=false,
        width=\linewidth,
        height=4cm,
        xmin=0.25,
        xmax=8.75,
        xtick=data,
        xtick={1,...,8},
        xticklabels from table={Figures/DE/data/illu.csv}{Illuminant},
        x tick label style={rotate=33,anchor=east,font=\tiny},
        y tick label style={font=\tiny},
        ylabel={\tiny{$\Delta E_{2000}^*$}},
        ylabel near ticks,
        ]
        \boxplot [
            thick,
            blue,
            fill=blue!30!white,
            forget plot,
            box plot whisker bottom index=2,
            box plot box bottom index=3,
            box plot median index=1,
            box plot box top index=5,
            box plot whisker top index=6
        ] {Figures/DE/data/with_diag/de_patches_1-measured_egsr_uv.csv}

        \boxplot [
            thick,
            red,
            fill=red!30!white,
            forget plot,
            box plot whisker bottom index=2,
            box plot box bottom index=3,
            box plot median index=1,
            box plot box top index=5,
            box plot whisker top index=6
        ] {Figures/DE/data/with_diag/de_patches_1-measured_gauss_uv.csv}

        \boxplot [
            thick,
            brown,
            fill=brown!30!white,
            forget plot,
            box plot whisker bottom index=2,
            box plot box bottom index=3,
            box plot median index=1,
            box plot box top index=5,
            box plot whisker top index=6
        ] {Figures/DE/data/with_diag/de_patches_1-analytic.csv}
    \end{axis}

    \node[xshift=.5cm, yshift=-0.15cm, anchor=north west] (key_g1) at (moustache.north west) {\scriptsize{Measurement reduction on Fichet et al. UV}};
    \node[xshift=.5cm, yshift=-0.45cm, anchor=north west] (key_g2) at (moustache.north west) {\scriptsize{Measurement reduction on Gauss UV}};
    \node[xshift=.5cm, yshift=-0.75cm, anchor=north west] (key_g3) at (moustache.north west) {\scriptsize{Our model analytic reduction}};

    \draw[blue,fill=blue!30!white]            ($(key_g1.west) - (.1cm, 0) + (.1cm, .1cm)$) rectangle ($(key_g1.west) - (.1cm, 0) - (.1cm, .1cm)$);
    \draw[red,fill=red!30!white]              ($(key_g2.west) - (.1cm, 0) + (.1cm, .1cm)$) rectangle ($(key_g2.west) - (.1cm, 0) - (.1cm, .1cm)$);
    \draw[brown!60!black,fill=brown!30!white] ($(key_g3.west) - (.1cm, 0) + (.1cm, .1cm)$) rectangle ($(key_g3.west) - (.1cm, 0) - (.1cm, .1cm)$);
\end{tikzpicture}}
    \vspace{-20pt}
    \caption{\textbf{Accuracy of our reduction.}
%	Comparison of our analytic reduction on our model with the reduction of the full reradiation matrix and the reduction of the tabulated spectral reradiation using our model.
	We compute $\Delta E^*_{2000}$ differences between a spectral evaluation of the original data and the XYZ evaluation of the reradiation based on three XYZU reduction methods.
	Two of these methods apply brute-force reduction on measured materials, using either our UV basis or the one from Fichet et al.~\shortcite{Fichet2024}.
	The third reduction method applies our analytical Gaussian-based integration, using a single 2D Gaussian for the normalized fluorescence component and six 1D Gaussians for the reflectance along the diagonal.
	The whiskers bottom and top represent resp. the minimum and the maximum, the box bottom and top represent resp. the first and third quartiles and the central line represent the average $\Delta E^*_{2000}$ over 143 reradiation matrices from \cite{Gonzalez00}.
    }
    \label{fig:reduction-accuracy}
%    \vspace{-10pt}
\end{figure}

\paragraph*{Reduction accuracy}
The accuracy of our Gaussian-based analytic reduction method is evaluated in Figure~\ref{fig:reduction-accuracy} on the Gonzalez and Fairchild's dataset~\shortcite{Gonzalez00} for a range of illuminants, using $\Delta E^*_{2000}$ differences with colors obtained with the full bispectral reradiation matrices.
A first difference of our approach compared to the one of Fichet et al.~\shortcite{Fichet2024} is that we use different UV bases.
We thus compare the two XYZU representations using a brute-force reduction on \emph{measured} materials, which shows that our representation (red ribbons) is on par with theirs (blue ribbons).
Yet the main advantage of our model is to grant analytic, on-the-fly reduction, provided the material is approximated with Gaussians: here we also fit $6$ Gaussians to the reradiation diagonal to grant full analytic integration.
In spite of these approximations, Figure~\ref{fig:reduction-accuracy} shows that the $\Delta E^*_{2000}$ differences obtained with our model (marroon ribbons) remain very close to those obtained with brute-force integration.

%In Figure~\ref{fig:reduction-accuracy}, we validated the accuracy of using Gaussian based CMFs when reducing fluorescence matrices. First, we compared the original CIE XYZ CMFs with the addition of Fichet et al.~\shortcite{Fichet2024} UV basis (blue ribbons) with our approximation with Gaussian based UV (red ribbons) using the brute-force reduction of Fichet et al.~\shortcite{Fichet2024} on the Gonzalez and Fairchild's dataset~\shortcite{Gonzalez00}. Our approximation permits to reproduce colors with the same level of accuracy than the classical CIE XYZs curves. We further validated that using our fitted Gaussian model and analytical integration (marroon ribbons) yeld the same level of accuracy that both methods.

% \begin{figure}
%     \centering
%     \resizebox{\linewidth}{!}{\subimport{Figures/DE/}{content_red_cmp_xyzu.tex}}
%     \caption{\textbf{Accuracy of our model.}
%     }
%     \label{fig:model-overview}
% \end{figure}

\section{Artist-friendly fluorescence model}
\label{sec:edit}

In this section, we show that besides accurately \emph{representing} measured fluorescent materials, a simplified Gaussian-based model is well adapted to the \emph{creation} of fluorescent materials from scratch.
We first detail our choices of model simplification (Section~\ref{sec:simple-model}) and explain how energy conservation is ensured (Section~\ref{sec:simple-energy}), before showcasing our fluorescence palette editing system (Section~\ref{sec:simple-palettes}).

\subsection{Simplified model}
\label{sec:simple-model}

Our first design choice is to reduce the number of material parameters to make direct artistic editing tractable.
We thus use a single Gaussian for the normalized fluorescence component $\bar{\mathcal{F}}$ (i.e., $Q=1$ in Equation~\ref{eqn:fluo-model}), yielding five parameters: the Gaussian intensity $\alpha$, its mean vector $\pmb{\mu} = [\mu_a, \mu_e]^{\top}$ and diagonal covariance matrix $\Sigma=\mbox{diag}(\sigma^2_a, \sigma^2_e)$, where we have used the $a$ and $e$ subscripts to denote the absorption and emission axes respectively.

\change{
We then specify the reduced reflectance matrix $R$ using an albedo color $\pmb{\rho}$ (three floats). This further permits to reuse workflows for albedo editing without change. We build $R$ by first upscaling $\pmb{\rho}$ to a spectrum using the dual basis, $\pmb{\rho} \times \tilde{S}$, and then reducing the diagonal matrix $\mathcal{R} = \mbox{diag}(\pmb{\rho} \times \tilde{S})$. Thanks to linearity of the reduction, we obtain $\mathrm{R}_{\pmb{\rho}} = \sum_k \pmb{\rho}_k \mathrm{R}_k$, where $R_k = S^{\top}  \mbox{diag}(\tilde{S}_k) \tilde{S}$. This form permits to compute $R$ from precomputed $R_k$ matrices and user specified albedo color  $\pmb{\rho}$.
The outgoing color is obtained by using $\mathrm{R}_{\pmb{\rho}}$ in Equation~\ref{eqn:reduced-out-col2}.
We compare this approach with the approximation $R = \mbox{diag}(\pmb{\rho})$ in Supplemental material and conclude the former works better at approximating multiple scattering.
}
% In non-spectral rendering, the reflectance component is seldom provided in spectral form.
% It is more common to work with an albedo color $\pmb{\rho}$, akin to a spectral albedo pre-integrated over sensitivity functions.
% Our second design choice is to directly build the reduced reflectance matrix $\mathrm{R}_{\pmb{\rho}}$ from this albedo color.
% We first build a reflectance matrix $\mathcal{R}_k = \mbox{diag}(\tilde{S}_k)$ for each color channel $k$, with $\tilde{S}_k$ the $k$th column of $\tilde{S}$ acting as a spectral upsampling operator.
% We then approximate the spectral albedo through $\mathcal{R}_{\pmb{\rho}} \approx \sum_k \rho_k \mathcal{R}_k$, with $\rho_k$ the $k$th albedo color channel.
% Thanks to the linearity of reduction, we then obtain $\mathrm{R}_{\pmb{\rho}} = \sum_k \pmb{\rho}_k \mathrm{R}_k$, with $\mathrm{R}_k = S^{\top} \mathcal{R}_k \tilde{S}$.
% The outgoing color is obtained by using $\mathrm{R}_{\pmb{\rho}}$ in Equation~\ref{eqn:reduced-out-col2}.
% We compare this approach with using a mere diagonal reduced reflectance matrix in Supplemental material.

For RGB rendering, we assume that the albedo color is first converted to XYZ color space to yield $\pmb{\rho}_{\scriptscriptstyle XYZ}$.
As shown in the work of Fichet et al.~\shortcite{Fichet2024} and confirmed in Section~\ref{sec:gauss-bases}, using an additional ultraviolet channel significantly increases accuracy for some materials.
When working with XYZU colors, we thus also need to decide of an ultraviolet channel value for the corresponding albedo color $\pmb{\rho}_{\scriptscriptstyle XYZU}$.
One option would be to let artists specify it, but this would likely be unintuitive since its effects are indirect: it reduces (resp. contributes to) fluorescent reradiation in direct (resp. indirect) lighting.
We thus make the choice to determine it automatically, using $\pmb{\rho}_{\scriptscriptstyle XYZU} = T_{\scriptscriptstyle U} \, \pmb{\rho}_{\scriptscriptstyle XYZ}$, with $T_{\scriptscriptstyle U}$ a $4 \times 3$ transfer matrix given by:
\begin{equation}
	\label{eqn:transfer-U}
	T_{\scriptscriptstyle U} =
	S_{\scriptscriptstyle XYZU}^{\top}\tilde{S}_{\scriptscriptstyle XYZ} =
	\left[
	\begin{array}{@{}ccc@{}}
	1 & 0 & 0 \\
	0 & 1 & 0 \\
	0 & 0 & 1 \\
	-0.0145415 & 0.0267372 & 0.397627
	\end{array}
	\right].
\end{equation}
It amounts to a spectral upsampling through $\tilde{S}_{\scriptscriptstyle XYZ}$, followed by a downsampling to the XYZU color space through $S_{\scriptscriptstyle XYZU}$.
As evident in Equation~\ref{eqn:transfer-U}, the X, Y and Z channels are left unchanged, while the UV channel is obtained as a linear combination of other channels in a way proportional to the overlap of their sensitivity functions.

%%
%\correct{
%% \begin{equation}
%%     \label{eqn:reduced-out-col-simple}
%%     \mathbf{c}_o = \mathbf{c}_{o,r} + \mathbf{c}_{o,f} \approx \pmb{\rho} \odot \mathbf{c}_i + \mathrm{\bar{F}} \ (\mathbf{1}-\pmb{\rho}) \odot \mathbf{c}_i,
%% \end{equation}
%}{
%    \begin{equation}
%        \label{eqn:reduced-out-col-simple}
%        \mathbf{c}_o = \mathbf{c}_{o,r} + \mathbf{c}_{o,f} \approx \mathrm{R}_{\pmb{\rho}} \ \mathbf{c}_i + \mathrm{\bar{F}} \ (\mathrm{I}-\mathrm{R}_{\pmb{\rho}}) \ \mathbf{c}_i,
%    \end{equation}
%}
%%
%\correct{where $\odot$ denotes component-wise vector multiplication.}{}

\subsection{Energy conservation}
\label{sec:simple-energy}

Thanks to the decomposition of Equation~\ref{eqn:reduced-out-col2}, energy conservation may be enforced separately on the reflectance and fluorescence components.
For the reflectance component, we enforce each coefficient of the albedo color $\pmb{\rho}$ to be in the $[0,1]$ range.

Energy conservation of the normalized fluorescence component is more involved, since the corresponding reduced normalized fluorescence matrix $\mathrm{\bar{F}}$ in Equation~\ref{eqn:reduced-out-col2} is obtained through the analytical integration process of Section~\ref{sec:gauss-int}.
In other words, constraints must be found on material parameters $\{\alpha, \mu_a, \sigma_a, \mu_e, \sigma_e\}$ to ensure the energy conservation constraint of Equation~\ref{eqn:energy-conservation2} is met.

\begin{figure}[t]
	\resizebox{\linewidth}{!}{
	\begin{tabular}{ccc}
	\includegraphics[width=0.3\linewidth]{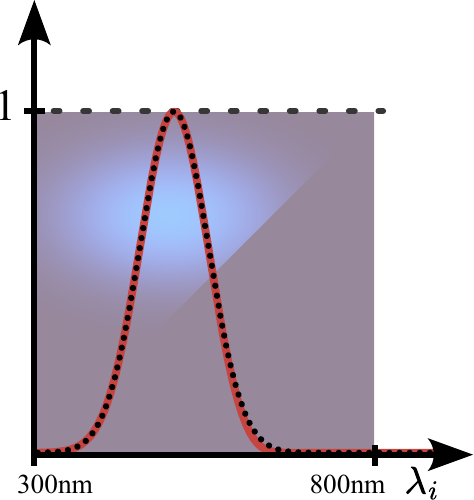} &
	\includegraphics[width=0.3\linewidth]{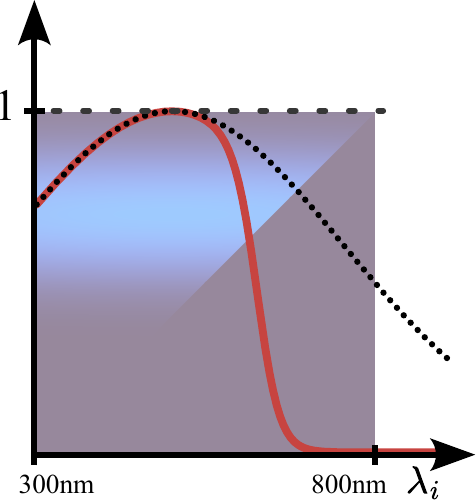} &
	\includegraphics[width=0.3\linewidth]{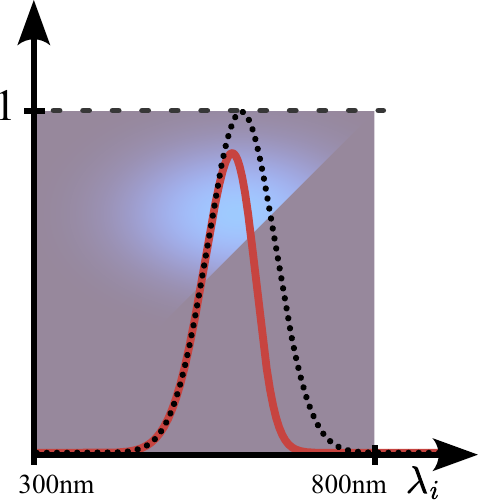} \\
	(a) $\scriptstyle \mu_a=500,~ \sigma_a=100$ &
	(b) $\scriptstyle \mu_a=500,~ \sigma_a=500$ &
	(c) $\scriptstyle \mu_a=600,~ \sigma_a=100$
	\end{tabular}
	}
	\vspace{-10pt}
	\caption{\textbf{Energy conservation.}
	We evaluate our conservative bound $\bar{\alpha}_{\max}$ for energy conservation by showing $\bar{\alpha}_{\max}		\mathcal{\bar{A}}(\lambda_i)$ (red solid line), which should reach $1$ at its maximum for perfect energy conservation.
	The reduced fluorescence component $\mathcal{\bar{F}}$ is shown in the background.
	We examine three absorption configurations, using $\mu_e=650$nm and $\sigma_e=60$nm in all cases.
	Our computation is based on a Gaussian approximation to $\mathcal{\bar{A}}$ (black dotted line), whose maximum departs from the true maximum when a significant portion of the Gaussian lies below the bispectral diagonal, as in c).
	\label{fig:energy}
	% \vspace{-15pt}
}
\end{figure}

For consision, we rewrite that constraint as $\mathcal{\bar{A}}(\lambda_i) \le 1 \ \forall \lambda_i$, with $\mathcal{\bar{A}}$ a reduced absorption spectrum shown in Figure~\ref{fig:energy}.
Energy conservation may then be ensured if the intensity parameter $\alpha \le \alpha_{\max}$, where the upperbound is a function of other material parameters:
\begin{equation}
    \label{eqn:alpha-max}
    \alpha_{\max}(\mu_a, \sigma_a, \mu_e, \sigma_e) = \frac{1}{\max_{\lambda_i} \mathcal{\bar{A}}(\lambda_i)}.
\end{equation}
The intensity parameter is then computed as $\alpha = \bar{\alpha} \, \alpha_{\max}$, where $\bar{\alpha} \in [0,1]$ controls the material fluorescence strength.
Even though $\mathcal{\bar{A}}$ has an analytical form, we have not found an analytical solution to Equation~\ref{eqn:alpha-max} due to the maximum in the denominator.

A conservative analytical solution is obtained by ignoring the Heaviside term in Equation~\ref{eqn:fluo-model}, leaving only the single Gaussian term $\mathcal{G}$.
Since the peak of $\mathcal{G}$ is located at $[\mu_a, \mu_e]^{\top}$, we only need to perform integration in $\mathcal{\bar{A}}$ at $\lambda_i=\mu_a$, yielding:
\begin{equation}
    \label{eqn:alpha-max-approx}
    \max_{\lambda_i} \mathcal{\bar{A}}(\lambda_i)
    \approx \int_0^{\infty} \mathcal{G}(\mu_a,\lambda_o) d\lambda_o
    = \sqrt{\tfrac{\pi}{2}} \sigma_e \left(1+\text{erf}\left(\frac{\mu_e}{\sqrt{2}\sigma_e}\right)\right).
\end{equation}
When used in Equation~\ref{eqn:alpha-max}, we obtain our approximate bound: $\hat{\alpha}_{\max}(\mu_e,\sigma_e)$. This bound only departs from the true bound when a significant portion of $\mathcal{G}$ lies below the bispectral diagonal (Figure~\ref{fig:energy}).
%However, we empirically observed that the fluorescence component of measured materials is consistently much weaker than what is afforded by this conservative bound, which is likely due to absorption.

% Alternatively, the exact bound may be computed as $\alpha_{\max}=\tfrac{1}{\mathcal{\bar{A}}(\lambda_i^{\star})}$ by numerically finding the wavelength $\lambda_i^{\star} = \arg \max_{\lambda_i} \mathcal{\bar{A}}(\lambda_i)$.
% We know that $\lambda_i^{\star} < \mu_a$ due to clamping by the bispectral diagonal.
% Since $\tfrac{d\mathcal{\bar{A}}}{d\lambda_i}$ has an analytical formula, we may use any numerical approach to find $\lambda_i^{\star} \in [0,\mu_a)$ such that $\tfrac{d\mathcal{\bar{A}}}{d\lambda_i}(\lambda_i^{\star})=0$.
% \todo{In practice, we use a dichotomy in the $[\mu_a-\sigma_a, \mu_a]$ range...}

\begin{figure*}[t]
	% \vspace{-5pt}
	\resizebox{\linewidth}{!}{
	\begin{tikzpicture}[font=\footnotesize]
		\setlength{\x}{0.13\linewidth}
		% \matcal{F}
		\node[inner sep=0pt                    ]   (rerad_0) {\includegraphics[width=\x]{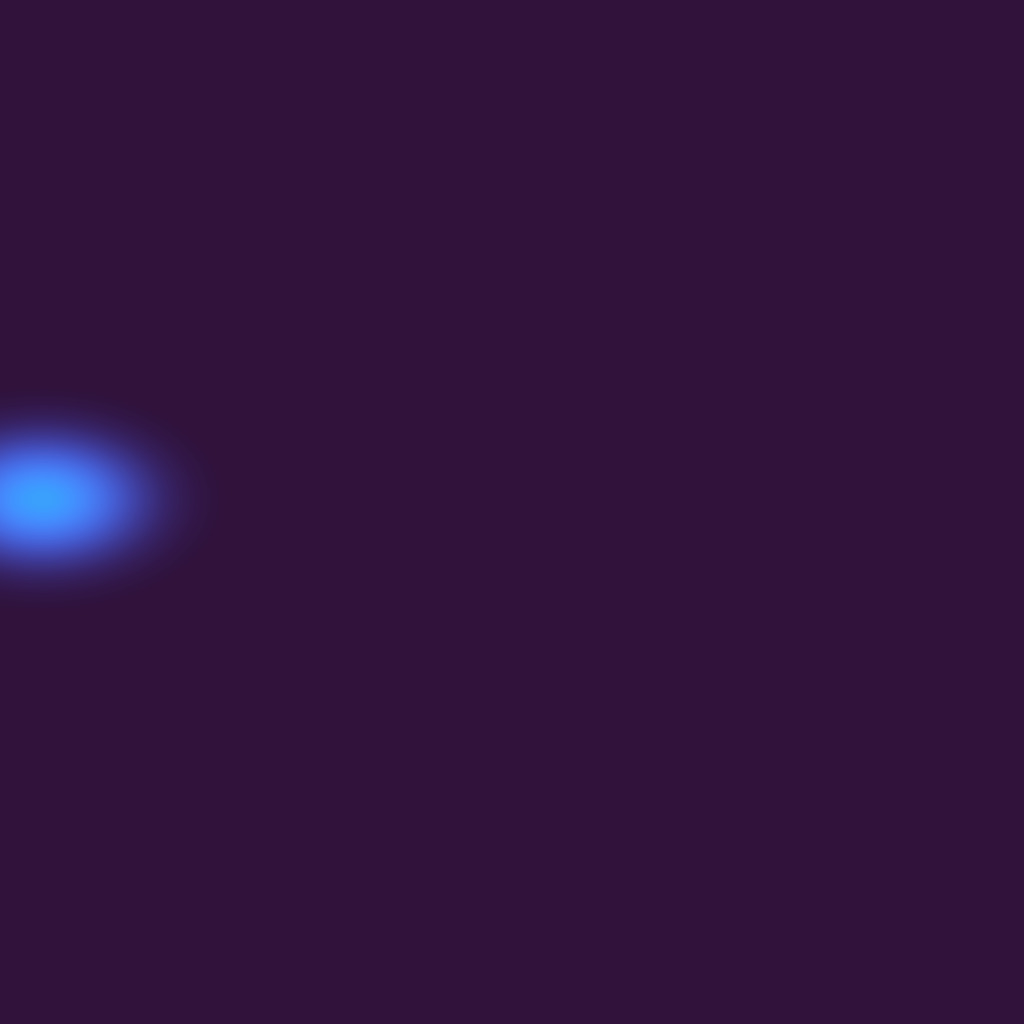}};
		\node[inner sep=0pt, below=2pt of rerad_0] (rerad_1) {\includegraphics[width=\x]{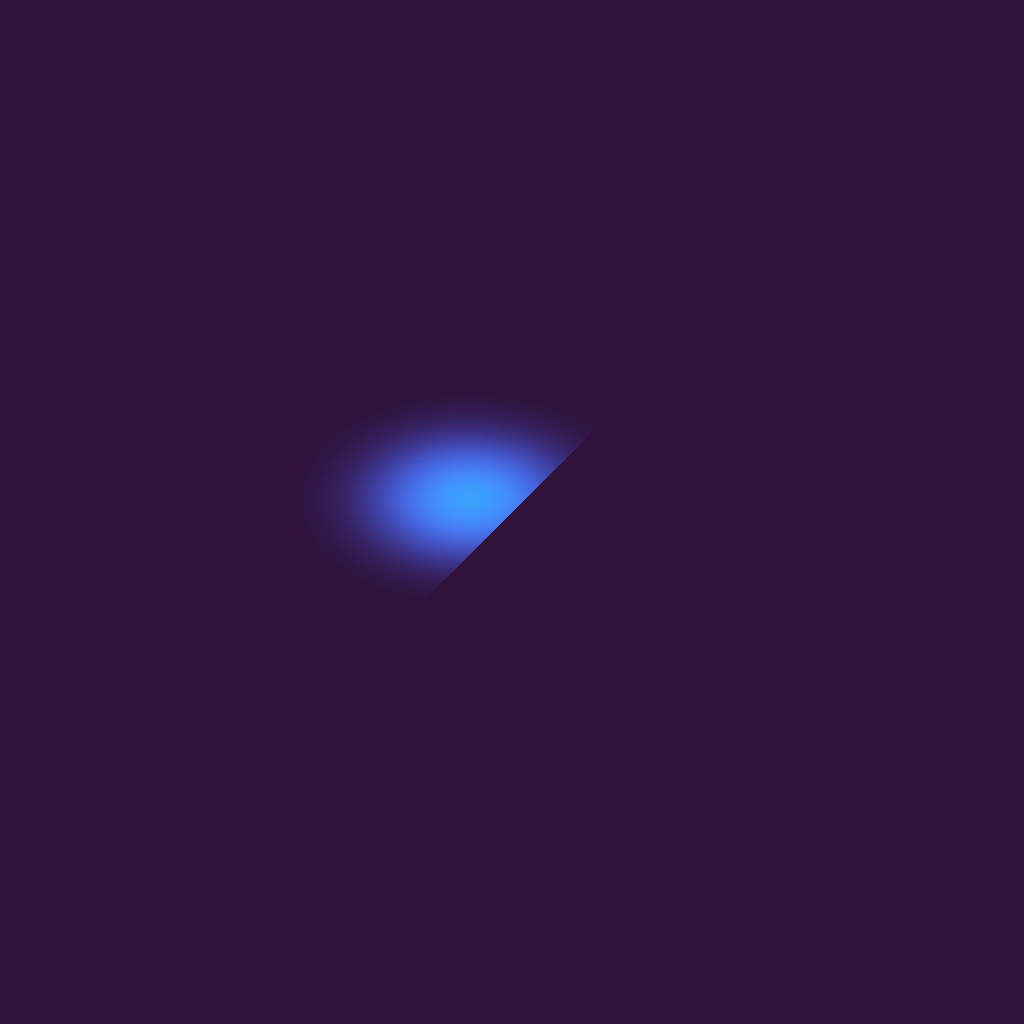}};
		\node[inner sep=0pt, below=2pt of rerad_1] (rerad_2) {\includegraphics[width=\x]{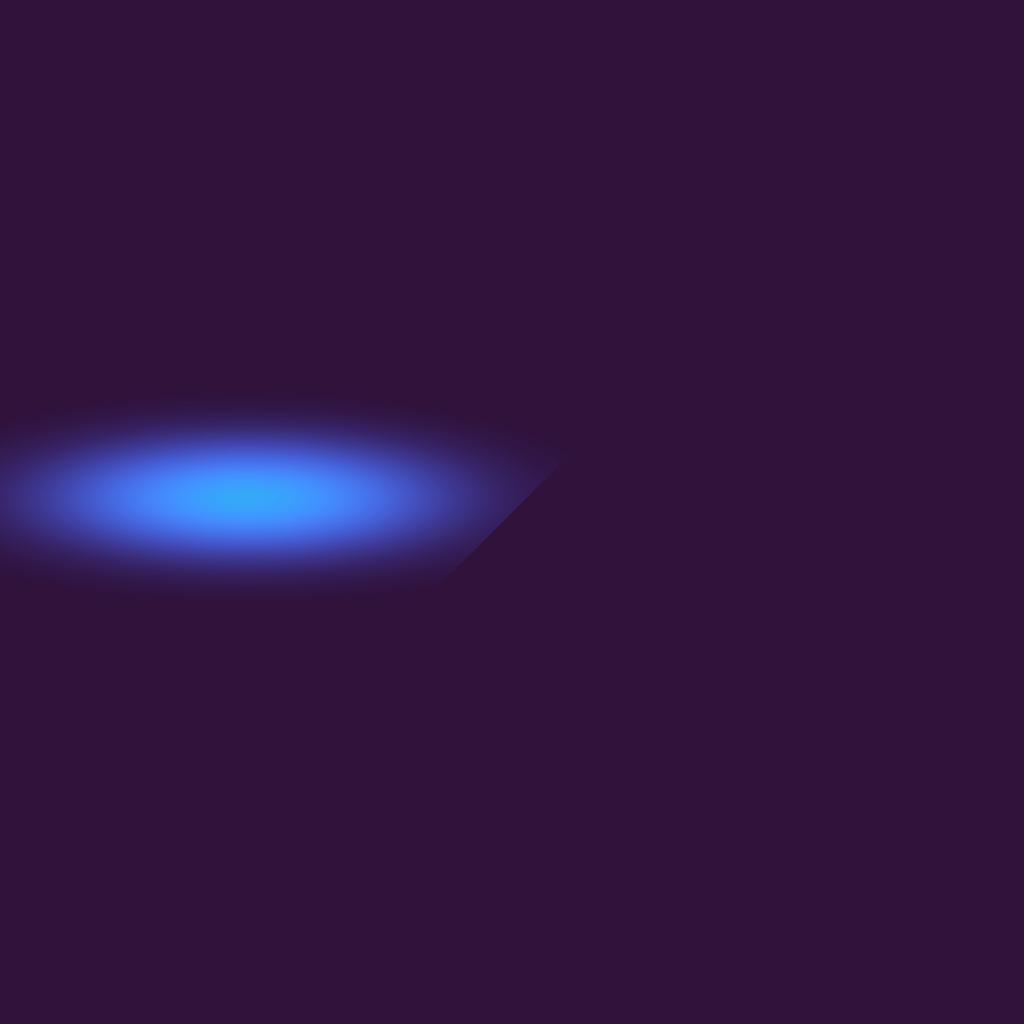}};

		\draw (rerad_0.south west) rectangle (rerad_0.north east);
		\draw (rerad_1.south west) rectangle (rerad_1.north east);
		\draw (rerad_2.south west) rectangle (rerad_2.north east);

		% Palette D65
		\node[inner sep=0pt, right=2pt of rerad_0] (paletteD65_0) {\includegraphics[width=\x]{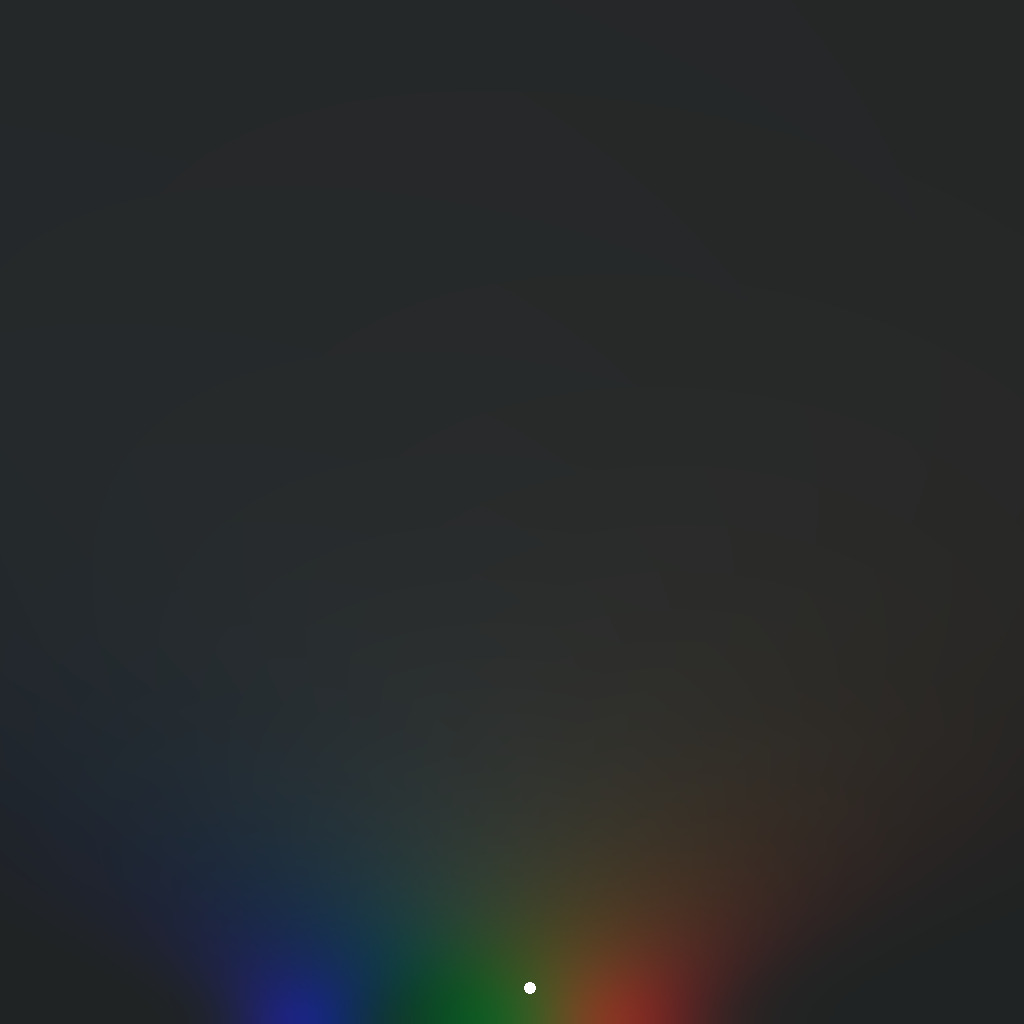}};
		\node[inner sep=0pt, right=2pt of rerad_1] (paletteD65_1) {\includegraphics[width=\x]{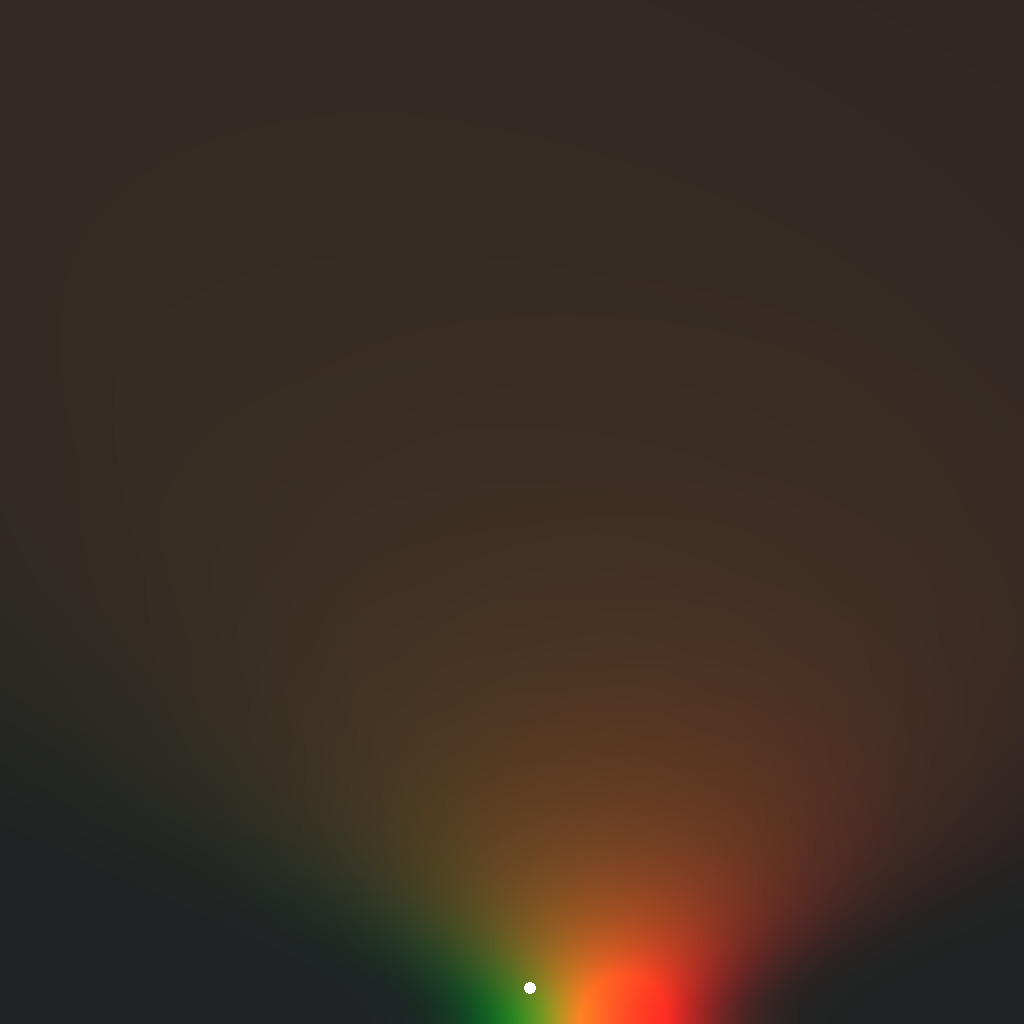}};
		\node[inner sep=0pt, right=2pt of rerad_2] (paletteD65_2) {\includegraphics[width=\x]{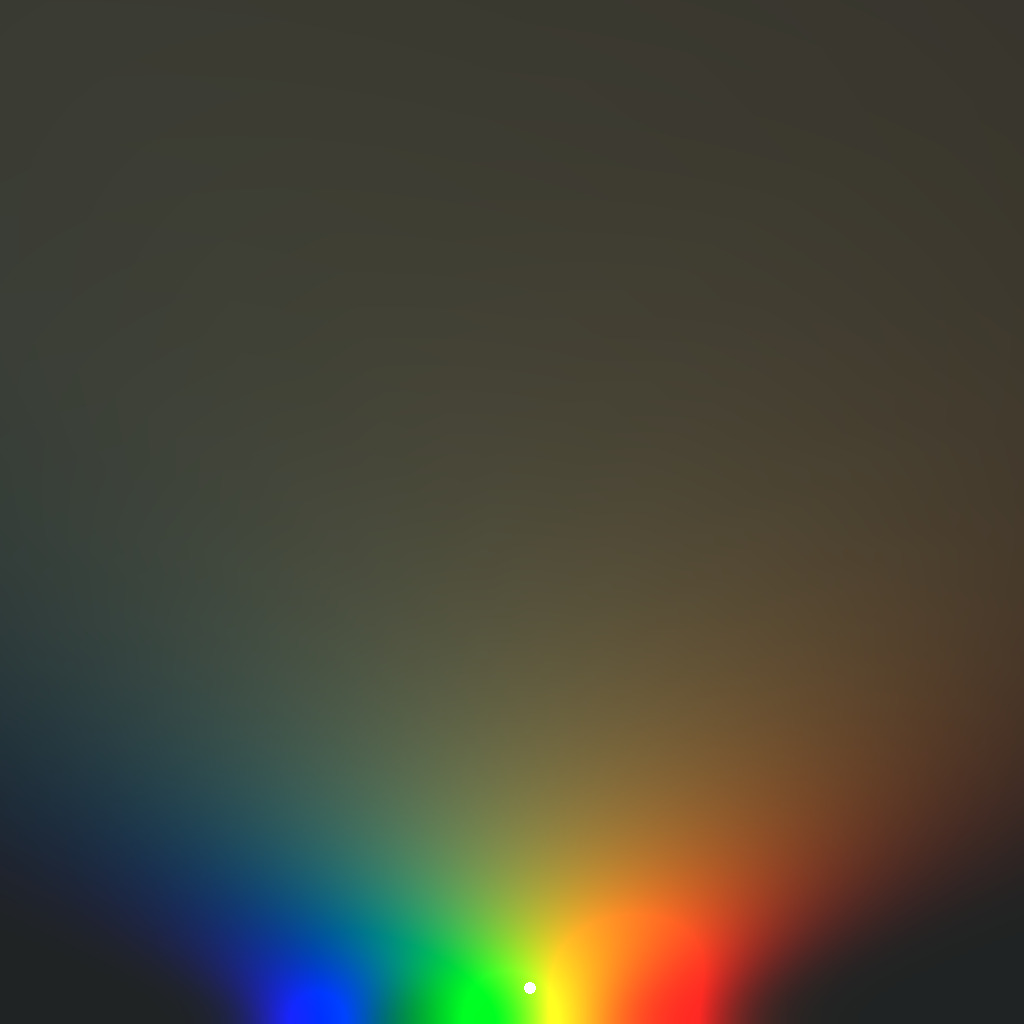}};

		\draw[fill=white] (paletteD65_0.center)++( 0.010\x, -0.460\x) circle[radius=1pt, color=black];
		\draw[white, <-, opacity=0.5] (paletteD65_0.center)++((0.01\x,-0.40\x) -- ++(0, 0.40\x) node[scale=0.6, yshift=5pt, opacity=1.0] { $\mu_e=560,\ \sigma_e=14$ };
		\draw[fill=white] (paletteD65_1.center)++( 0.010\x, -0.460\x) circle[radius=1pt, color=black];
		\draw[white, <-, opacity=0.5] (paletteD65_1.center)++((0.01\x,-0.40\x) -- ++(0, 0.40\x) node[scale=0.6, yshift=5pt, opacity=1.0] { $\mu_e=560,\ \sigma_e=14$ };
		\draw[fill=white] (paletteD65_2.center)++( 0.010\x, -0.460\x) circle[radius=1pt, color=black];
		\draw[white, <-, opacity=0.5] (paletteD65_2.center)++((0.01\x,-0.40\x) -- ++(0, 0.40\x) node[scale=0.6, yshift=5pt, opacity=1.0] { $\mu_e=560,\ \sigma_e=14$ };

		\draw[white, <->] (paletteD65_0.south west)++(0.02\x,0.04\x)
			node[anchor=south east, rotate=-90,font=\tiny,scale=0.6]{$1$} -- ++(0, 0.9\x)
			node[anchor=south west, rotate=-90,font=\tiny,scale=0.6]{$500$};% -- ++(0.95\x, 0) ;
		\draw[white, <->] (paletteD65_0.north west)++(0.04\x, -0.02\x)
			node[anchor=north west, font=\tiny,scale=0.6]{$300$} -- ++(0.9\x, 0)
			node[anchor=north east, font=\tiny,scale=0.6]{$800$};% -- ++(0.95\x, 0) ;
		\node[anchor=north, font=\tiny,scale=0.7] at (paletteD65_0.north) {{\color{white}$\mu_e\:[\mathrm{nm}]$}};
		\node[anchor=south, font=\tiny,scale=0.7, rotate=-90] at (paletteD65_0.west)  {{\color{white}$\sigma_e\:[\mathrm{nm}]$}};

		\draw (paletteD65_0.south west) rectangle (paletteD65_0.north east);
		\draw (paletteD65_1.south west) rectangle (paletteD65_1.north east);
		\draw (paletteD65_2.south west) rectangle (paletteD65_2.north east);

		% Palette A
		\node[inner sep=0pt, right=2pt of paletteD65_0] (paletteA_0) {\includegraphics[width=\x]{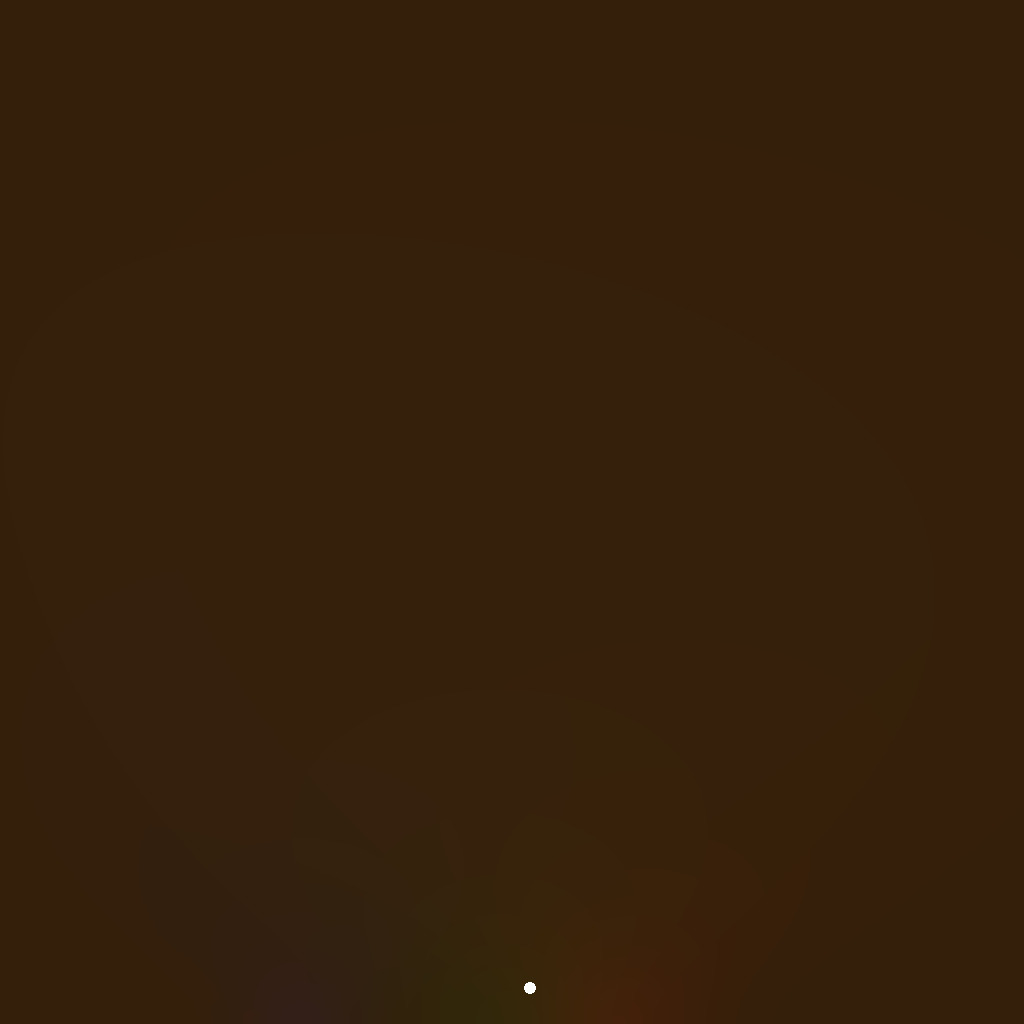}};
		\node[inner sep=0pt, right=2pt of paletteD65_1] (paletteA_1) {\includegraphics[width=\x]{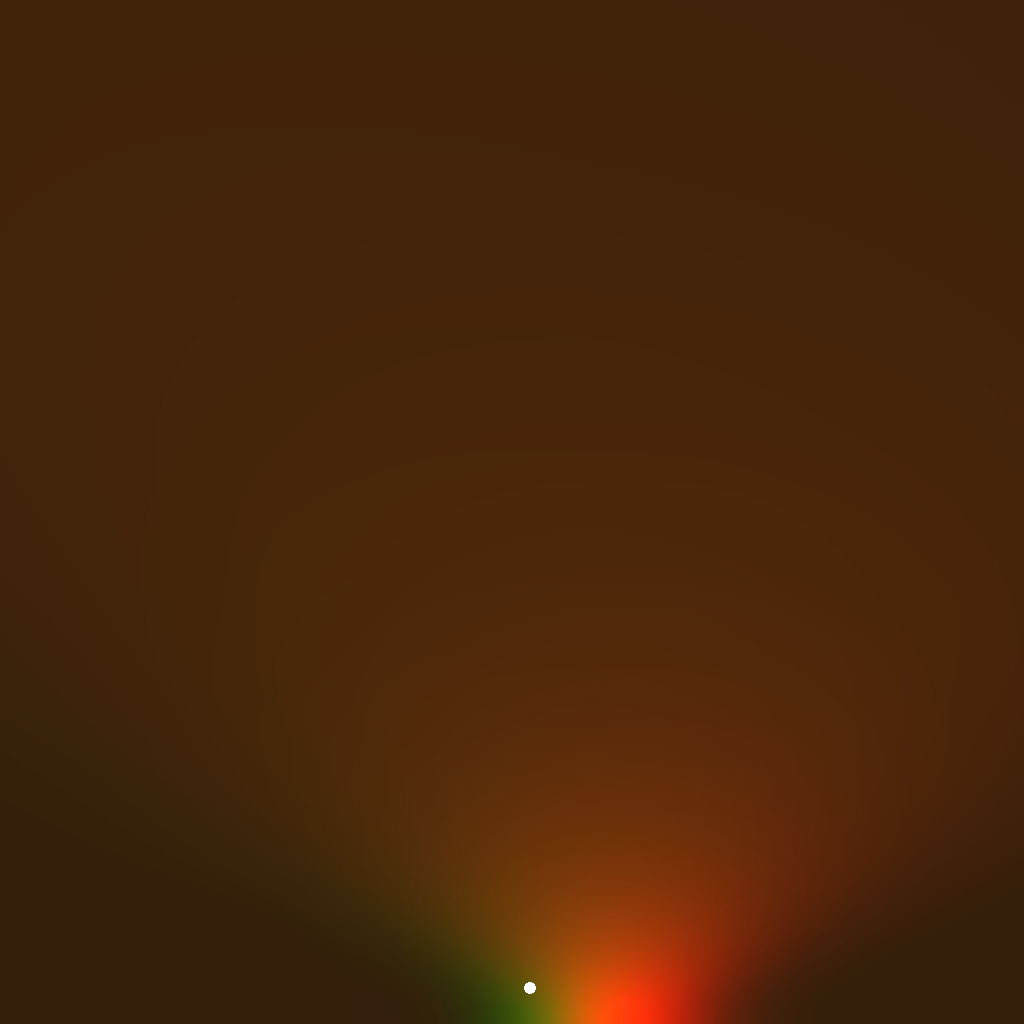}};
		\node[inner sep=0pt, right=2pt of paletteD65_2] (paletteA_2) {\includegraphics[width=\x]{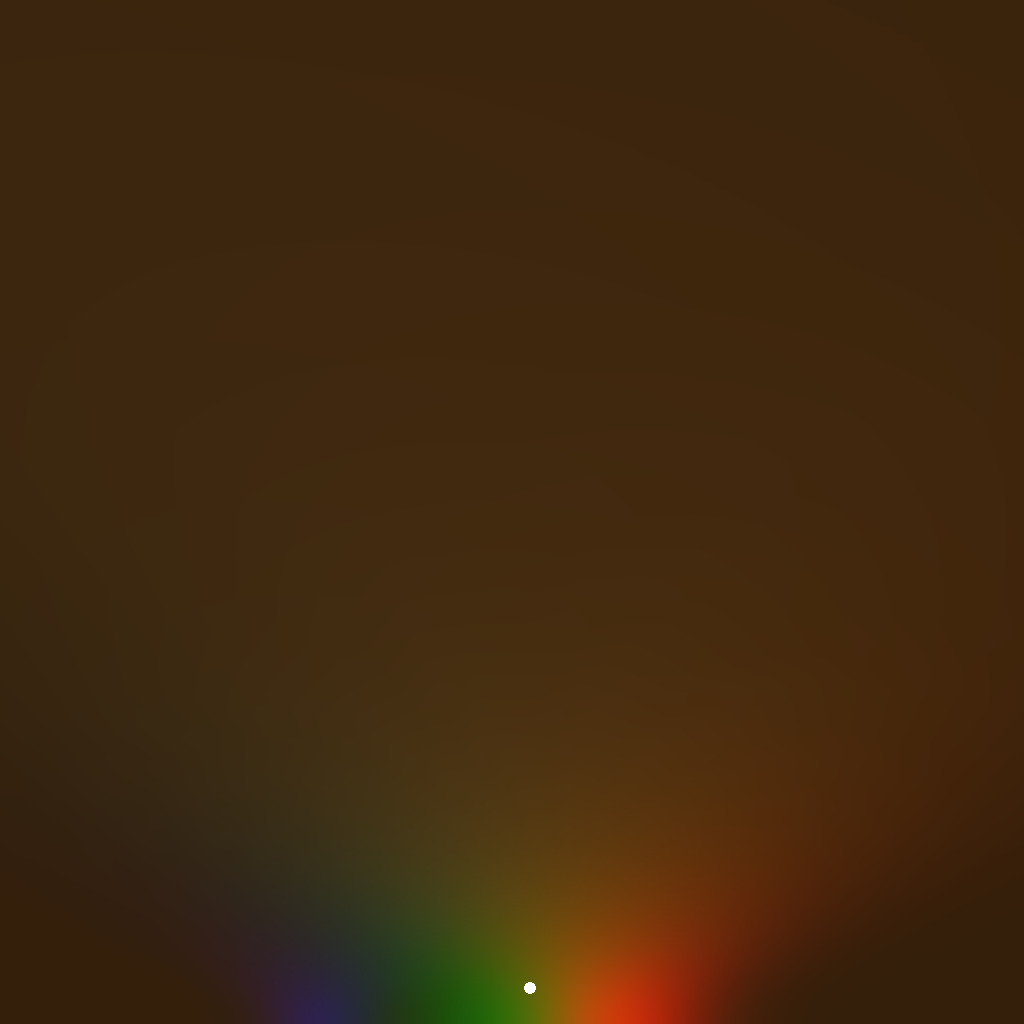}};

		\draw (paletteA_0.south west) rectangle (paletteA_0.north east);
		\draw (paletteA_1.south west) rectangle (paletteA_1.north east);
		\draw (paletteA_2.south west) rectangle (paletteA_2.north east);

		\draw[fill=white] (paletteA_0.center)++( 0.010\x, -0.460\x) circle[radius=1pt, color=black];
		\draw[fill=white] (paletteA_1.center)++( 0.010\x, -0.460\x) circle[radius=1pt, color=black];
		\draw[fill=white] (paletteA_2.center)++( 0.010\x, -0.460\x) circle[radius=1pt, color=black];

		% Palette UV
		\node[inner sep=0pt, right=2pt of paletteA_0] (paletteUV_0) {\includegraphics[width=\x]{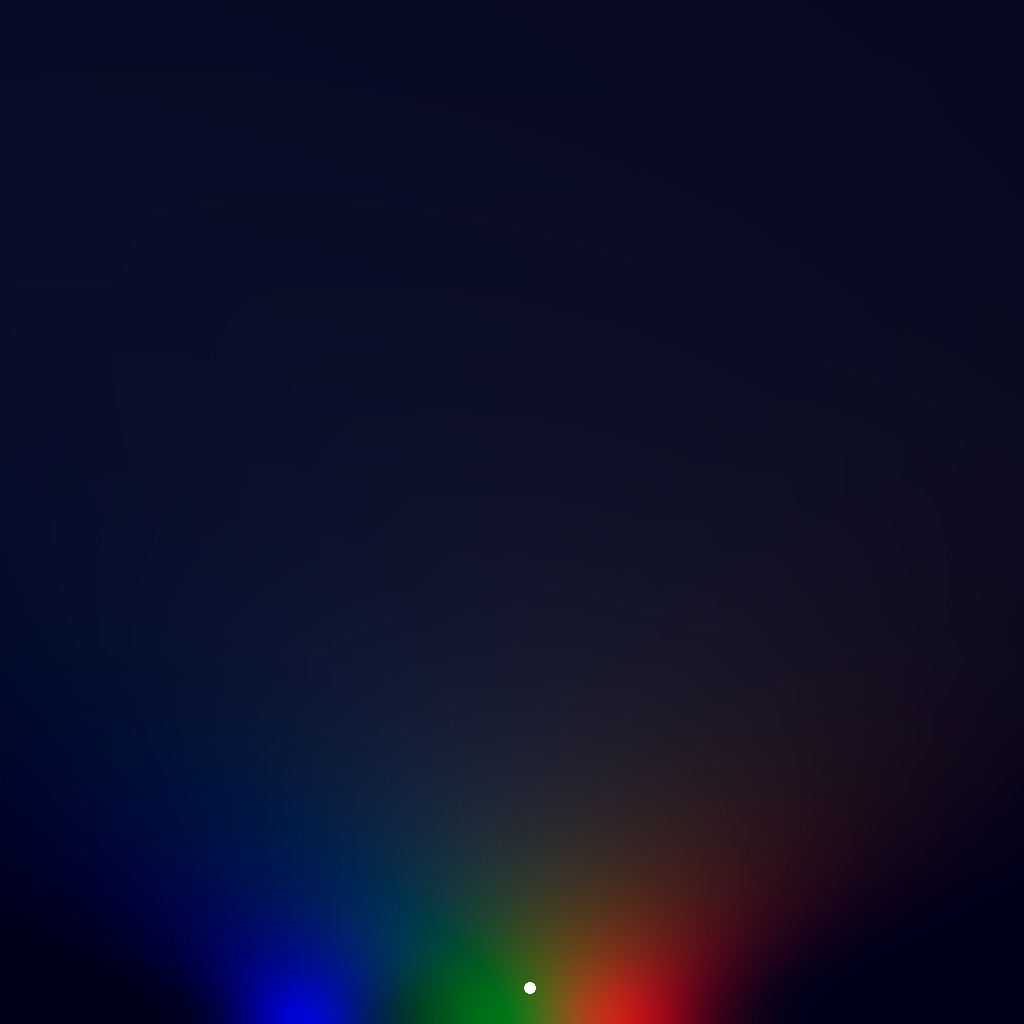}};
		\node[inner sep=0pt, right=2pt of paletteA_1] (paletteUV_1) {\includegraphics[width=\x]{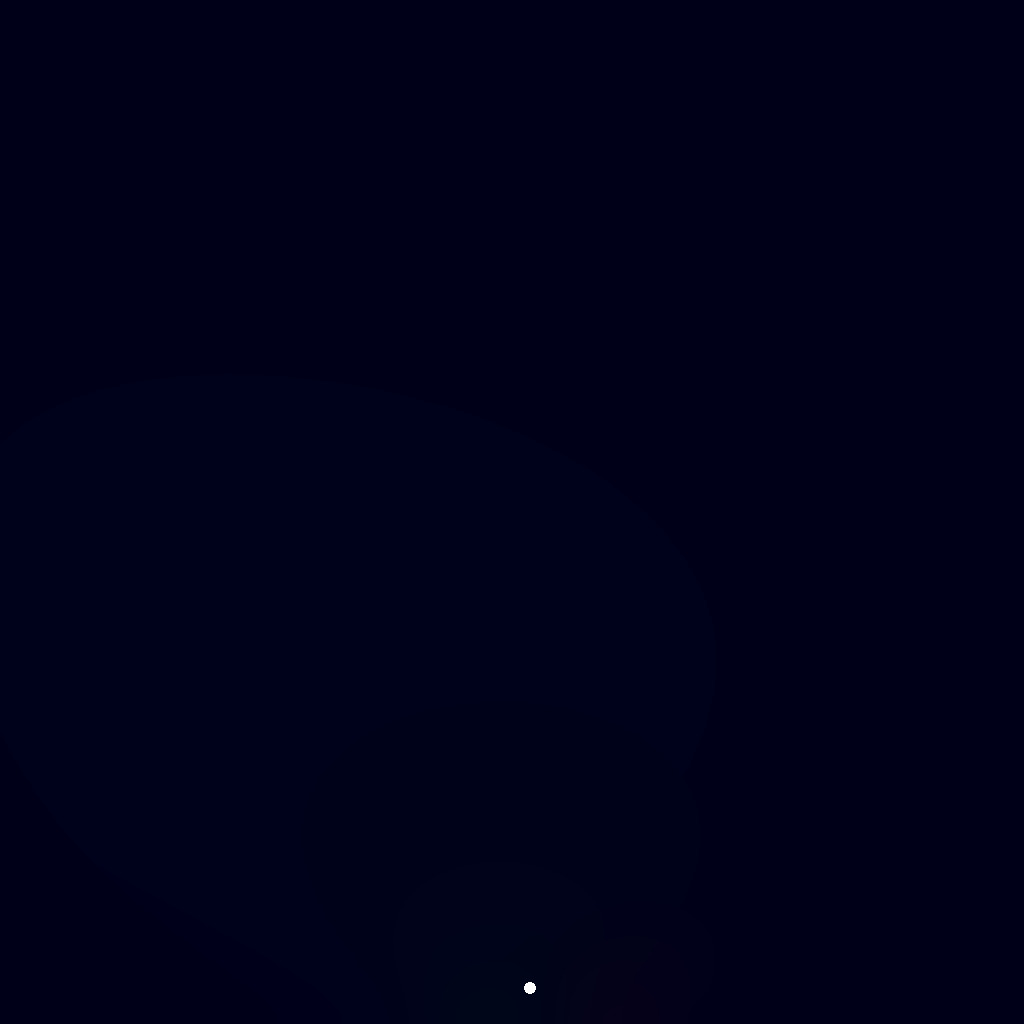}};
		\node[inner sep=0pt, right=2pt of paletteA_2] (paletteUV_2) {\includegraphics[width=\x]{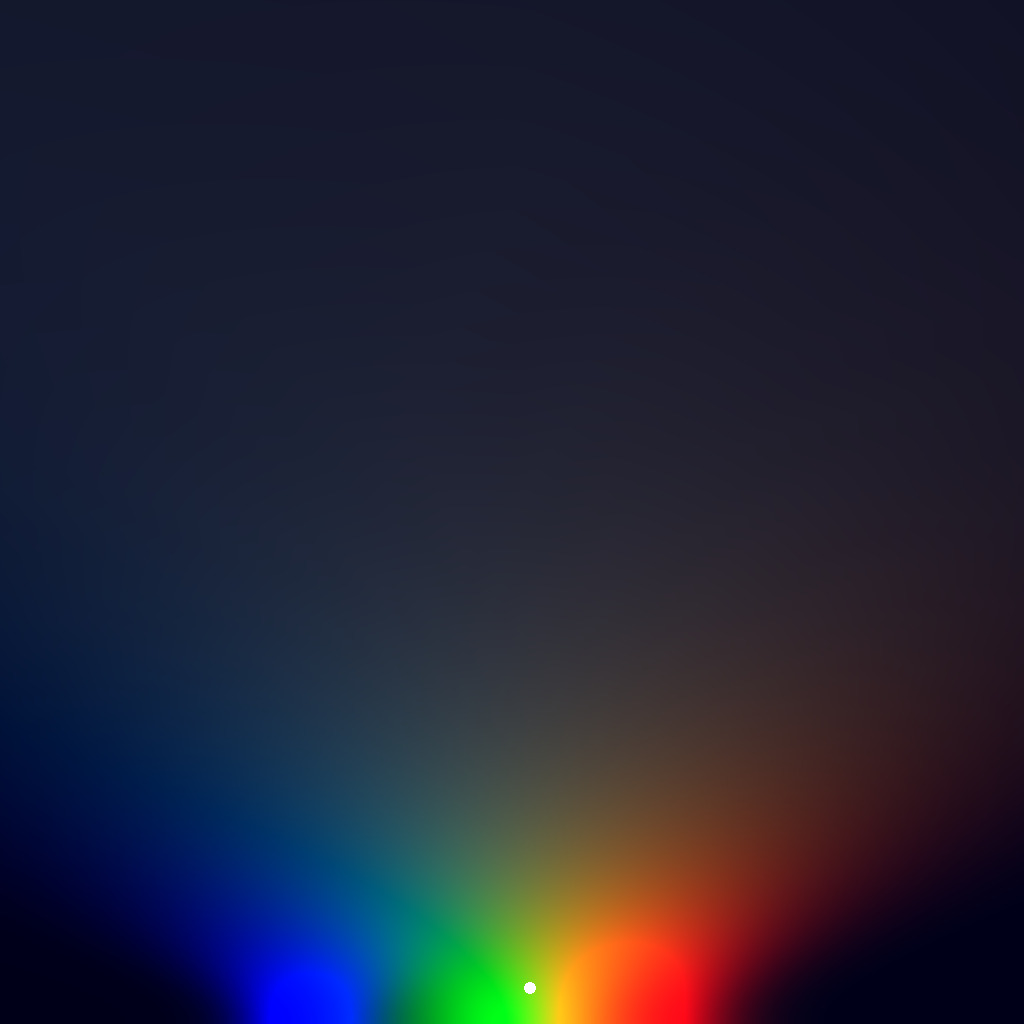}};

		\draw (paletteUV_0.south west) rectangle (paletteUV_0.north east);
		\draw (paletteUV_1.south west) rectangle (paletteUV_1.north east);
		\draw (paletteUV_2.south west) rectangle (paletteUV_2.north east);

		\draw[fill=white] (paletteUV_0.center)++( 0.010\x, -0.460\x) circle[radius=1pt, color=black];
		\draw[fill=white] (paletteUV_1.center)++( 0.010\x, -0.460\x) circle[radius=1pt, color=black];
		\draw[fill=white] (paletteUV_2.center)++( 0.010\x, -0.460\x) circle[radius=1pt, color=black];

		% Render D65
		\node[inner sep=0pt, right=2pt of paletteUV_0] (renderD65_0) {\includegraphics[width=\x]{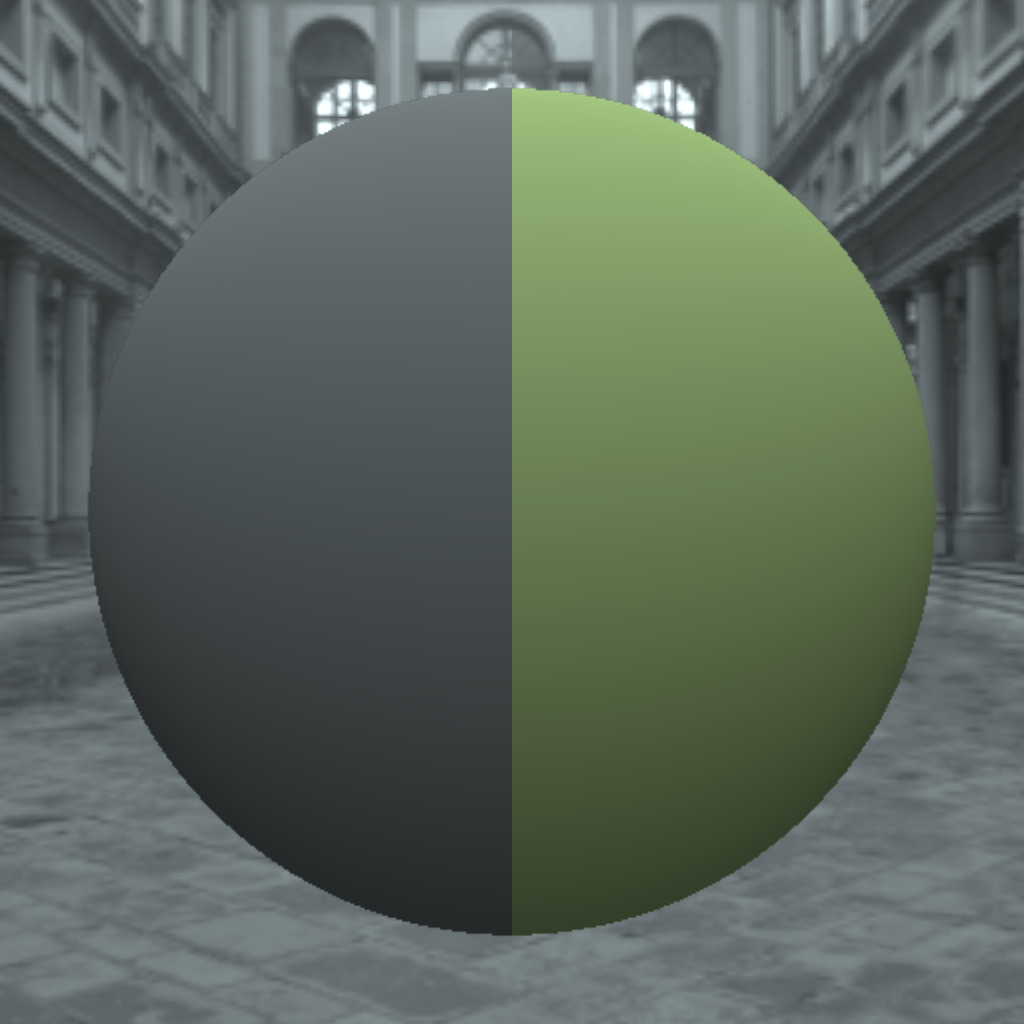}};
		\node[inner sep=0pt, right=2pt of paletteUV_1] (renderD65_1) {\includegraphics[width=\x]{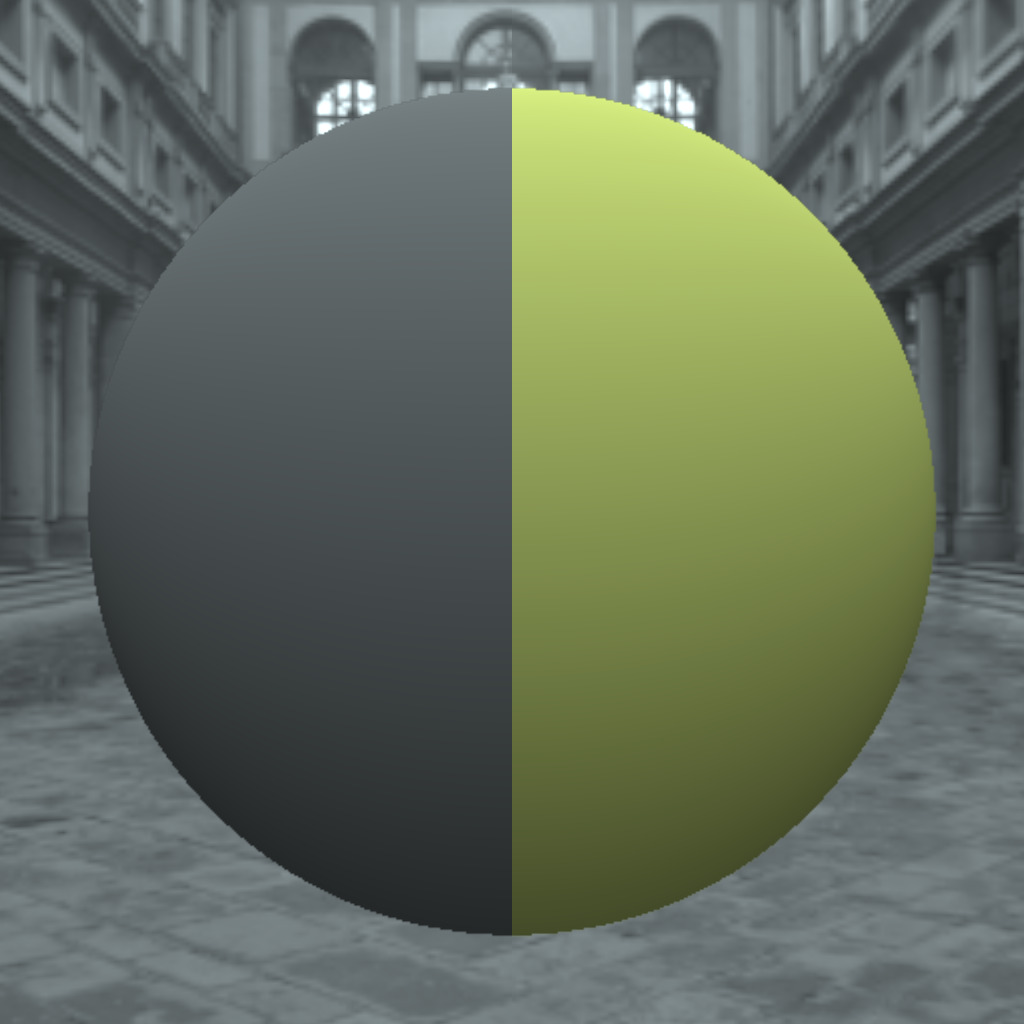}};
		\node[inner sep=0pt, right=2pt of paletteUV_2] (renderD65_2) {\includegraphics[width=\x]{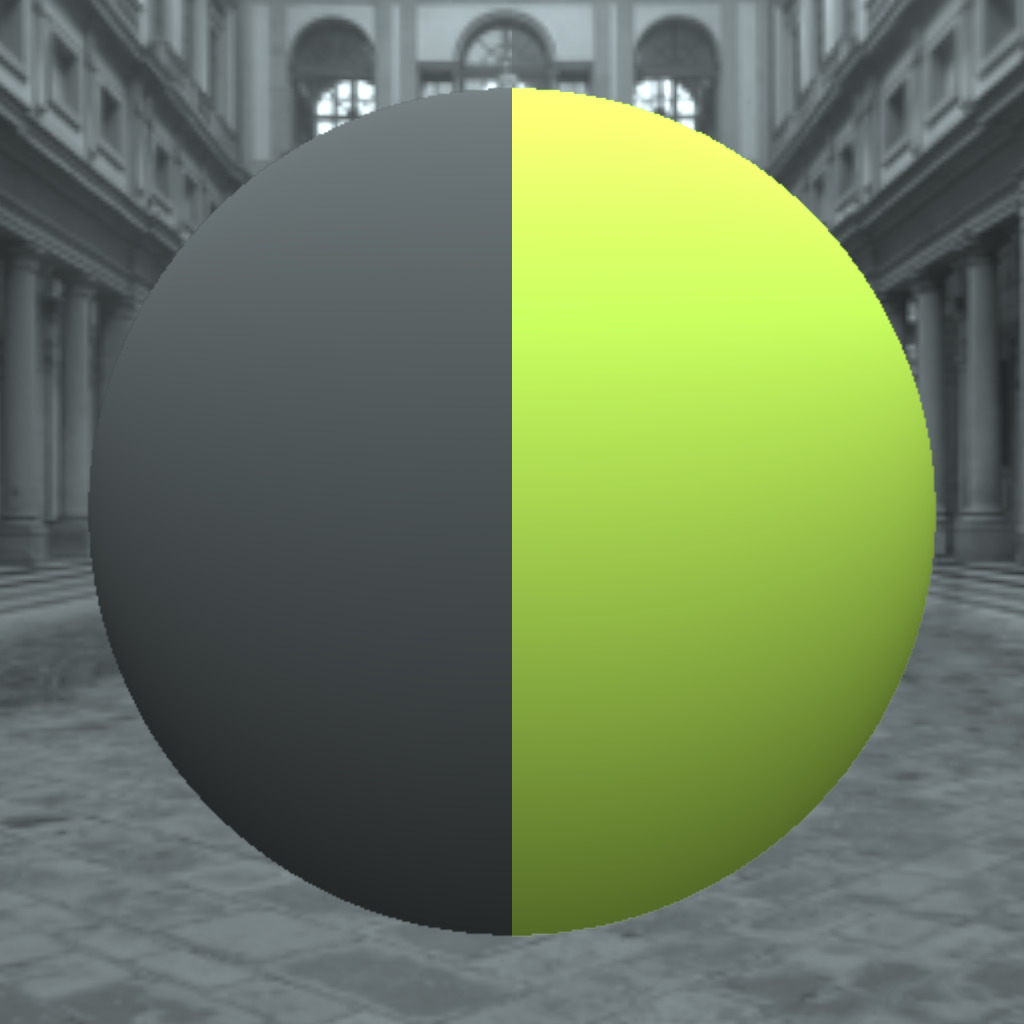}};

		\draw (renderD65_0.south west) rectangle (renderD65_0.north east);
		\draw (renderD65_1.south west) rectangle (renderD65_1.north east);
		\draw (renderD65_2.south west) rectangle (renderD65_2.north east);

		\draw[very thin] (renderD65_0.north) -- (renderD65_0.south);
		\draw[very thin] (renderD65_1.north) -- (renderD65_1.south);
		\draw[very thin] (renderD65_2.north) -- (renderD65_2.south);

		% Render A
		\node[inner sep=0pt, right=2pt of renderD65_0] (renderA_0) {\includegraphics[width=\x]{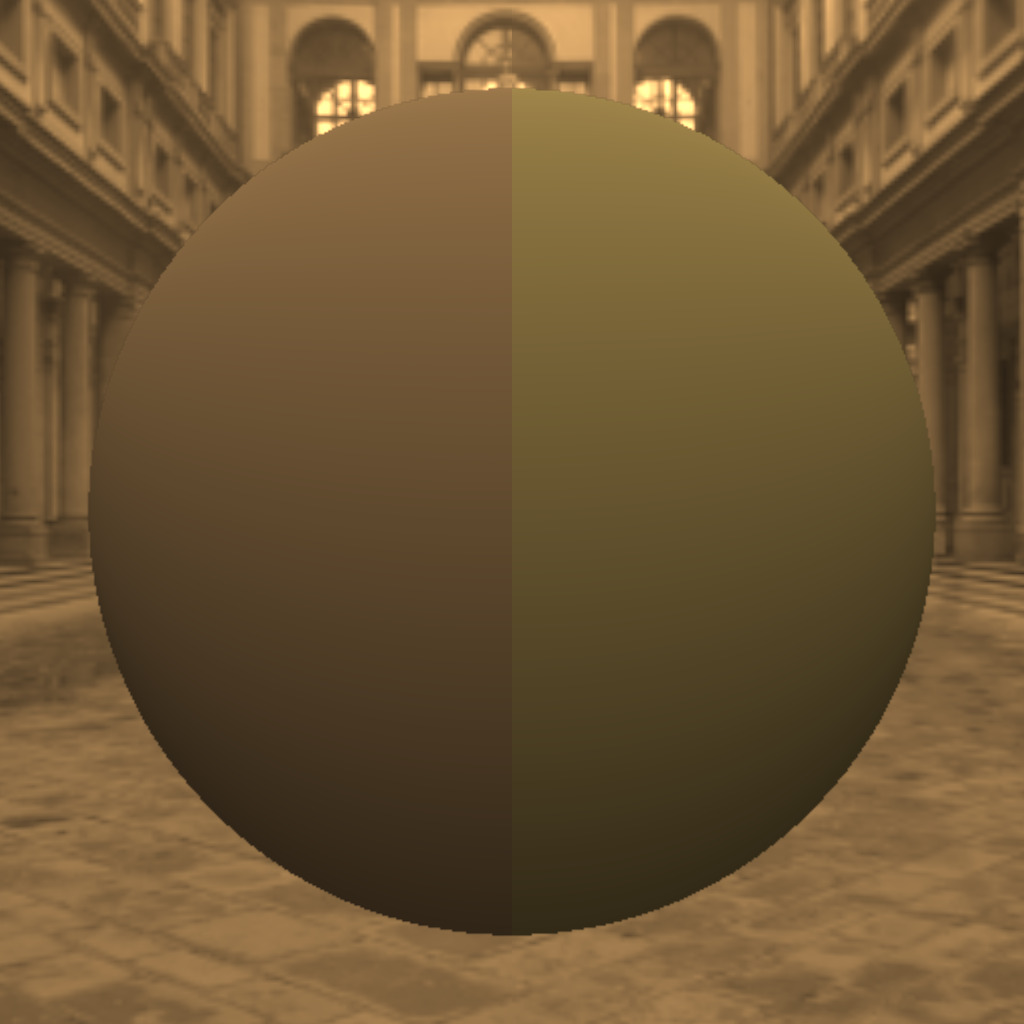}};
		\node[inner sep=0pt, right=2pt of renderD65_1] (renderA_1) {\includegraphics[width=\x]{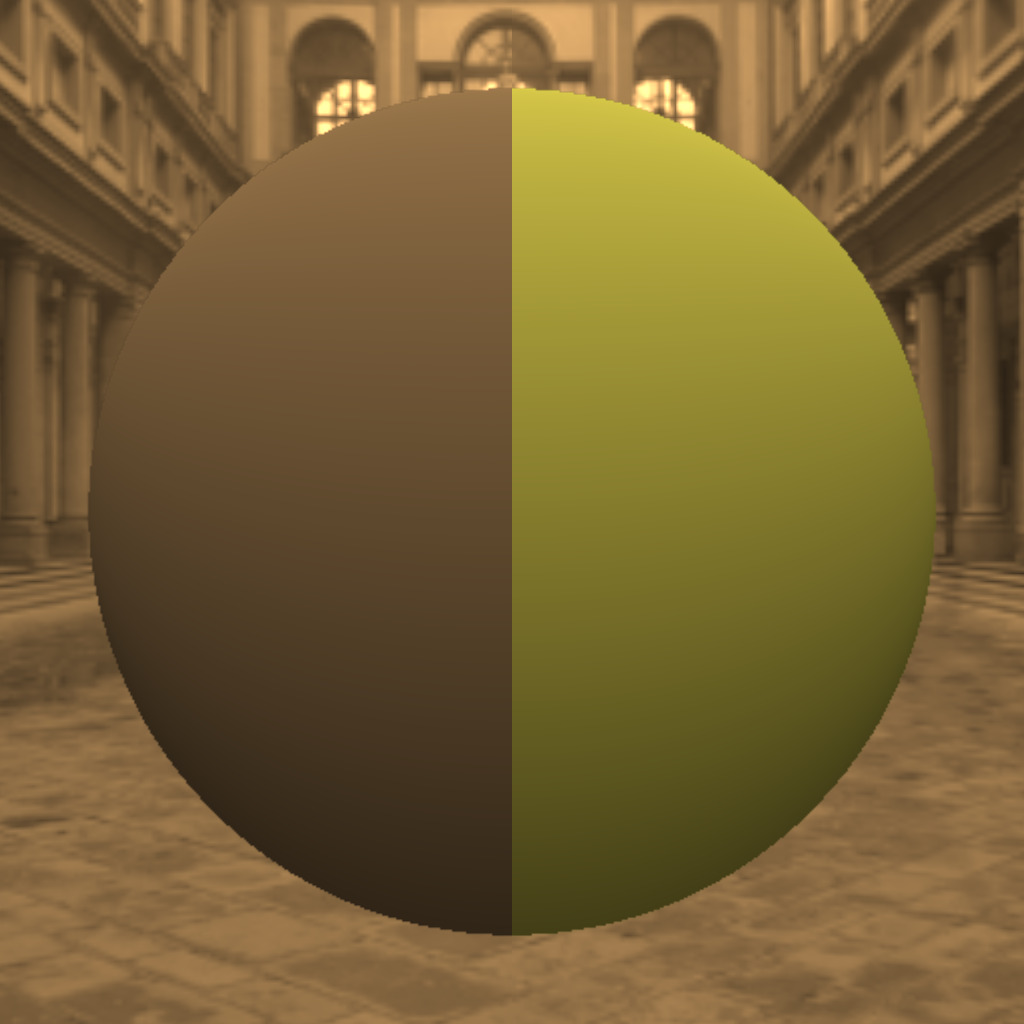}};
		\node[inner sep=0pt, right=2pt of renderD65_2] (renderA_2) {\includegraphics[width=\x]{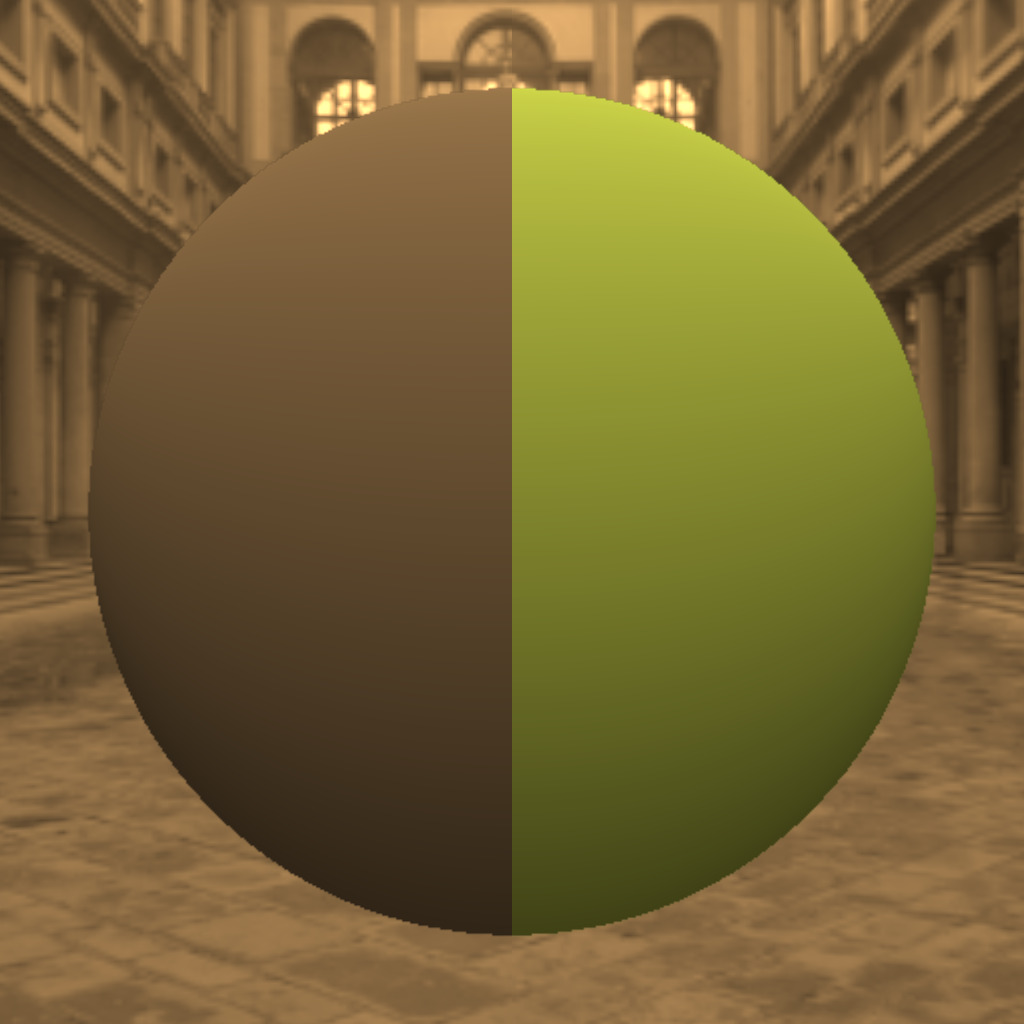}};

		\draw (renderA_0.south west) rectangle (renderA_0.north east);
		\draw (renderA_1.south west) rectangle (renderA_1.north east);
		\draw (renderA_2.south west) rectangle (renderA_2.north east);

		\draw[very thin] (renderA_0.north) -- (renderA_0.south);
		\draw[very thin] (renderA_1.north) -- (renderA_1.south);
		\draw[very thin] (renderA_2.north) -- (renderA_2.south);

		% Rennder UV
		\node[inner sep=0pt, right=2pt of renderA_0] (renderUV_0) {\includegraphics[width=\x]{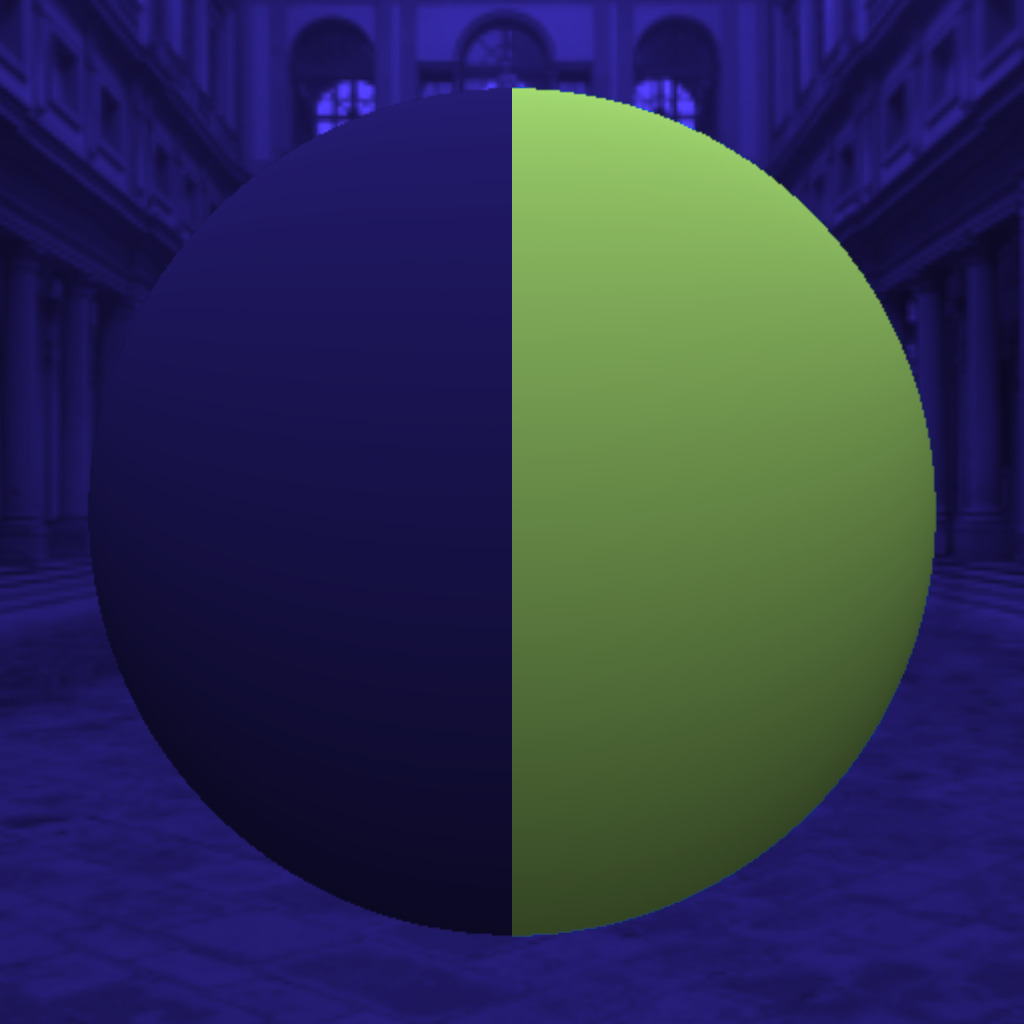}};
		\node[inner sep=0pt, right=2pt of renderA_1] (renderUV_1) {\includegraphics[width=\x]{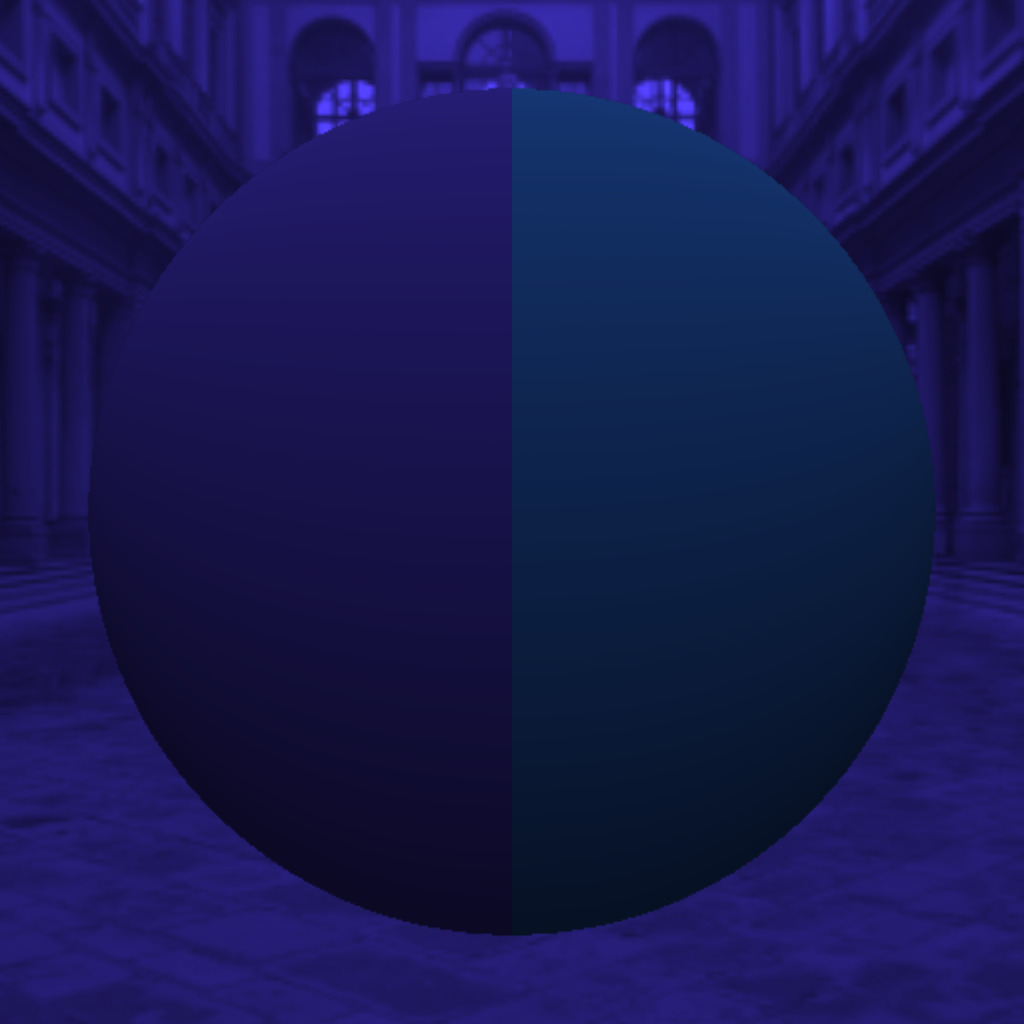}};
		\node[inner sep=0pt, right=2pt of renderA_2] (renderUV_2) {\includegraphics[width=\x]{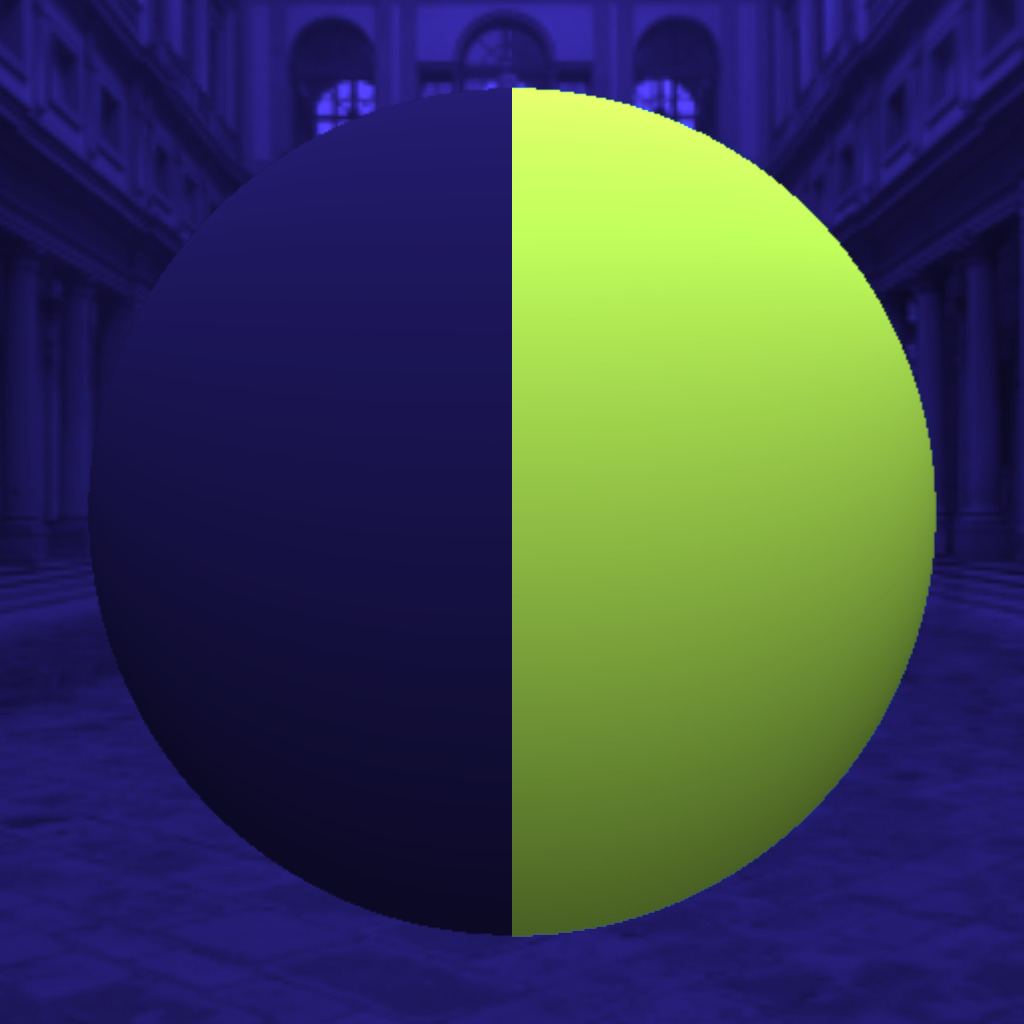}};

		\draw (renderUV_0.south west) rectangle (renderUV_0.north east);
		\draw (renderUV_1.south west) rectangle (renderUV_1.north east);
		\draw (renderUV_2.south west) rectangle (renderUV_2.north east);

		\draw[very thin] (renderUV_0.north) -- (renderUV_0.south);
		\draw[very thin] (renderUV_1.north) -- (renderUV_1.south);
		\draw[very thin] (renderUV_2.north) -- (renderUV_2.south);

		% Labels
		\node[inner sep=0pt, above=2pt of rerad_0     ] { Fluorescence \protect{$\bar{\mathcal{F}}$} };
		\node[inner sep=0pt, above=1pt of paletteD65_0] { D65 palette };
		\node[inner sep=0pt, above=1pt of paletteA_0  ] { A palette };
		\node[inner sep=0pt, above=1pt of paletteUV_0 ] { UV palette };
		\node[inner sep=0pt, above=3pt of renderD65_0] { D65 render};
		\node[inner sep=0pt, above=3pt of renderA_0  ] { A render };
		\node[inner sep=0pt, above=3pt of renderUV_0 ] { UV render };

		\node[inner sep=0pt, left=3pt of rerad_0, rotate=90, minimum width=\x, anchor=south] { $\mu_a=320, \sigma_a=30$ };
		\node[inner sep=0pt, left=5pt of rerad_1, rotate=90, minimum width=\x, anchor=south] { $\mu_a=530, \sigma_a=30$ };
		\node[inner sep=0pt, left=3pt of rerad_2, rotate=90, minimum width=\x, anchor=south] { $\mu_a=420, \sigma_a=60$ };
	\end{tikzpicture}
	}
	\vspace{-18pt}
	\caption{\textbf{Fluorescence palettes with varying absorption.}
	We show fluorescence palettes for three types of illuminants (daylight D65, incandescent A, and a Gaussian illumination spectrum in the UV-blue range), and three absorption configurations as seen in $\mathcal{\bar{F}}$ (first column): narrow absorption in the UV range (first row), in the visible range (second row) or wide absorption (third row).
	We use an achromatic albedo $\rho=0.1$ throughout.
	In the last three columns, we show renderings of spheres, where the left and right sides use a non-fluorescent VS fluorescent material, obtained by setting $\alpha_n=0$ and $\alpha_n=1$ respectively.
	The fluorescence color has been chosen by picking a point in the corresponding palette (white dots), in this case yielding $\mu_e=560$nm  and $\sigma_e=14$nm.
	This shows that adjusting the absorption parameters allows for selective fluorescence behaviors: a narrow absorption in the UV (resp. in the visual) range significantly reduces fluorescence under the A (resp. UV) illuminant, while a wide absorption produces fluorescence in all lighting conditions.
	\label{fig:paletteAbsorb}
	\vspace{-5pt}
	}
\end{figure*}
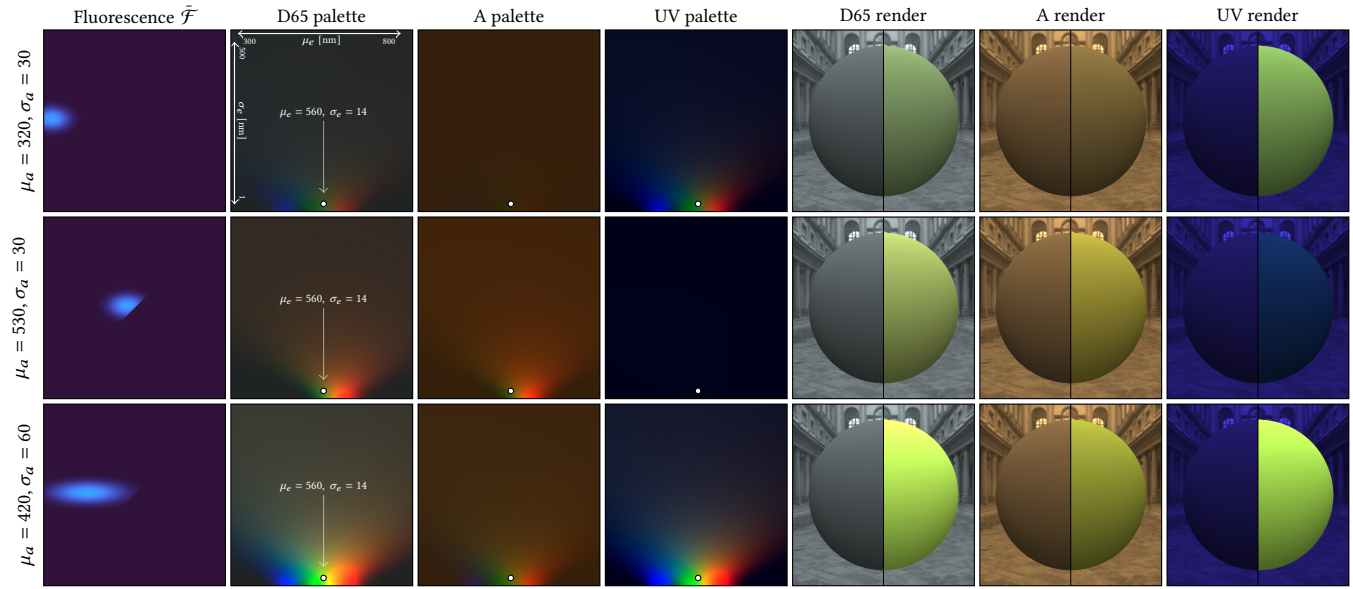

\subsection{Editing with fluorescence palettes}
\label{sec:simple-palettes}

\begin{figure}[h]
	% \vspace{-5pt}
	\resizebox{\linewidth}{!}{
	\begin{tikzpicture}[font=\footnotesize]
		\setlength{\x}{0.28\linewidth}

		\node[inner sep=0pt                      ] (rerad_0) {\includegraphics[width=\x]{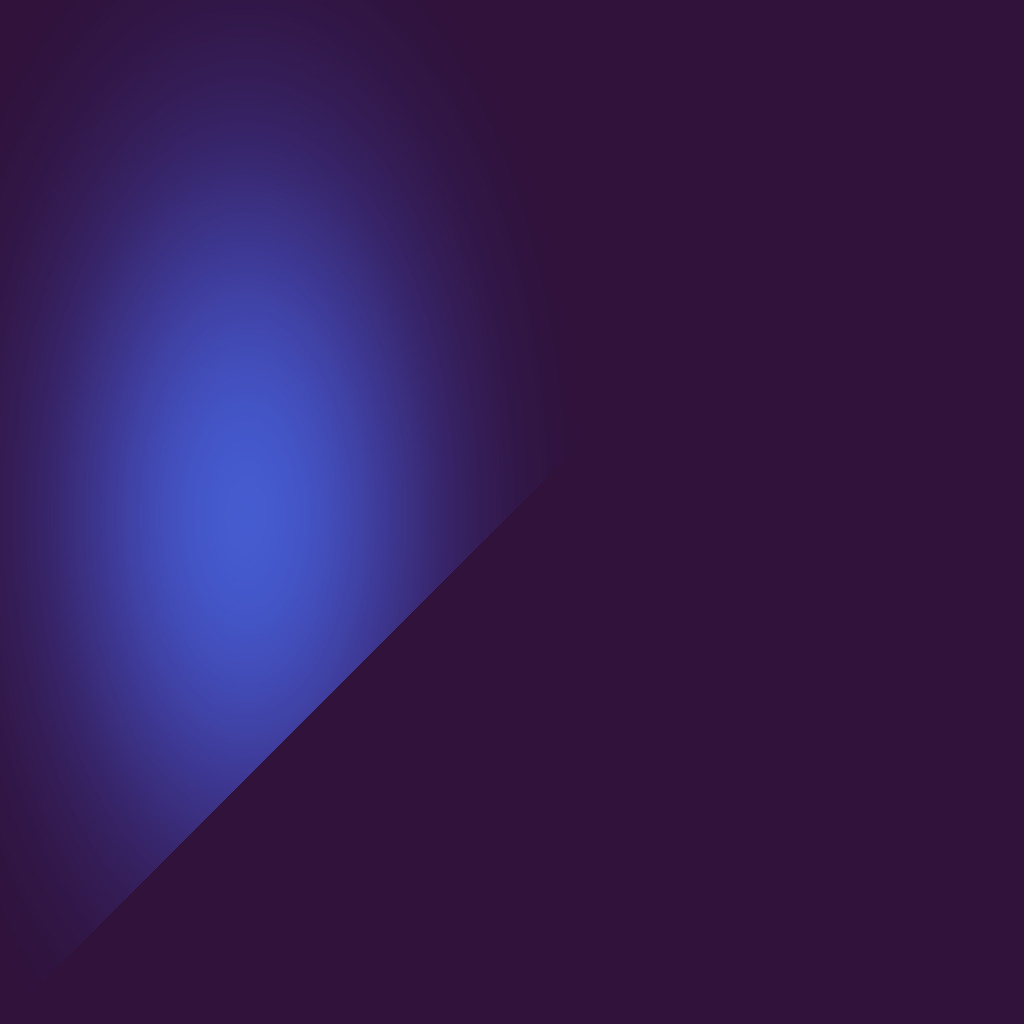}};
		\node[inner sep=0pt, below=2pt of rerad_0] (rerad_1) {\includegraphics[width=\x]{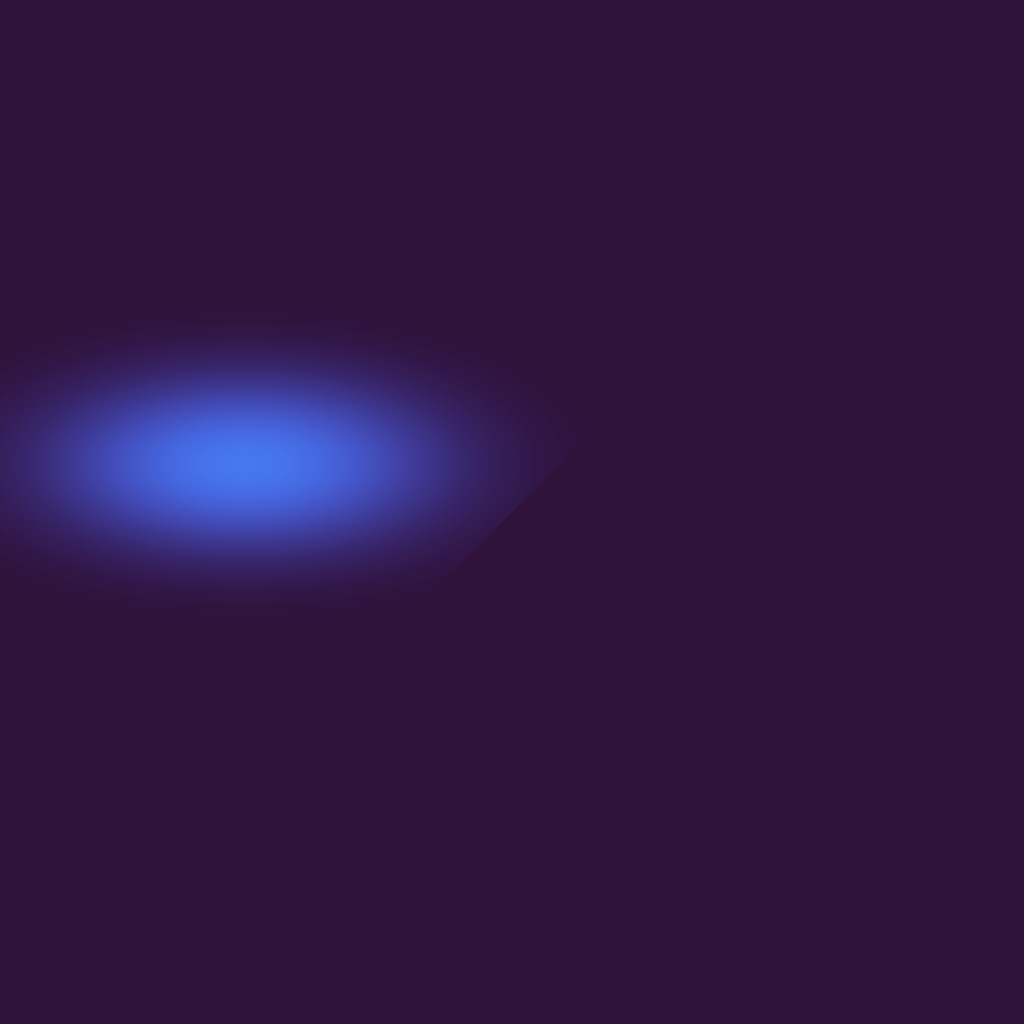}};
		\node[inner sep=0pt, below=2pt of rerad_1] (rerad_2) {\includegraphics[width=\x]{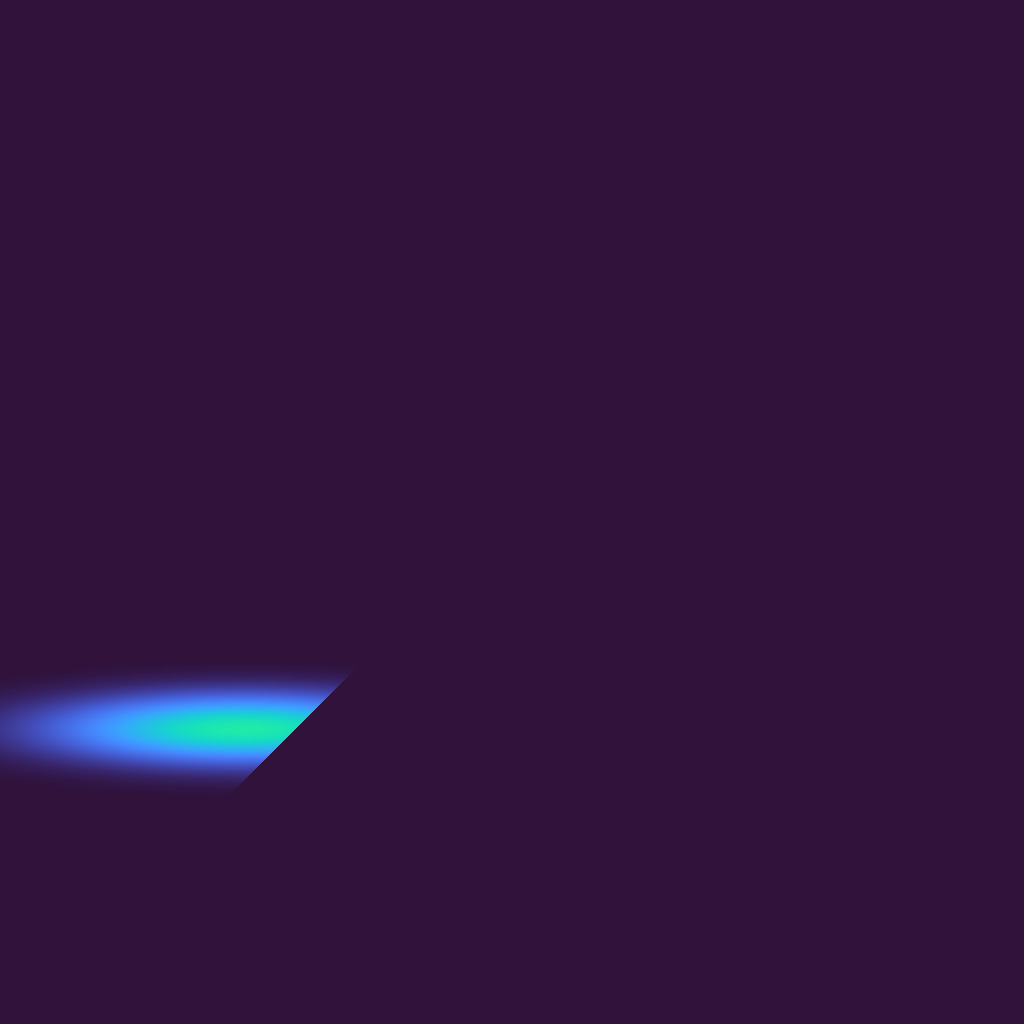}};

		\draw (rerad_0.south west) rectangle (rerad_0.north east);
		\draw (rerad_1.south west) rectangle (rerad_1.north east);
		\draw (rerad_2.south west) rectangle (rerad_2.north east);

		\node[inner sep=0pt, right=2pt of rerad_0] (palette_0) {\includegraphics[width=\x]{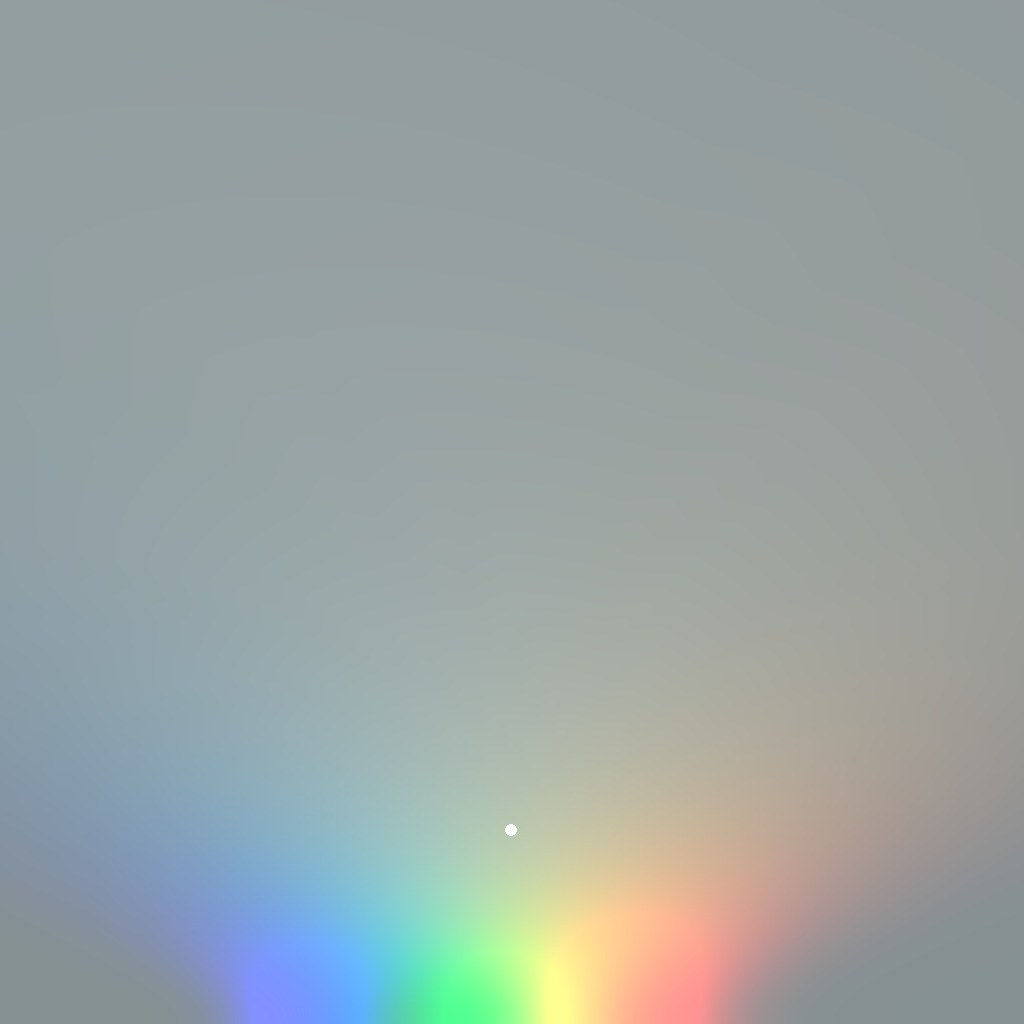}};
		\node[inner sep=0pt, right=2pt of rerad_1] (palette_1) {\includegraphics[width=\x]{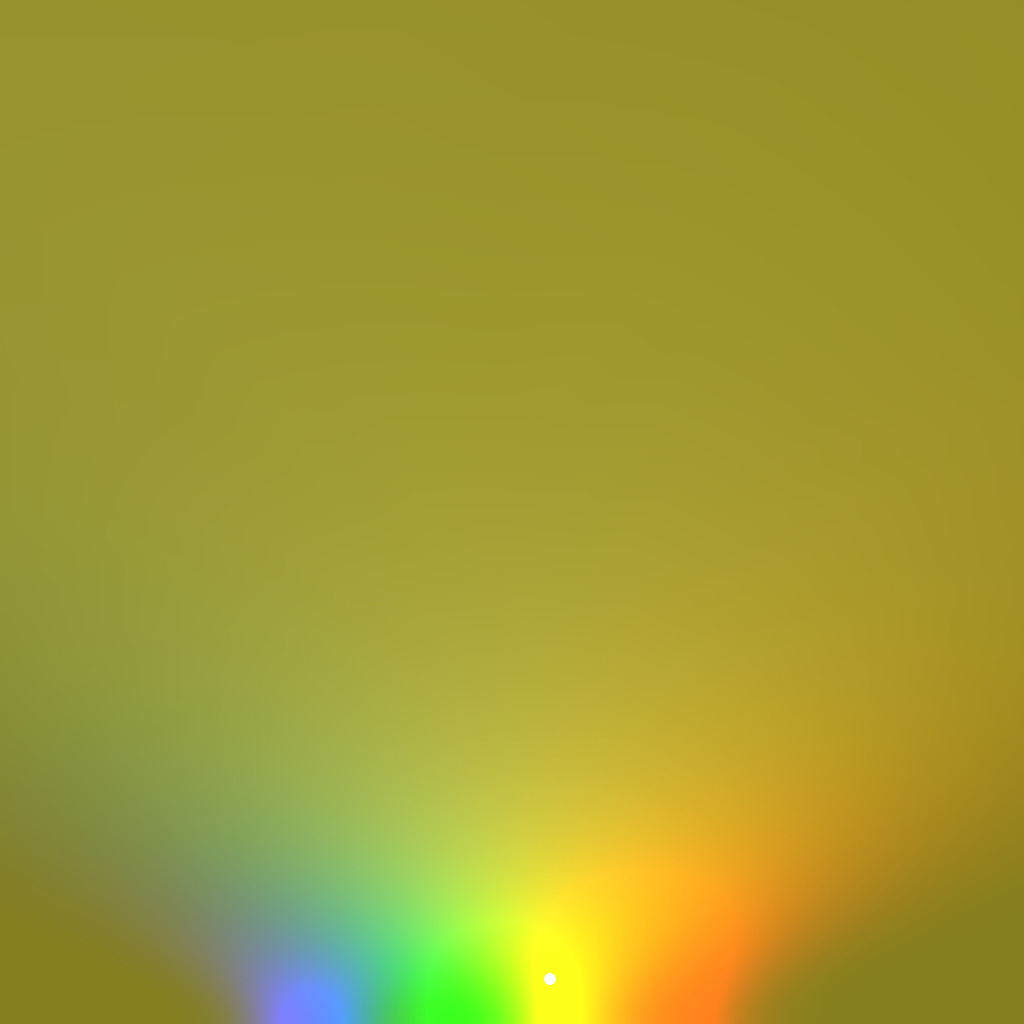}};
		\node[inner sep=0pt, right=2pt of rerad_2] (palette_2) {\includegraphics[width=\x]{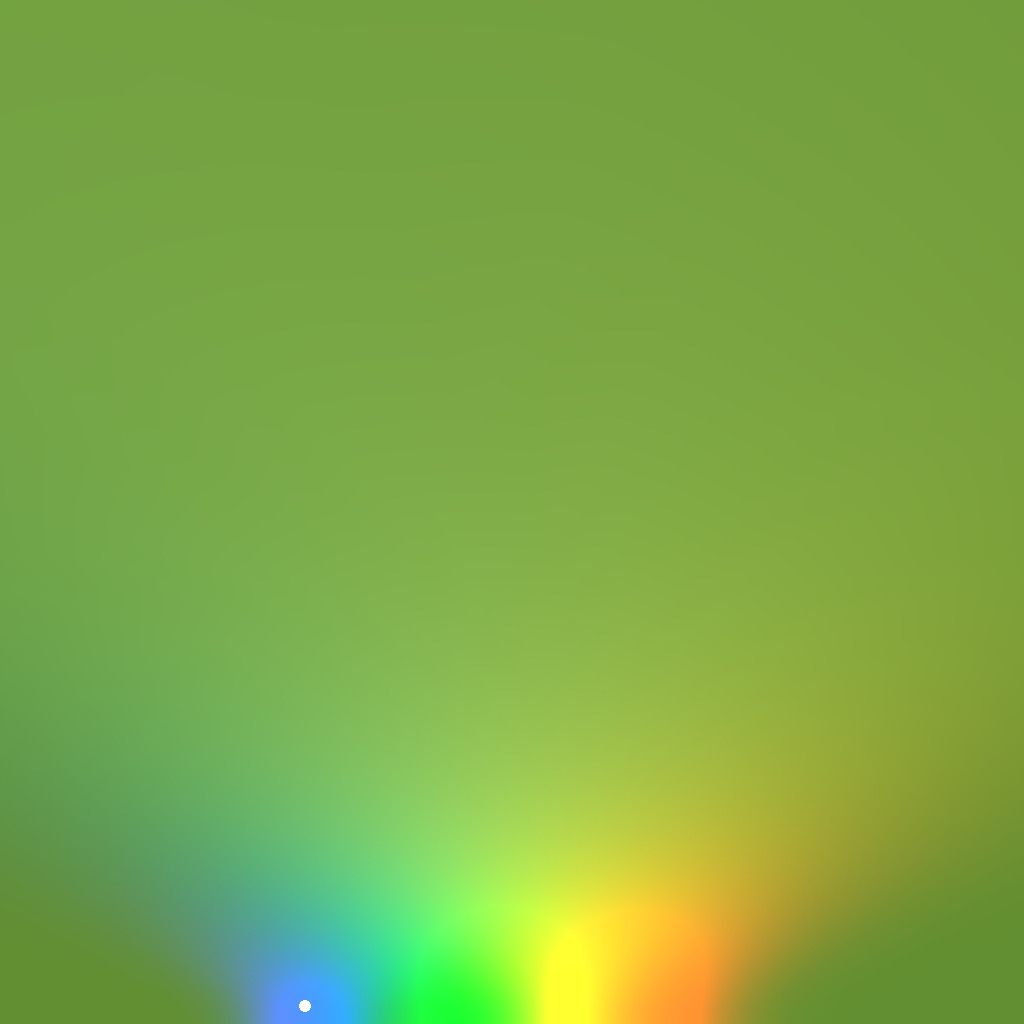}};

		\draw[fill=white] (palette_0.center)++( 0.000\x, -0.310\x) circle[radius=1pt, color=black];
		\draw[fill=white] (palette_1.center)++( 0.035\x, -0.455\x) circle[radius=1pt, color=black];
		\draw[fill=white] (palette_2.center)++(-0.200\x, -0.485\x) circle[radius=1pt, color=black];

		\draw (palette_0.south west) rectangle (palette_0.north east);
		\draw (palette_1.south west) rectangle (palette_1.north east);
		\draw (palette_2.south west) rectangle (palette_2.north east);

		\draw[white, <-, opacity=0.5] (palette_0.center)++(( 0.000\x,-0.270\x) -- ++(0, 0.40\x) node[scale=0.6, yshift=5pt, opacity=1.0] { $\mu_e=550, \sigma_e=95$ };
		\draw[white, <-, opacity=0.5] (palette_1.center)++(( 0.035\x,-0.415\x) -- ++(0, 0.50\x) node[scale=0.6, yshift=5pt, opacity=1.0] { $\mu_e=570, \sigma_e=22$ };
		\draw[white, <-, opacity=0.5] (palette_2.center)++((-0.200\x,-0.445\x) -- ++(0, 0.60\x) node[scale=0.6, yshift=5pt, opacity=1.0] { $\mu_e=450, \sigma_e=8$ };

		\node[inner sep=0pt, right=2pt of palette_0] (render_0) {\includegraphics[width=\x]{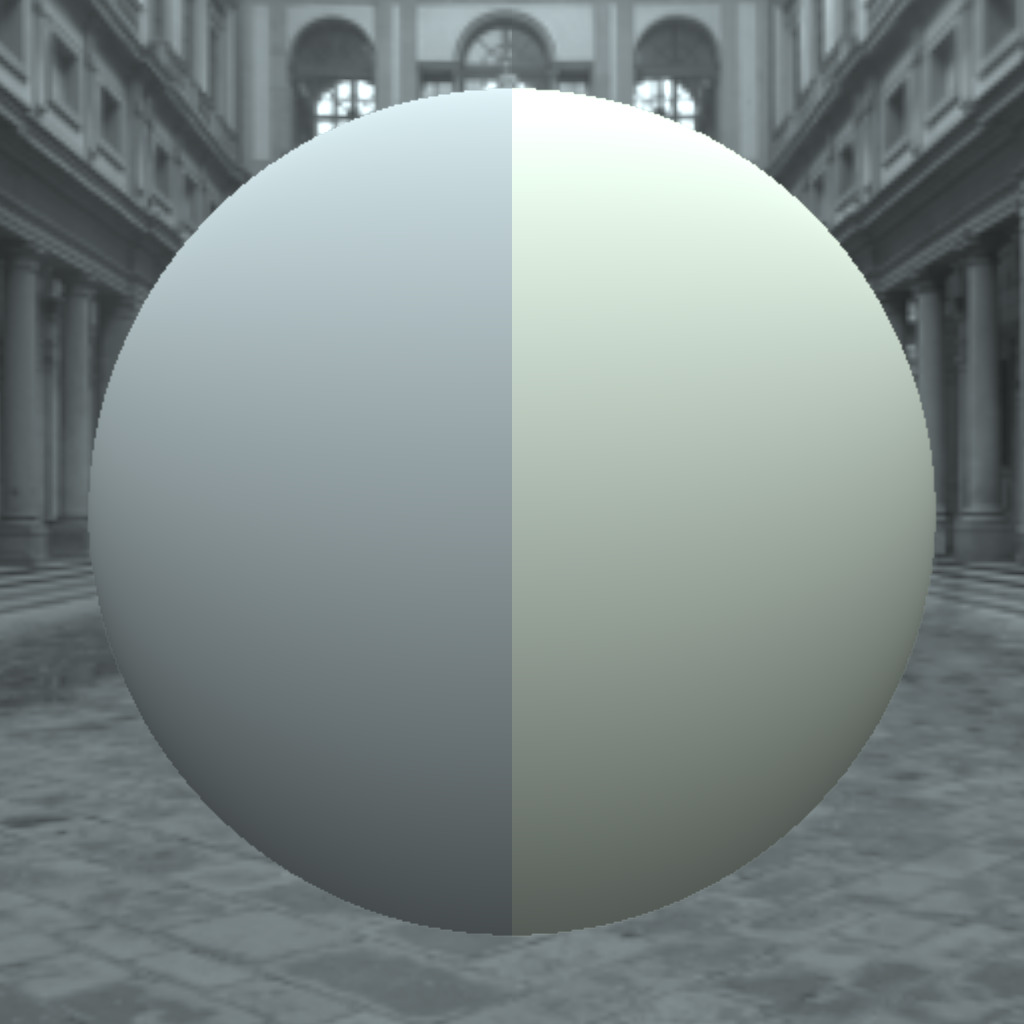}};
		\node[inner sep=0pt, right=2pt of palette_1] (render_1) {\includegraphics[width=\x]{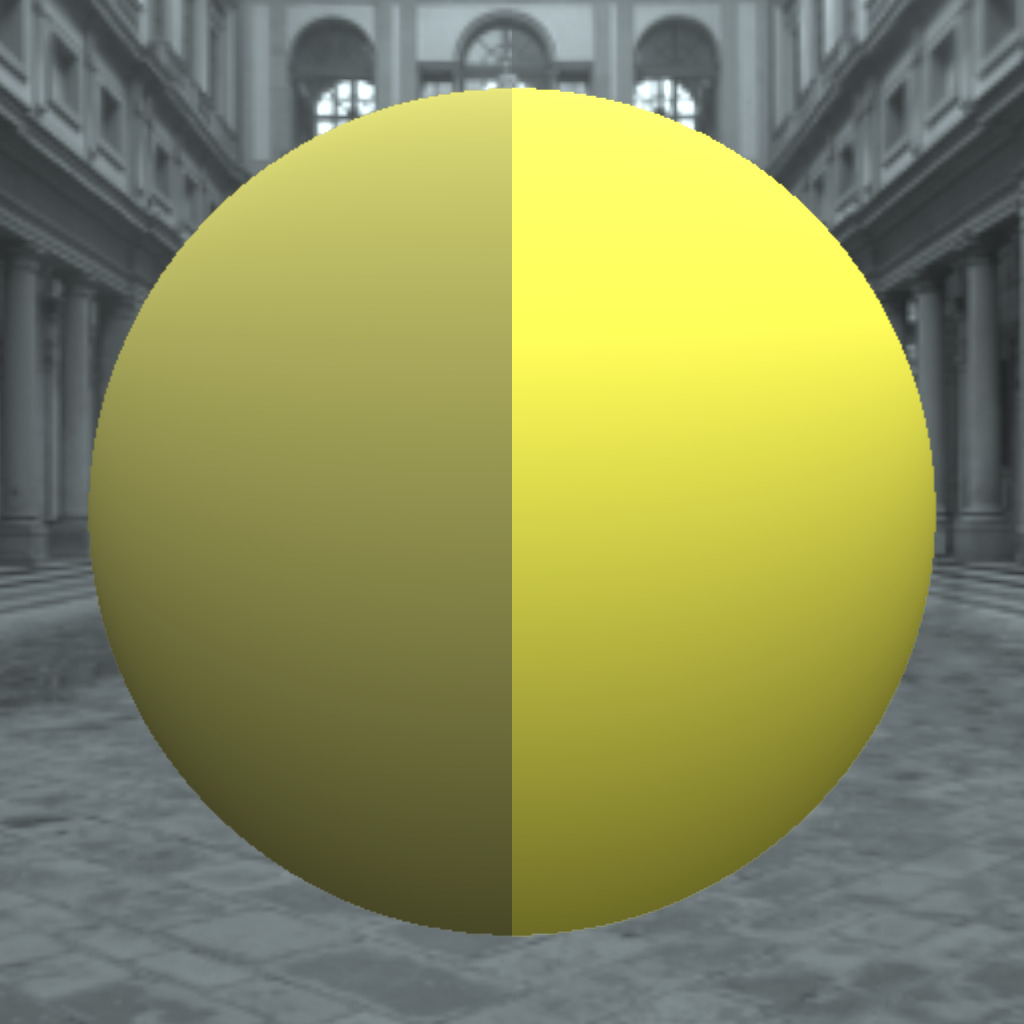}};
		\node[inner sep=0pt, right=2pt of palette_2] (render_2) {\includegraphics[width=\x]{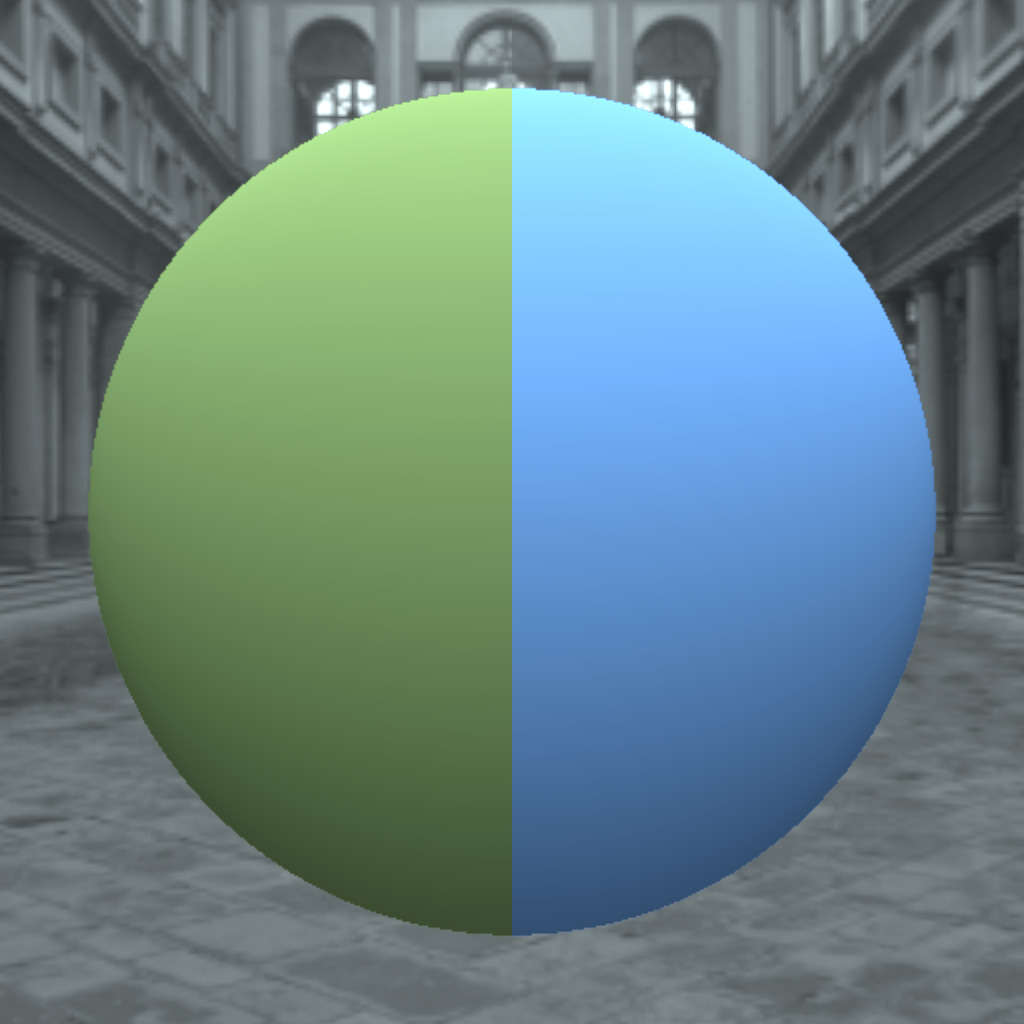}};

		\draw[very thin] (render_0.north) -- (render_0.south);
		\draw[very thin] (render_1.north) -- (render_1.south);
		\draw[very thin] (render_2.north) -- (render_2.south);

		\draw (render_0.south west) rectangle (render_0.north east);
		\draw (render_1.south west) rectangle (render_1.north east);
		\draw (render_2.south west) rectangle (render_2.north east);

		% Labels
		\node[inner sep=0pt, above=2pt of rerad_0  ] { Fluorescence \protect{$\bar{\mathcal{F}}$} };
		\node[inner sep=0pt, above=1pt of palette_0] { D65 palette };
		\node[inner sep=0pt, above=2pt of render_0 ] { D65 render };

		\node[inner sep=0pt, left=11pt of rerad_0, rotate=90, minimum width=\x, anchor=south] { inc. Brightness };
		\node[inner sep=0pt, left=3pt of rerad_0, rotate=90, minimum width=\x, anchor=south] { $\rho = 0.55$ };
		\node[inner sep=0pt, left=11pt of rerad_1, rotate=90, minimum width=\x, anchor=south] { inc. Saturation };
		\node[inner sep=0pt, left=3pt of rerad_1, rotate=90, minimum width=\x, anchor=south] { $\pmb{\rho}=(0.4,0.55,0.25)$ };
		\node[inner sep=0pt, left=11pt of rerad_2, rotate=90, minimum width=\x, anchor=south] { change Hue };
		\node[inner sep=0pt, left=3pt of rerad_2, rotate=90, minimum width=\x, anchor=south] { $\pmb{\rho}=(0.55,0.51,0.25)$ };
	\end{tikzpicture}
	}
	\vspace{-15pt}
	\caption{\textbf{Fluorescence palettes with varying emission.}
	We show fluorescence palettes for a single illuminant (D65) and fixed absorption parameters ($\mu_a=420$nm, $\sigma_a=60$nm), but with varying albedo and emission parameters.
	% In the first row, we boost the outgoing luminance of an achromatic albedo $\rho=0.55$ with a wide emission reradtiation.
	% In the second row, we use fluorescence to saturate a  low saturation albedo color ($\pmb{\rho}=(0.4,0.55,0.25)$).
	% In the third row, we modify the hue of a green albedo color ($\pmb{\rho}=(0.55,0.51,0.25)$) to obtain a vivid blue.
	From first to last row, we use fluorescence to boost the outgoing luminance, saturate the albedo color, and modify the albedo's hue.
	\label{fig:paletteEmit}
	\vspace{-5pt}
	}
\end{figure}
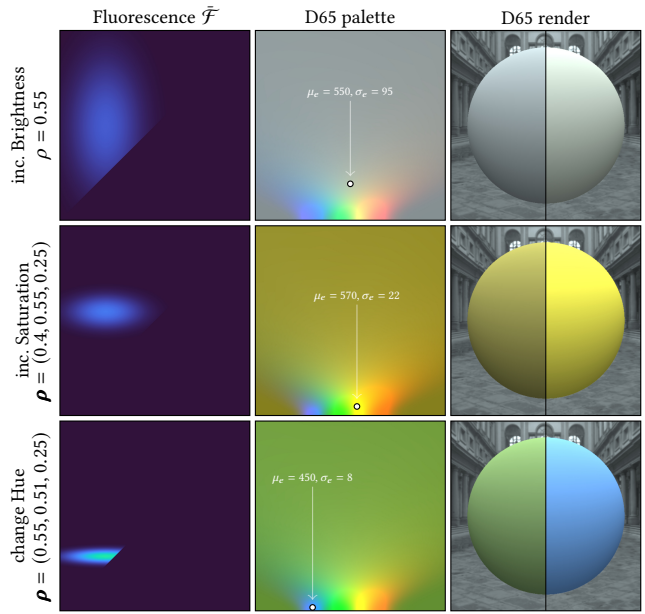

In order to let artists choose values for the mean and standard deviation paramers, we present them with color palettes that show a range of achievable colors (see Figures~\ref{fig:paletteAbsorb} and~\ref{fig:paletteEmit}).
For each palette, we assume that the albedo color $\pmb{\rho}$ is known, and that the incoming color $\mathbf{c}_i$ is determined by a choice of spectral illuminant integrated over Gaussian basis functions in pre-process.
In addition, we assume that the maximum fluorescent strength is achieved (i.e., $\alpha = \alpha_{\max}$ or $\bar{\alpha}=1$) when displaying fluorescence palettes.

We let artists control the parameters $(\mu_a,\sigma_a)$, which light absorbtion for fluorescent reradiation.
Intuitively, a large value of $\sigma_a$ reradiates from as many wavelengths as possible, while a smaller value permits to select a subset of wavelengths around $\mu_a$.
A fluorescence palette is then obtained by computing $\mathbf{c}_o$ using Equation~\ref{eqn:reduced-out-col2} for every pair $(\mu_e,\sigma_e)$.
When an artist picks a color in the palette, this determines the parameter pair $(\mu_e,\sigma_e)$, from which the final intensity is computed through $\alpha = \bar{\alpha} \hat{\alpha}_{\max}(\mu_e,\sigma_e)$.
%The vertical line -- located at $\mu_a$ -- identifies configurations where the Gaussian lies on the bispectral diagonal.
%Colors on its left side should not be considered physically-realistic, since they correspond to materials seldom encountered in measured materials (see Section~\ref{sec:gauss-model}).

\pagebreak
\paragraph*{Editing absorption}
In Figure~\ref{fig:paletteAbsorb}, we show several palettes in different absorption configurations and for different illuminants, keeping the $(\mu_e,\sigma_e)$ parameter pair constant throughout.
We choose to visualize the corresponding normalized fluorescence matrix $\mathcal{\bar{F}}$ instead of the full reradiation matrix $\mathcal{P}$ in the first column as it directly conveys the manipulated Gaussian.
We also show rendered spheres, with lighting environments made to emit a particular illuminant spectrum by first desaturating them then multiplying them by the XYZU color version of the illuminant.
As seen in the figure, focusing absorption in the visible (resp. UV) range has a selective effect in incandescent (resp. UV) lighting, while a daylight illuminant is much less affected since it has large values over both visible and UV wavelength ranges.
In contrast, with a wide absorption, the material exhibits marked fluorescent behaviors with all three types of illuminant.

\paragraph*{Editing re-emission}
In Figure~\ref{fig:paletteEmit}, we keep the $(\mu_a,\sigma_a)$ parameter pair constant but vary both $\pmb{\rho}$ and $(\mu_e,\sigma_e)$, using a D65 illuminant throughout.
Here we show that fluorescence may serve different purposes: it may increase the brightness of an achromatic material, increase its saturation, or change its hue.
Thanks to fluorescence palettes, finding the relevant $(\mu_e,\sigma_e)$ pair for a choice of $\pmb{\rho}$ is immediate.
We observe in particular that the hue of the fluorescence component varies with $\mu_e$, while its saturation varies with $\sigma_e$.
This is to be expected since for smaller $\sigma_e$ values, reradiated chromaticities should tend to the spectral locii.

\vfill

\section{Additional results}
\label{sec:results}

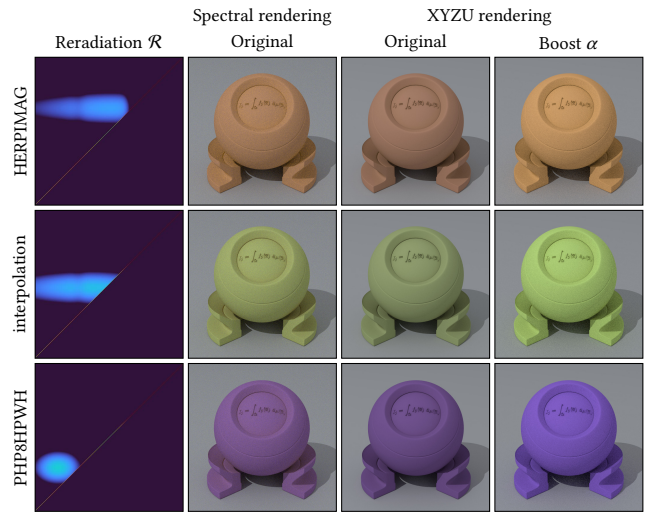
\begin{figure}[b]
	\vspace{-5pt}
	\resizebox{\linewidth}{!}{
	\begin{tikzpicture}[font=\footnotesize]
		\setlength{\x}{0.23\linewidth}
		\node[inner sep=0pt                      ] (rerad_0) {\includegraphics[width=\x]{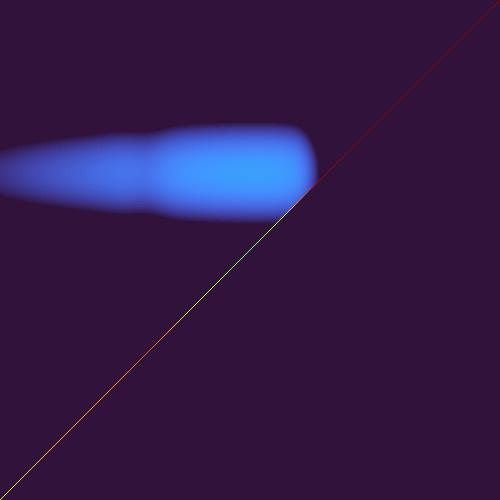}};
		\node[inner sep=0pt, below=2pt of rerad_0] (rerad_1) {\includegraphics[width=\x]{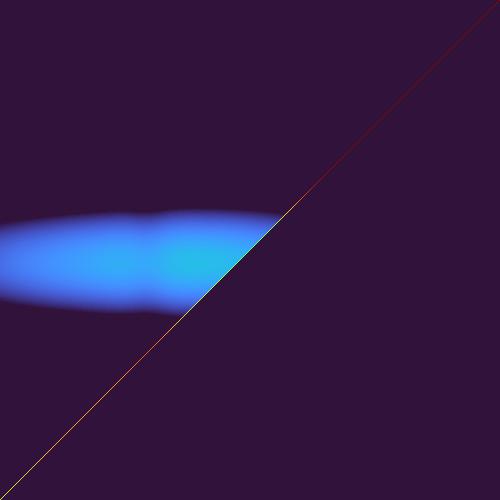}};
		\node[inner sep=0pt, below=2pt of rerad_1] (rerad_2) {\includegraphics[width=\x]{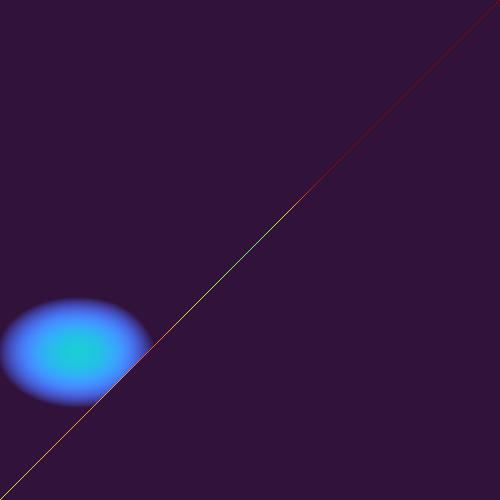}};

		\draw (rerad_0.south west) rectangle (rerad_0.north east);
		\draw (rerad_1.south west) rectangle (rerad_1.north east);
		\draw (rerad_2.south west) rectangle (rerad_2.north east);

		\node[inner sep=0pt, right=2pt of rerad_0] (spectral_0) {\includegraphics[width=\x]{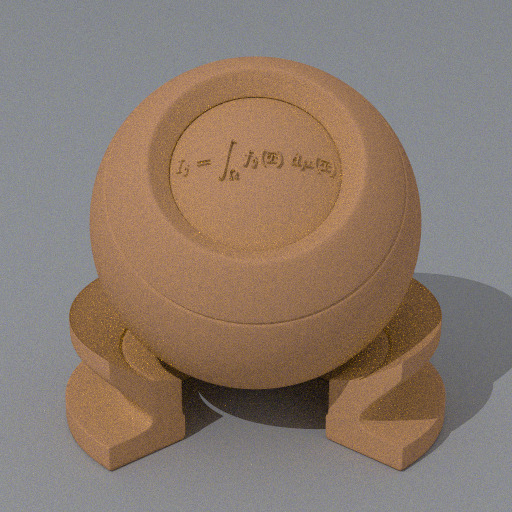}};
		\node[inner sep=0pt, right=2pt of rerad_1] (spectral_1) {\includegraphics[width=\x]{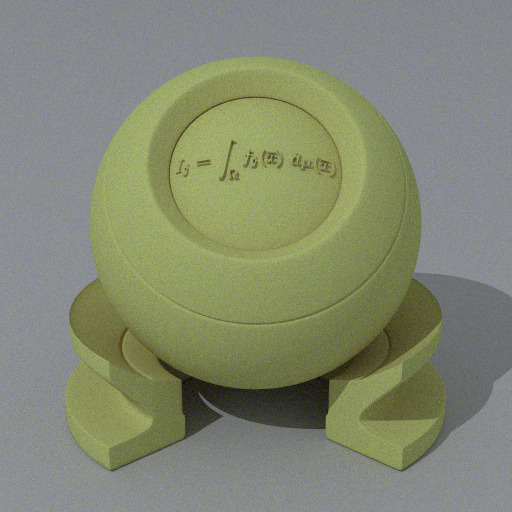}};
		\node[inner sep=0pt, right=2pt of rerad_2] (spectral_2) {\includegraphics[width=\x]{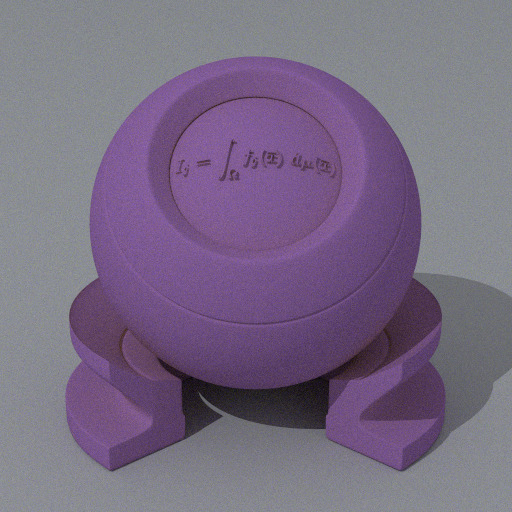}};

		\draw (spectral_0.south west) rectangle (spectral_0.north east);
		\draw (spectral_1.south west) rectangle (spectral_1.north east);
		\draw (spectral_2.south west) rectangle (spectral_2.north east);

		\node[inner sep=0pt, right=2pt of spectral_0] (render_0) {\includegraphics[width=\x]{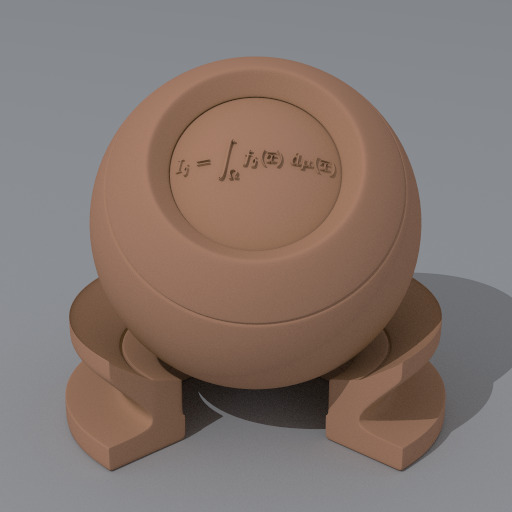}};
		\node[inner sep=0pt, right=2pt of spectral_1] (render_1) {\includegraphics[width=\x]{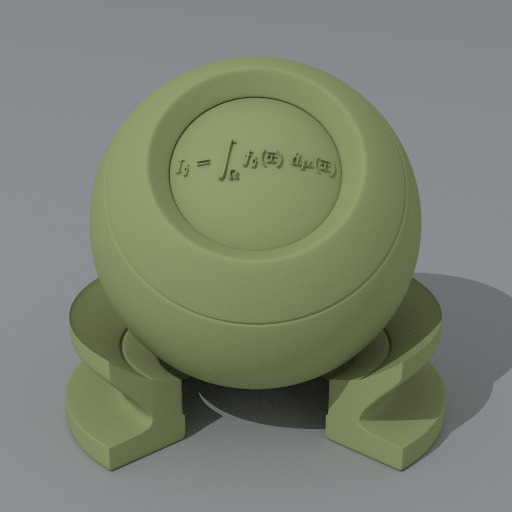}};
		\node[inner sep=0pt, right=2pt of spectral_2] (render_2) {\includegraphics[width=\x]{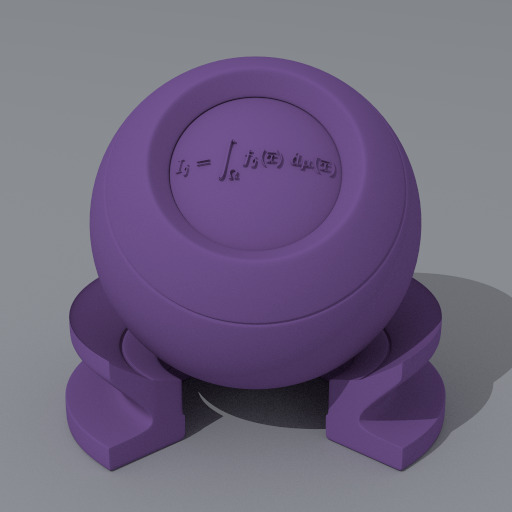}};

		\draw (render_0.south west) rectangle (render_0.north east);
		\draw (render_1.south west) rectangle (render_1.north east);
		\draw (render_2.south west) rectangle (render_2.north east);

		\node[inner sep=0pt, right=2pt of render_0] (boost_0) {\includegraphics[width=\x]{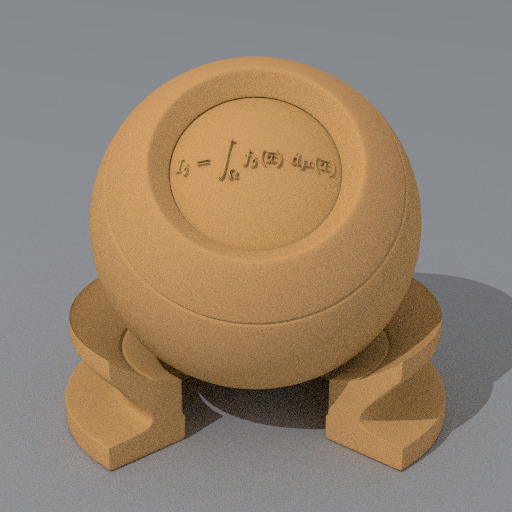}};
		\node[inner sep=0pt, right=2pt of render_1] (boost_1) {\includegraphics[width=\x]{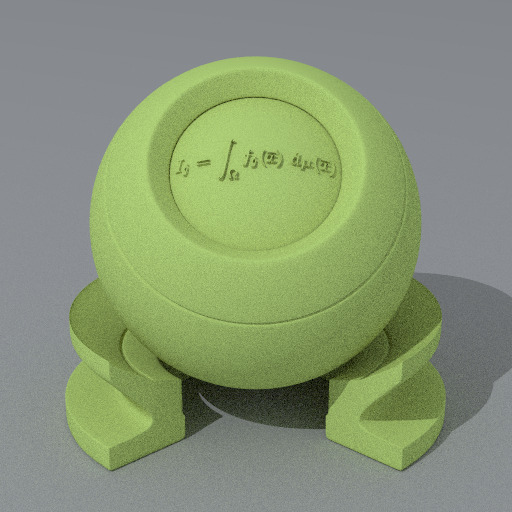}};
		\node[inner sep=0pt, right=2pt of render_2] (boost_2) {\includegraphics[width=\x]{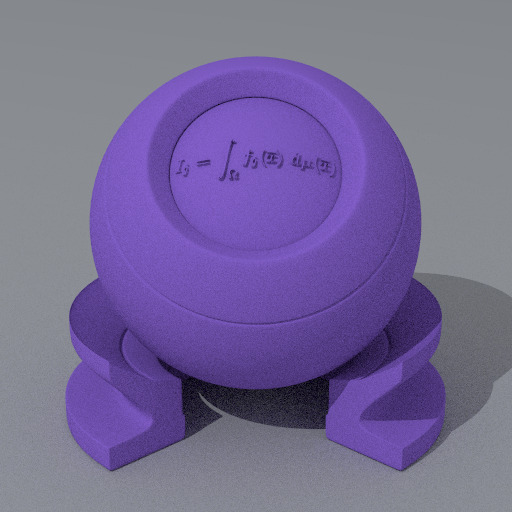}};

		\draw (boost_0.south west) rectangle (boost_0.north east);
		\draw (boost_1.south west) rectangle (boost_1.north east);
		\draw (boost_2.south west) rectangle (boost_2.north east);

		% Labels
		\node[inner sep=0pt, above=3pt of rerad_0  ] { Reradiation \protect{${\mathcal{R}}$} };
		\node[inner sep=0pt, above=12pt of spectral_0] { Spectral rendering };
		\node[inner sep=0pt, above=2pt of spectral_0] { Original };
		\node[inner sep=0pt, above=12pt of render_0.north east ] { XYZU rendering };
		\node[inner sep=0pt, above=2pt of render_0] { Original };
		\node[inner sep=0pt, above=3pt of boost_0 ] { Boost $\alpha$ };

		\node[inner sep=0pt, left=3pt of rerad_0, rotate=90, minimum width=\x, anchor=south] { \textsc{HERPIMAG} };
		\node[inner sep=0pt, left=2pt of rerad_1, rotate=90, minimum width=\x, anchor=south] { interpolation };
		\node[inner sep=0pt, left=3pt of rerad_2, rotate=90, minimum width=\x, anchor=south] { \textsc{PHP8HPWH} };
	\end{tikzpicture}
	}
	\vspace{-12pt}
	\caption{\textbf{Interpolation of Fitted Parameters.}
	Our parametric model permits to explore a space of Fluorescent material from fits of the Gonzalez and Fairchild's~\shortcite{Gonzalez00} database.
	In this example we take the fluorescence $\bar{\mathcal{F}}$ of \textsc{HERPIMAG} and \textsc{PHP8HPWH} samples, and blend their fitted parameters
   (strength, mean and variance) with half/half weights. Using the diagonal of \textsc{HERPIMAG} in each three case, we can build new plausible
   fluorescent materials. By changing the strength (last column), we can alter the appearance of all materials. All images were rendered using
   a D65 illuminant.
	\label{fig:fitInterp}
	}
\end{figure}

\begin{figure*}
	% \vspace{-7pt}
	\resizebox{0.95\linewidth}{!}{
		\begin{tikzpicture}
			\node[inner sep=0pt] 											   (albedo){\includegraphics[width=0.1\linewidth]{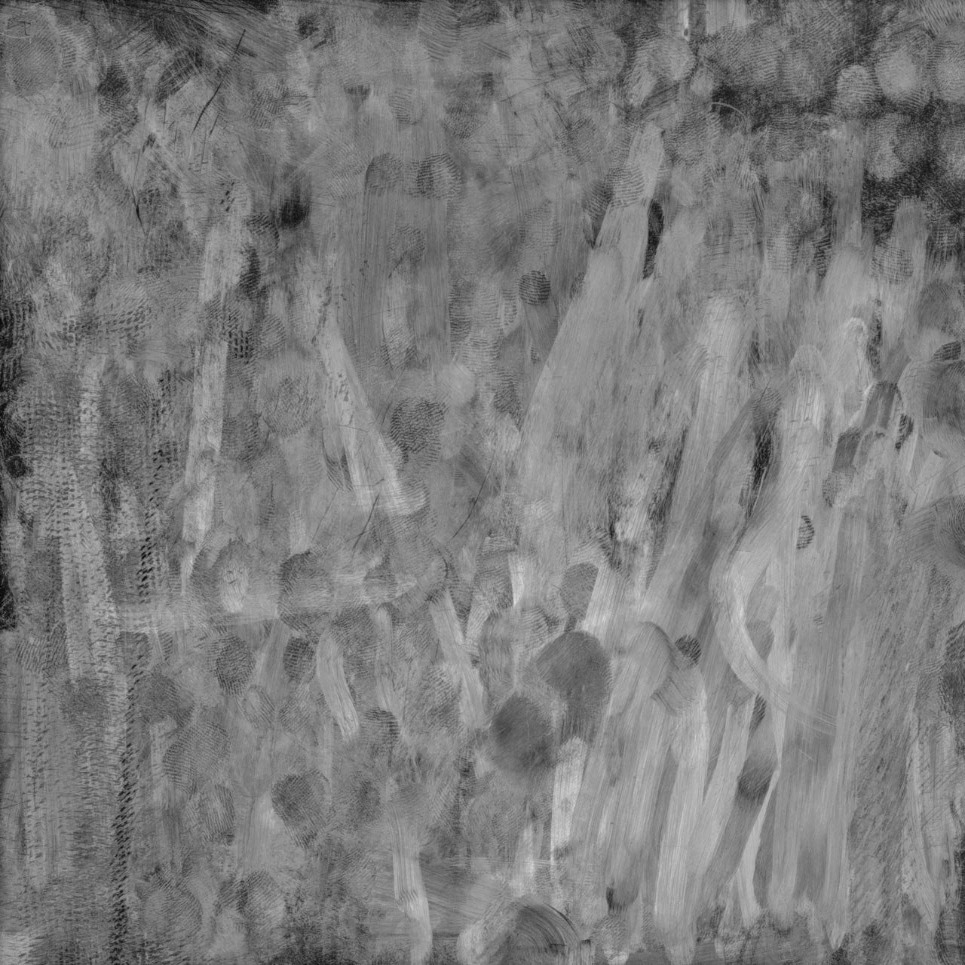}};
			\draw (albedo.south west) rectangle (albedo.north east);

			\node[inner sep=0pt,anchor=north west]       at (albedo.center)    (alpha) {\includegraphics[width=0.1\linewidth]{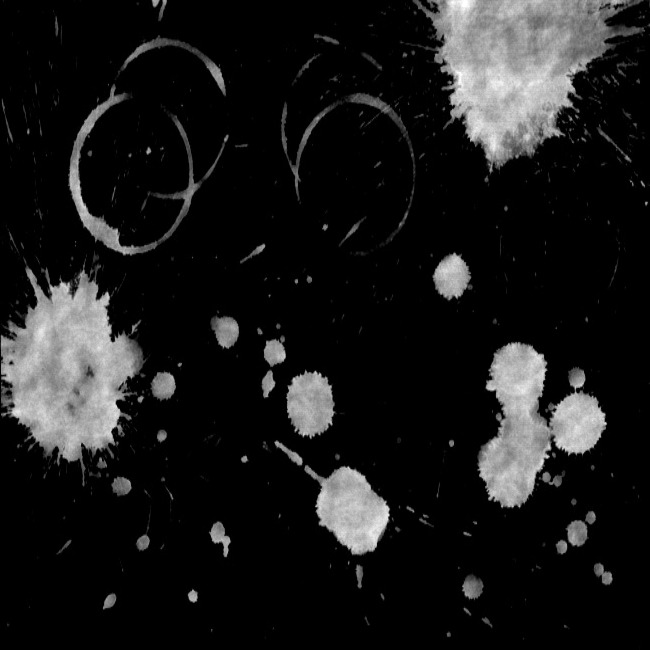}};
			\draw (alpha.south west) rectangle (alpha.north east);

			\node[yshift=0.05\linewidth, xshift=2pt, inner sep=0pt,anchor=north west] at (alpha.north east)   (a) {\includegraphics[width=0.15\linewidth]{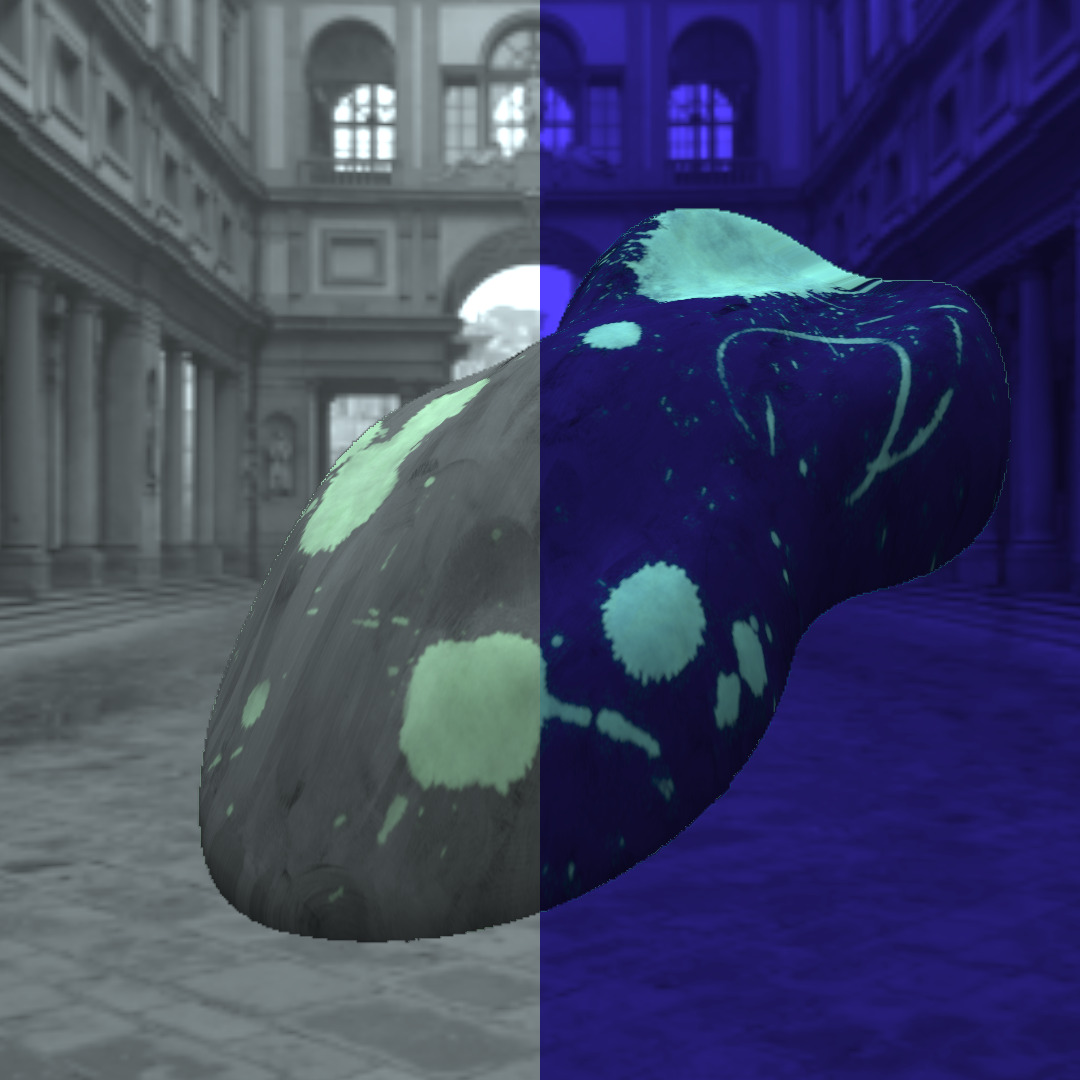}};
			\node[xshift=2pt, inner sep=0pt,anchor=west] at (a.east)                                          (b) {\includegraphics[width=0.15\linewidth]{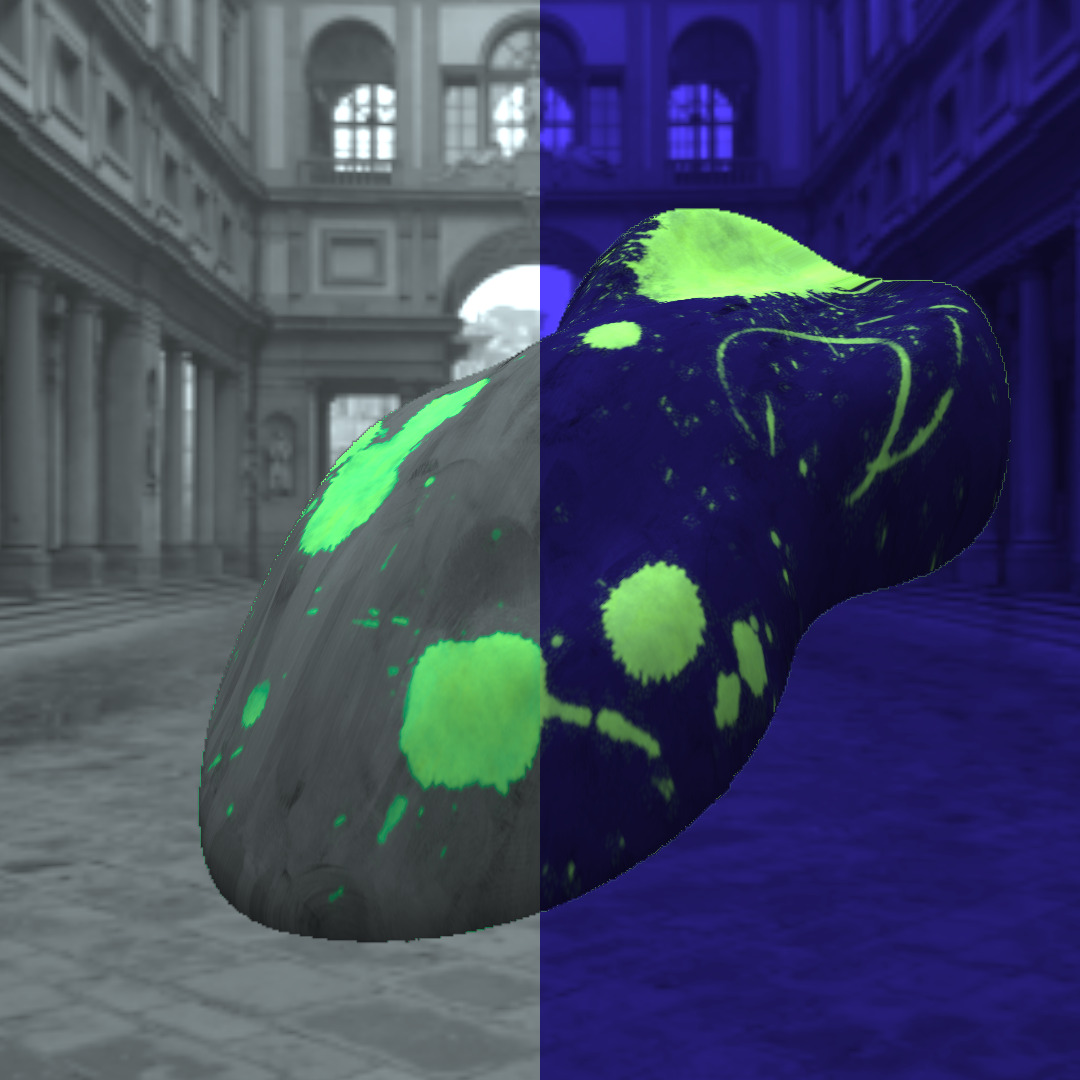}};
			\node[xshift=2pt, inner sep=0pt,anchor=west] at (b.east)                                          (c) {\includegraphics[width=0.15\linewidth]{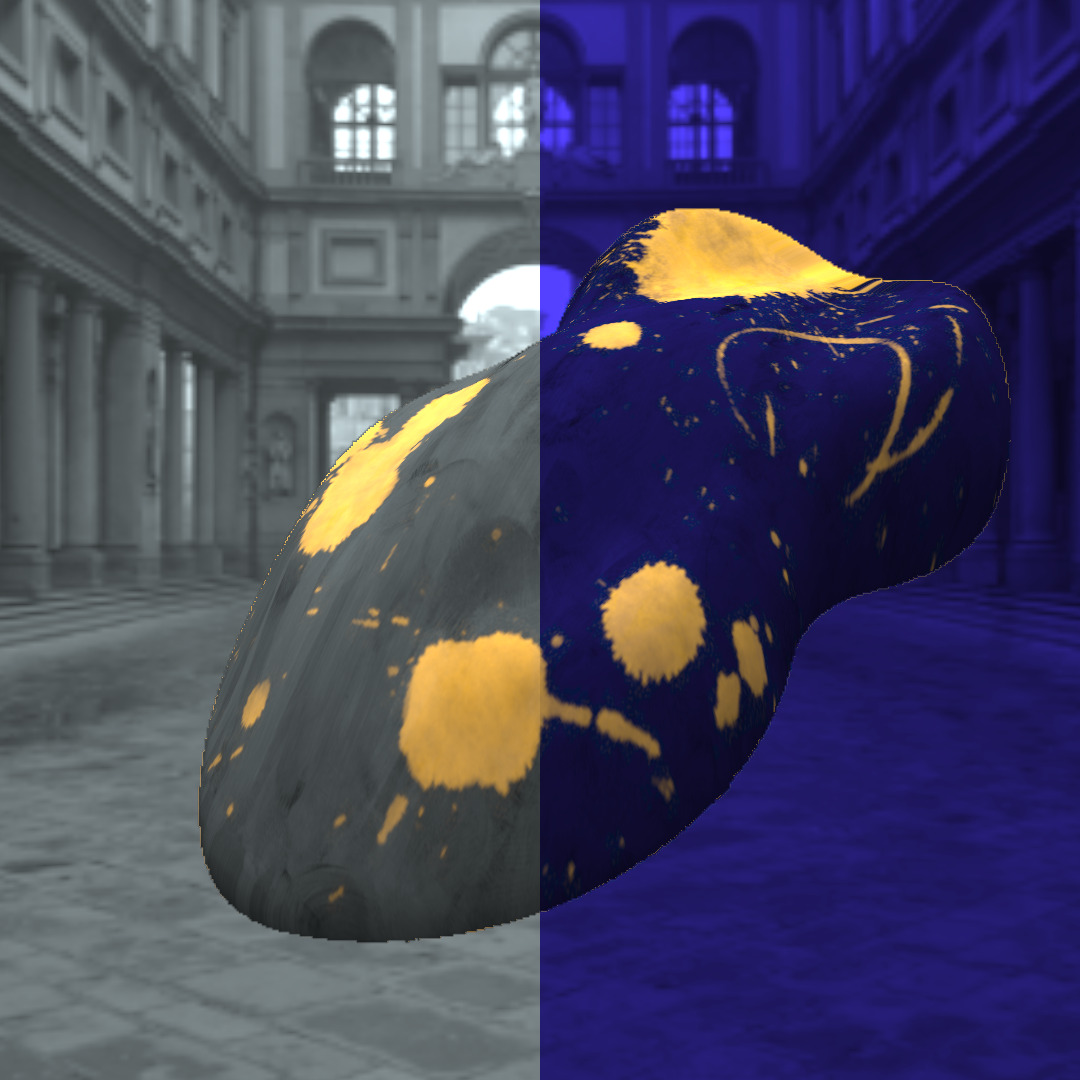}};
			\node[xshift=2pt, inner sep=0pt,anchor=west] at (c.east)                                          (d) {\includegraphics[width=0.15\linewidth]{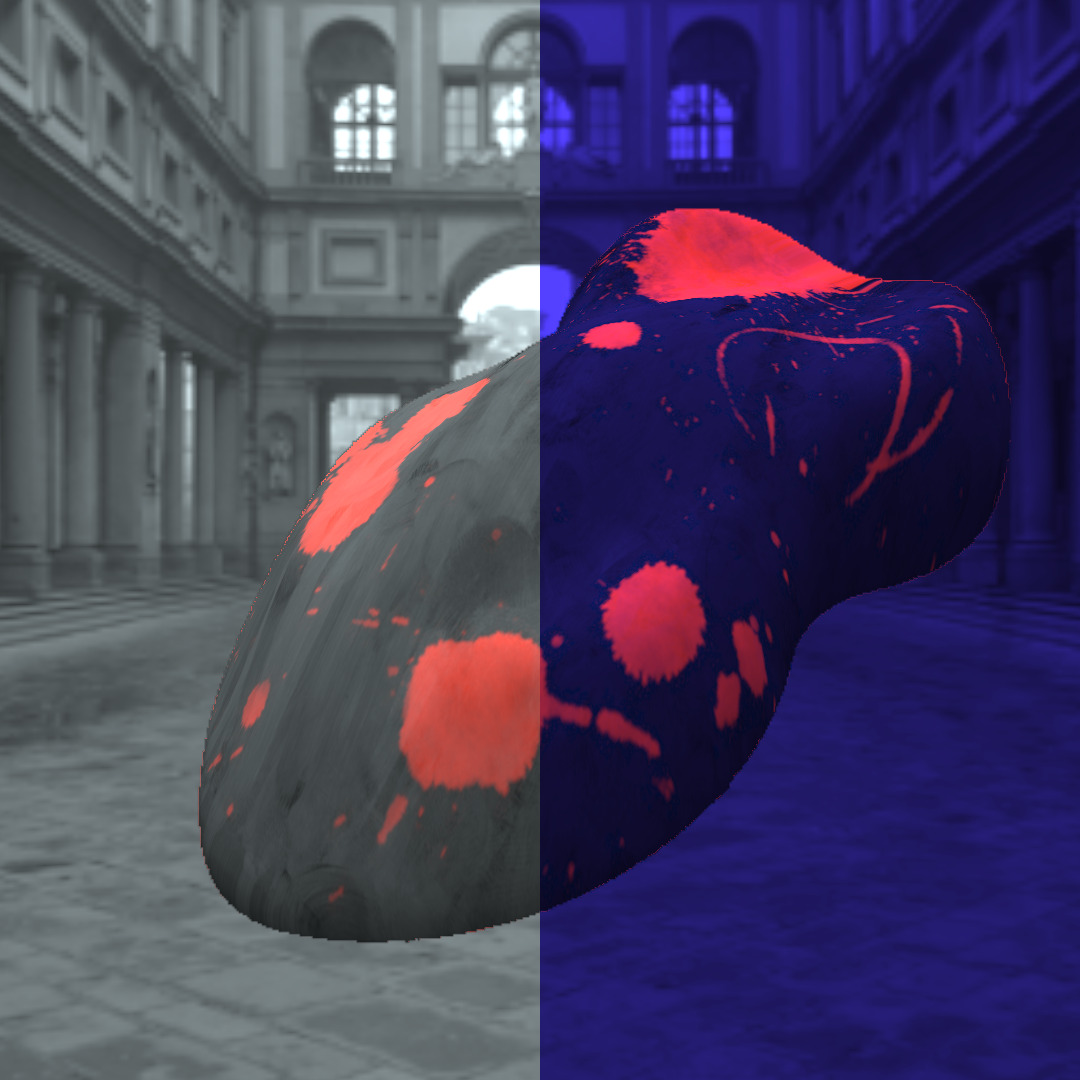}};

			\draw[very thin] (a.north) -- (a.south);
			\draw[very thin] (b.north) -- (b.south);
			\draw[very thin] (c.north) -- (c.south);
			\draw[very thin] (d.north) -- (d.south);

			\draw (a.south west) rectangle (a.north east);
			\draw (b.south west) rectangle (b.north east);
			\draw (c.south west) rectangle (c.north east);
			\draw (d.south west) rectangle (d.north east);
		\end{tikzpicture}
	}
	\vspace{-10pt}
	\caption{\textbf{Spatial variations of fluorescence strength.}
	The leftmost images show the textures used to vary the achromatic albedo $\rho \in [0, 0.2]$ and the fluorescence strength $\alpha_n \in [0,1]$ respectively.
	Rendering results are shown in the remaining four images, each under two illuminants (D65 at left, UV at right), with fixed $\mu_a=420$nm and $\sigma_a=100$nm, and $(\mu_e,\sigma_e) \in \{(419,69), (593,48), (515,21), (645,10)\}$ from left to right.
	\label{fig:var-alpha}}
\end{figure*}
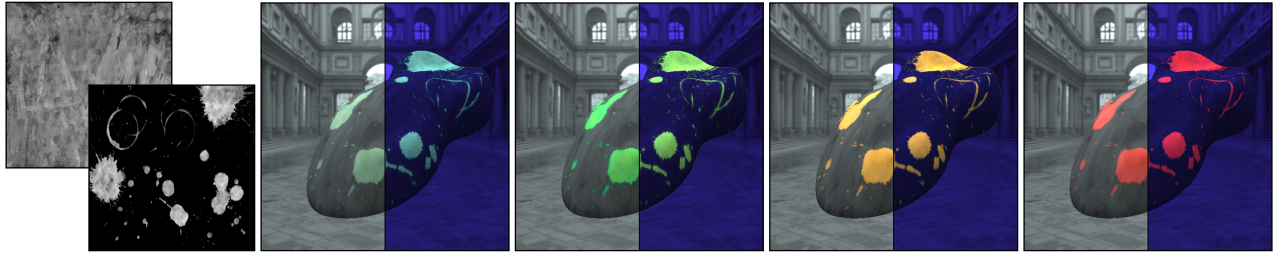

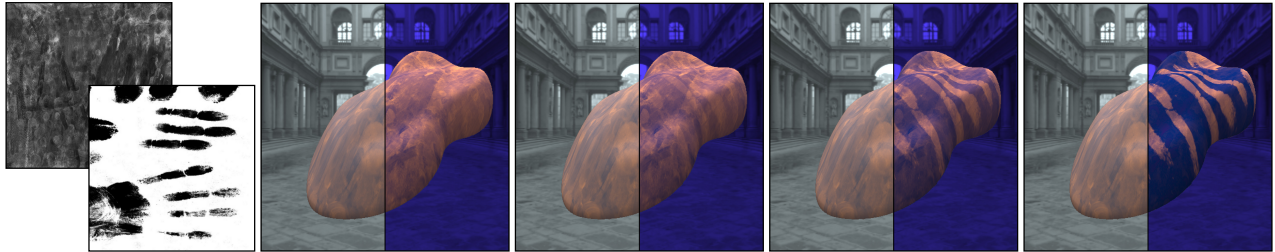
\begin{figure*}
	\vspace{-7pt}
	\resizebox{0.95\linewidth}{!}{
		\begin{tikzpicture}
			\node[inner sep=0pt]                                      (albedo){\includegraphics[width=0.1\linewidth]{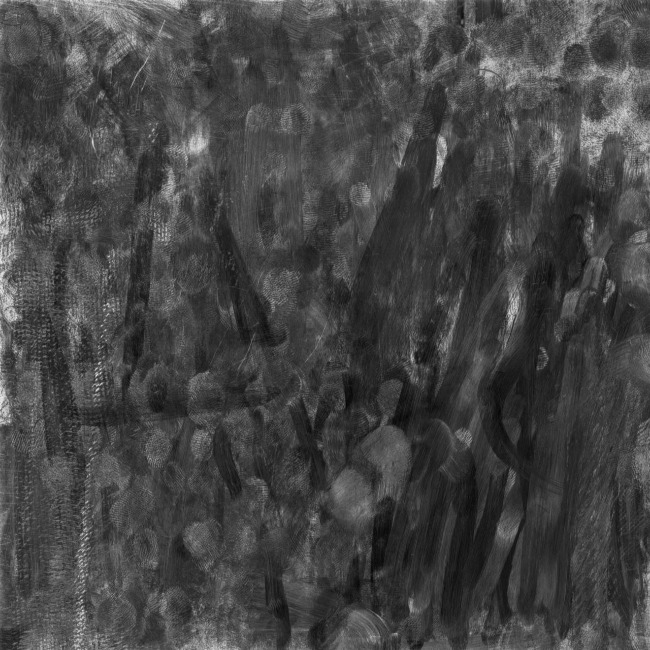}};
			\draw (albedo.south west) rectangle (albedo.north east);

			\node[inner sep=0pt,anchor=north west] at (albedo.center) (alpha) {\includegraphics[width=0.1\linewidth]{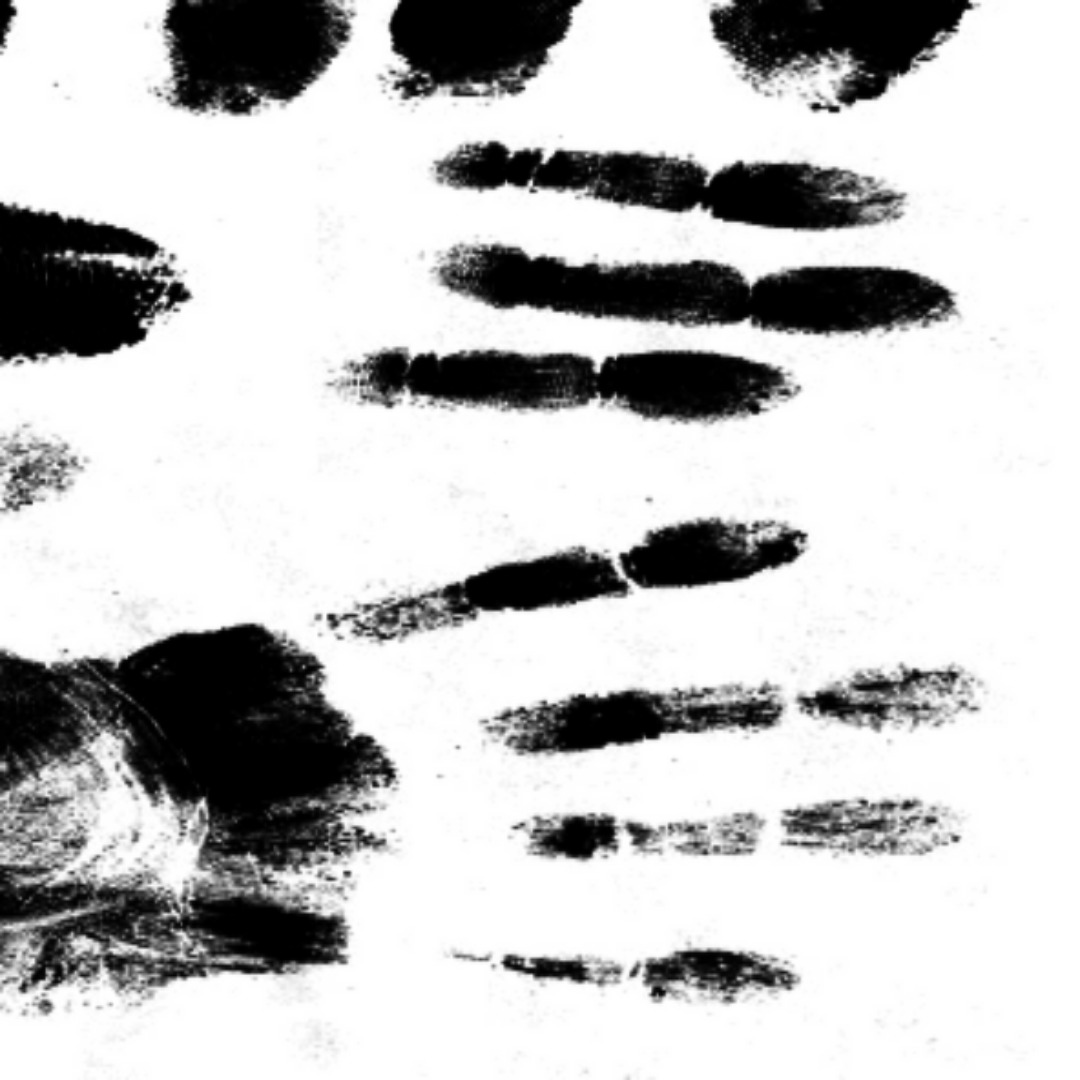}};
			\draw (alpha.south west) rectangle (alpha.north east);

			\node[yshift=0.05\linewidth, xshift=2pt, inner sep=0pt,anchor=north west] at (alpha.north east)   (a) {\includegraphics[width=0.15\linewidth]{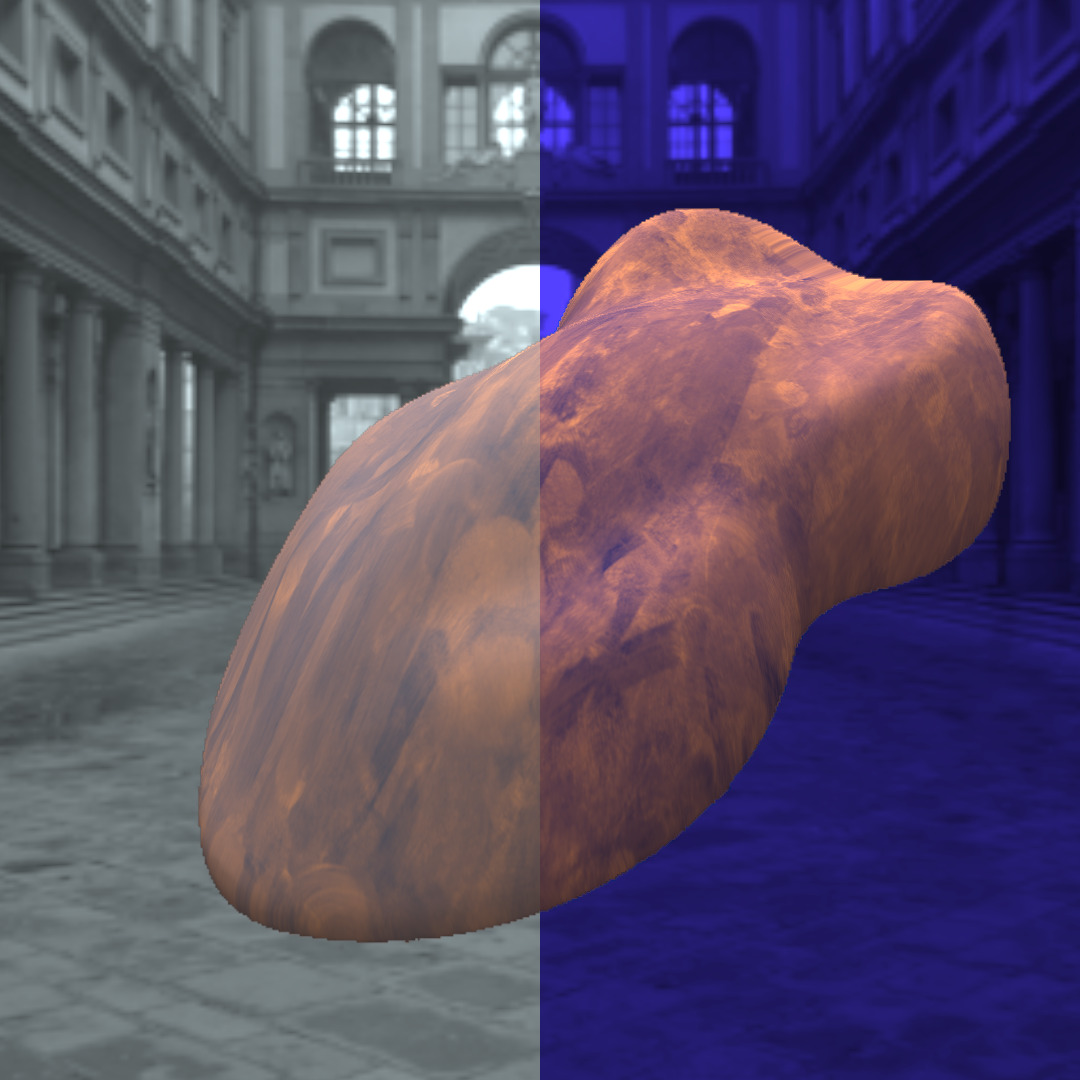}};
			\node[xshift=2pt, inner sep=0pt,anchor=west] at (a.east)                                          (b) {\includegraphics[width=0.15\linewidth]{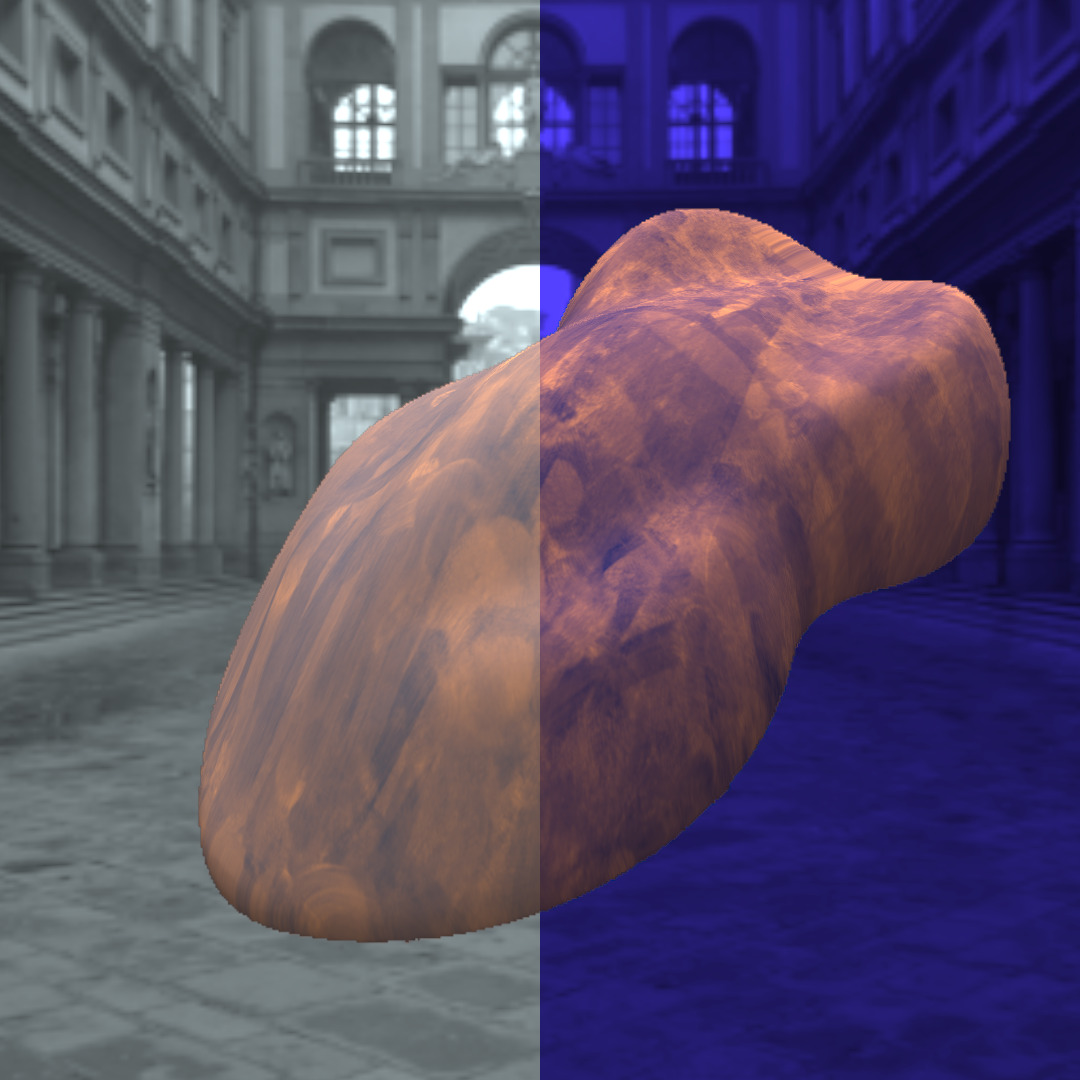}};
			\node[xshift=2pt, inner sep=0pt,anchor=west] at (b.east)                                          (c) {\includegraphics[width=0.15\linewidth]{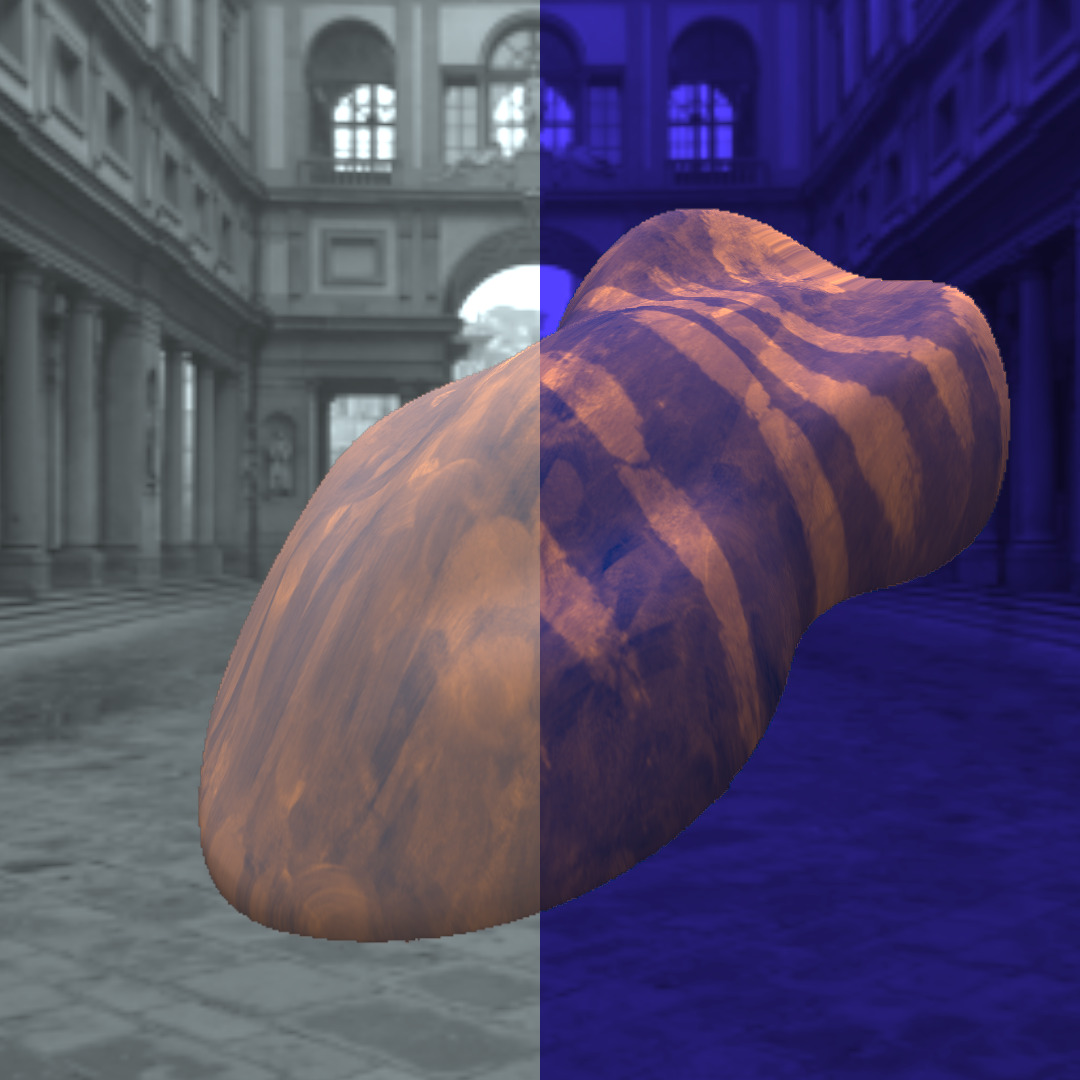}};
			\node[xshift=2pt, inner sep=0pt,anchor=west] at (c.east)                                          (d) {\includegraphics[width=0.15\linewidth]{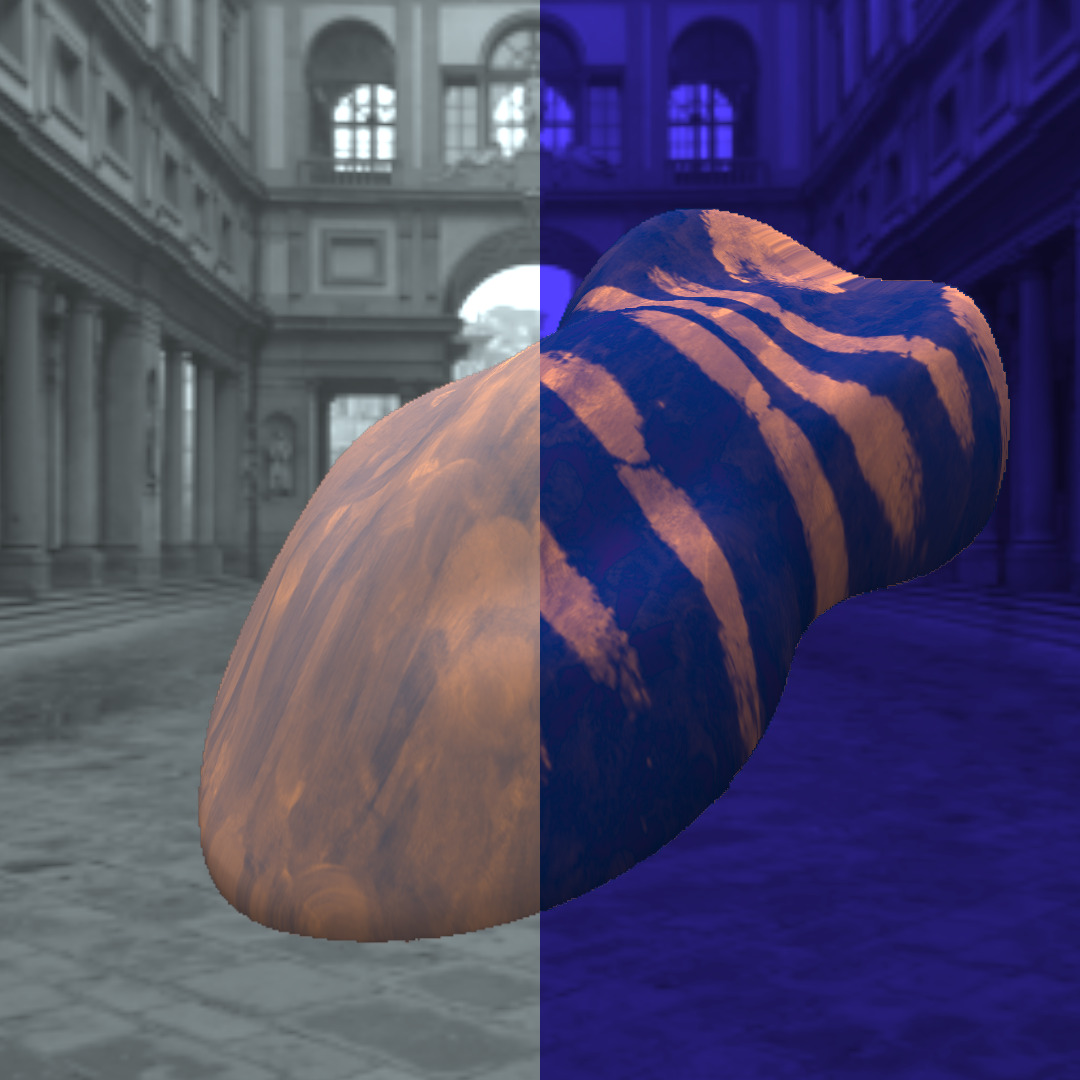}};

			\draw[very thin] (a.north) -- (a.south);
			\draw[very thin] (b.north) -- (b.south);
			\draw[very thin] (c.north) -- (c.south);
			\draw[very thin] (d.north) -- (d.south);

			\draw (a.south west) rectangle (a.north east);
			\draw (b.south west) rectangle (b.north east);
			\draw (c.south west) rectangle (c.north east);
			\draw (d.south west) rectangle (d.north east);
		\end{tikzpicture}
	}
	% \centering
	% \begin{tabular}{@{}c@{\hspace{2pt}}c@{\hspace{2pt}}c@{\hspace{2pt}}c@{\hspace{2pt}}c@{}}
	% \includegraphics[width=0.2\linewidth]{Figures/SpatialVars/var_abs/biTex.jpg} &
	% \includegraphics[width=0.2\linewidth]{Figures/SpatialVars/var_abs/var_abs_fluo0.jpg} &
	% \includegraphics[width=0.2\linewidth]{Figures/SpatialVars/var_abs/var_abs_fluo1.jpg} &
	% \includegraphics[width=0.2\linewidth]{Figures/SpatialVars/var_abs/var_abs_fluo2.jpg} &
	% \includegraphics[width=0.2\linewidth]{Figures/SpatialVars/var_abs/var_abs_fluo3.jpg}
	% \end{tabular}
	\vspace{-10pt}
	\caption{\textbf{Spatial variations of fluorescence strength and absorption.}
	The leftmost images here show the textures used to vary the fluorescence strength $\alpha_n \in [0,1]$ and $\mu_a$ respectively.
	In the second image, $\mu_a$ is fixed to $400$nm, while in the remaining images, it varies in different intervals: from left to right, $\mu_a \in [400,440]$, $\mu_a \in [400,480]$, $\mu_a \in [400,520]$.
	The wider the interval, the more pronounced the difference in appearance between the D65 and UV illuminants (left and right halves respectively).
	All other parameters are kept fixed: $\sigma_a=33$nm, $\mu_e=419$nm, $\sigma_e=69$nm, and $\pmb{\rho} = (0.14, 0.14, 0.2)$.
	\label{fig:var-abs}}
\end{figure*}

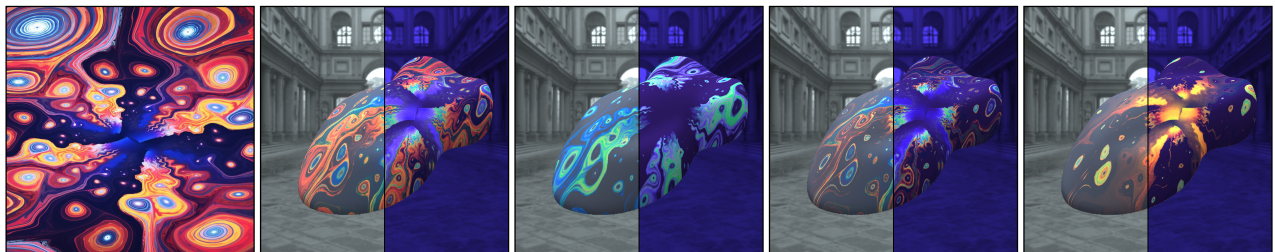
\begin{figure*}
	\vspace{-7pt}
	\resizebox{0.95\linewidth}{!}{
		\begin{tikzpicture}
			\node[inner sep=0pt]                                       (tex){\includegraphics[width=0.15\linewidth]{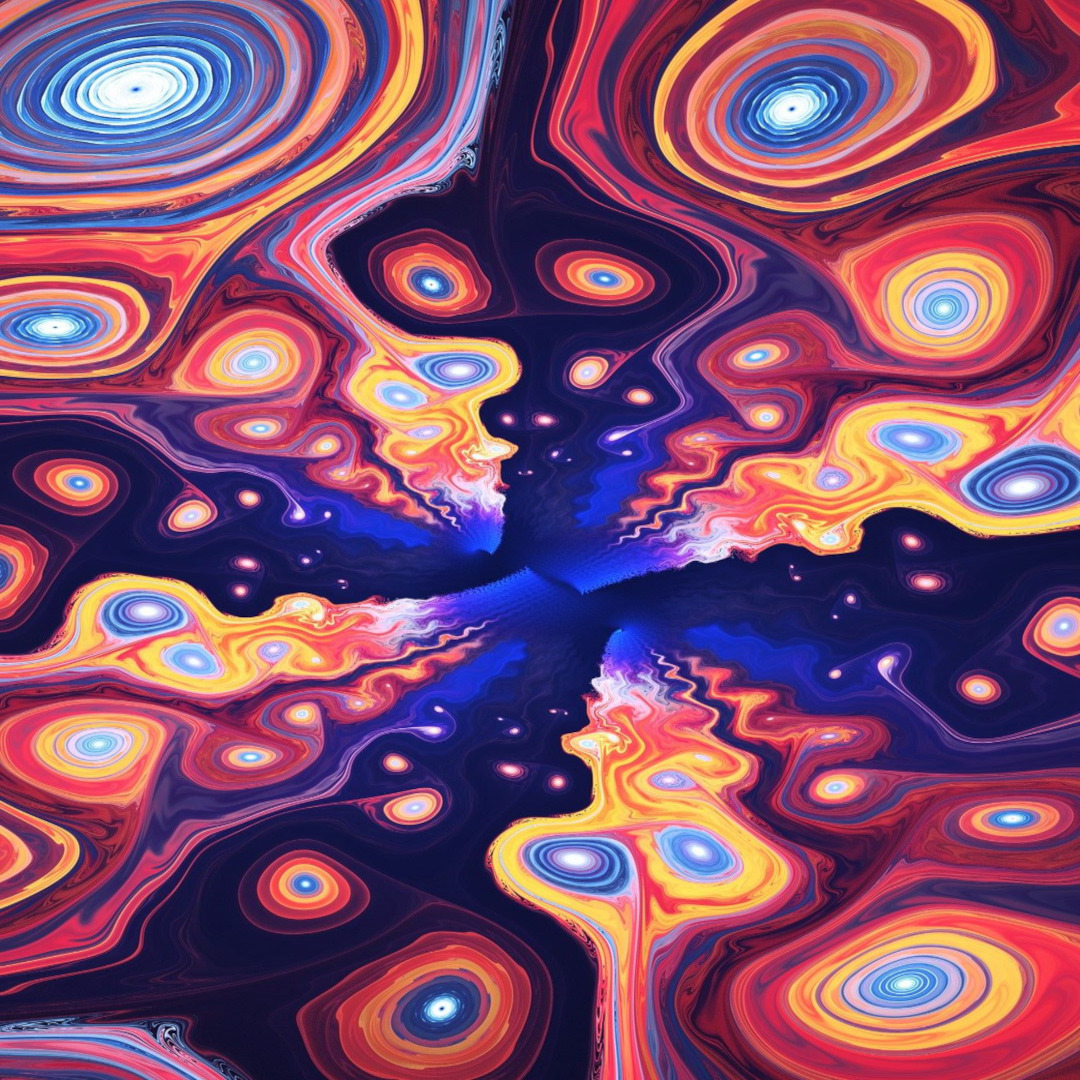}};
			\draw (tex.south west) rectangle (tex.north east);

			\node[xshift=2pt, inner sep=0pt,anchor=west] at (tex.east) (a) {\includegraphics[width=0.15\linewidth]{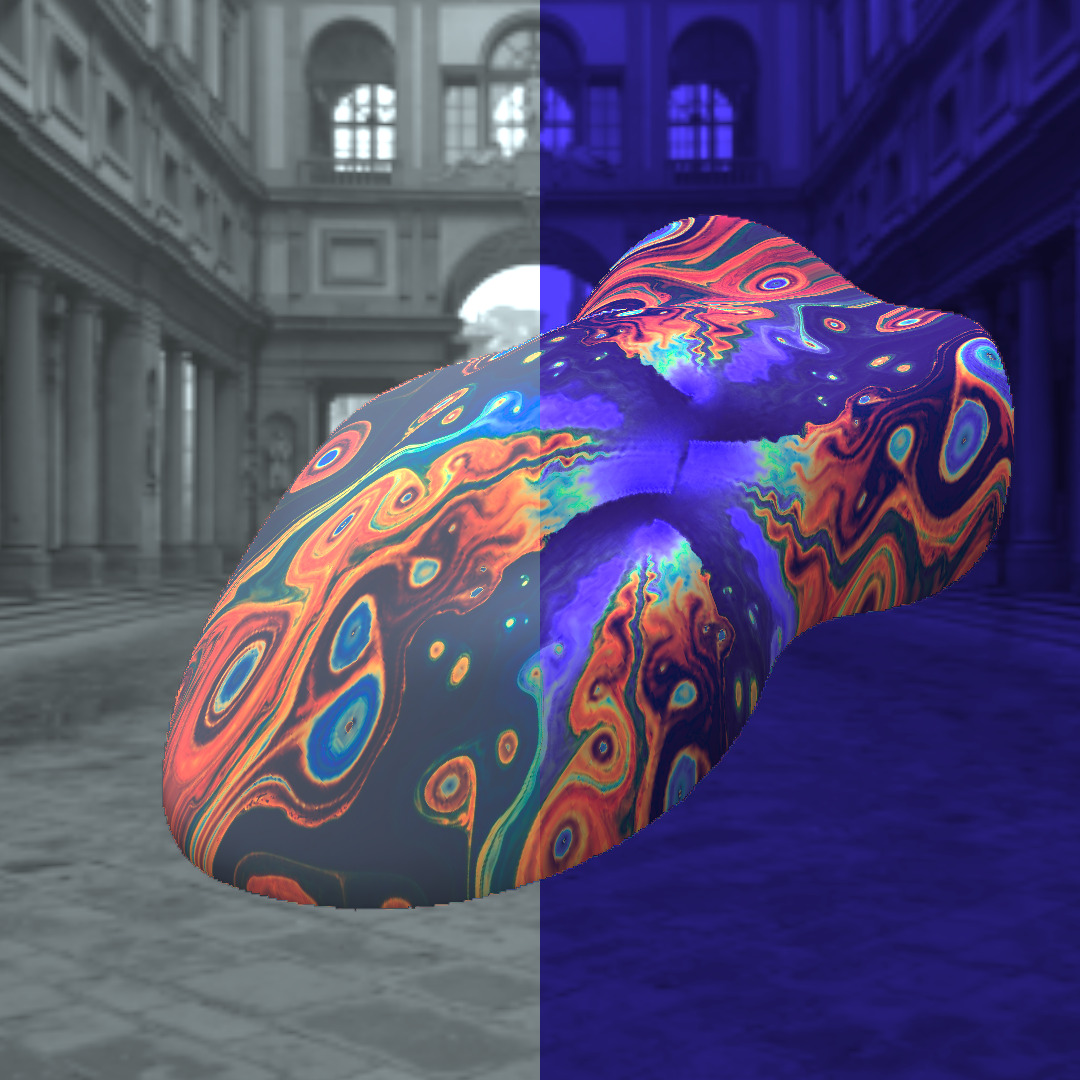}};
			\node[xshift=2pt, inner sep=0pt,anchor=west] at (a.east)   (b) {\includegraphics[width=0.15\linewidth]{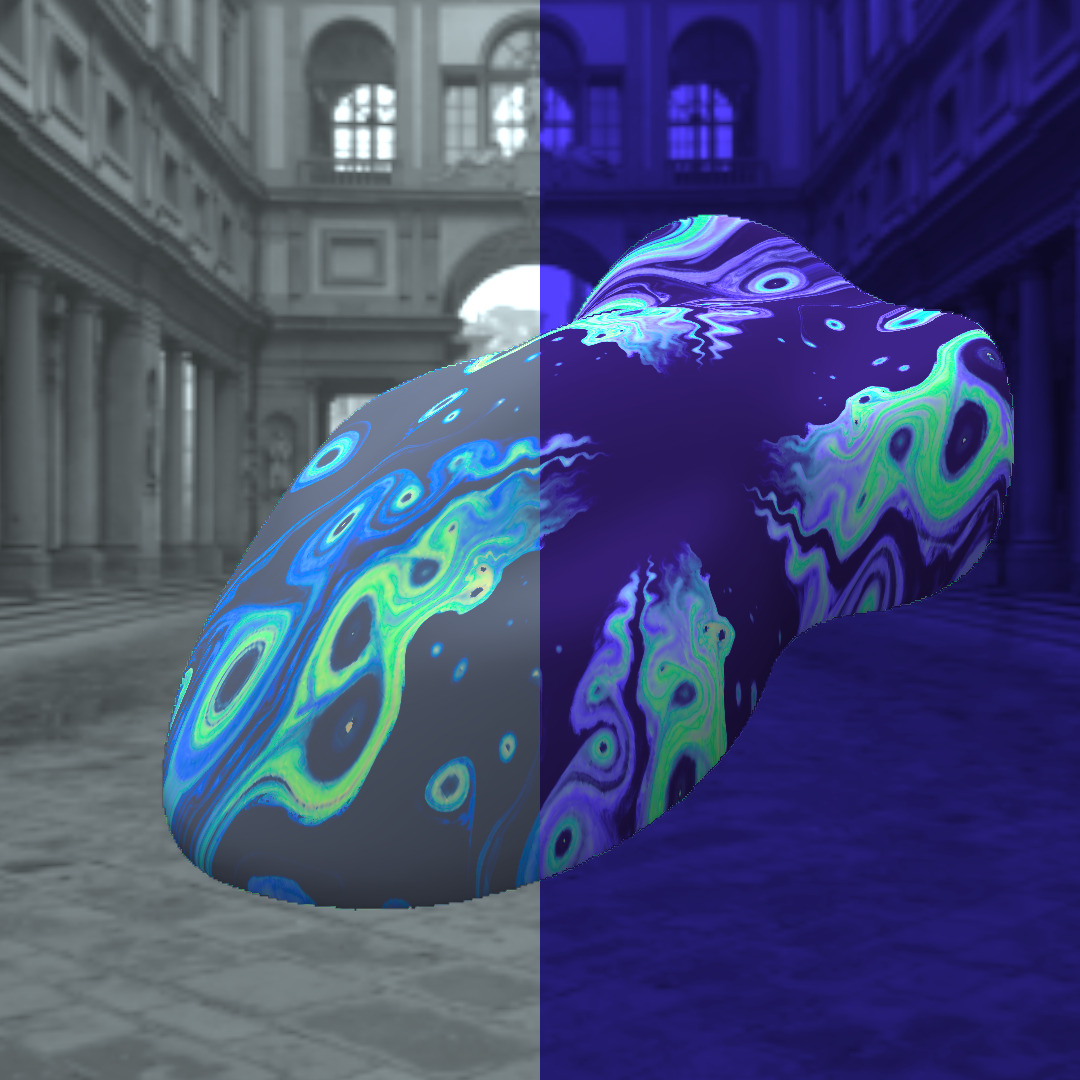}};
			\node[xshift=2pt, inner sep=0pt,anchor=west] at (b.east)   (c) {\includegraphics[width=0.15\linewidth]{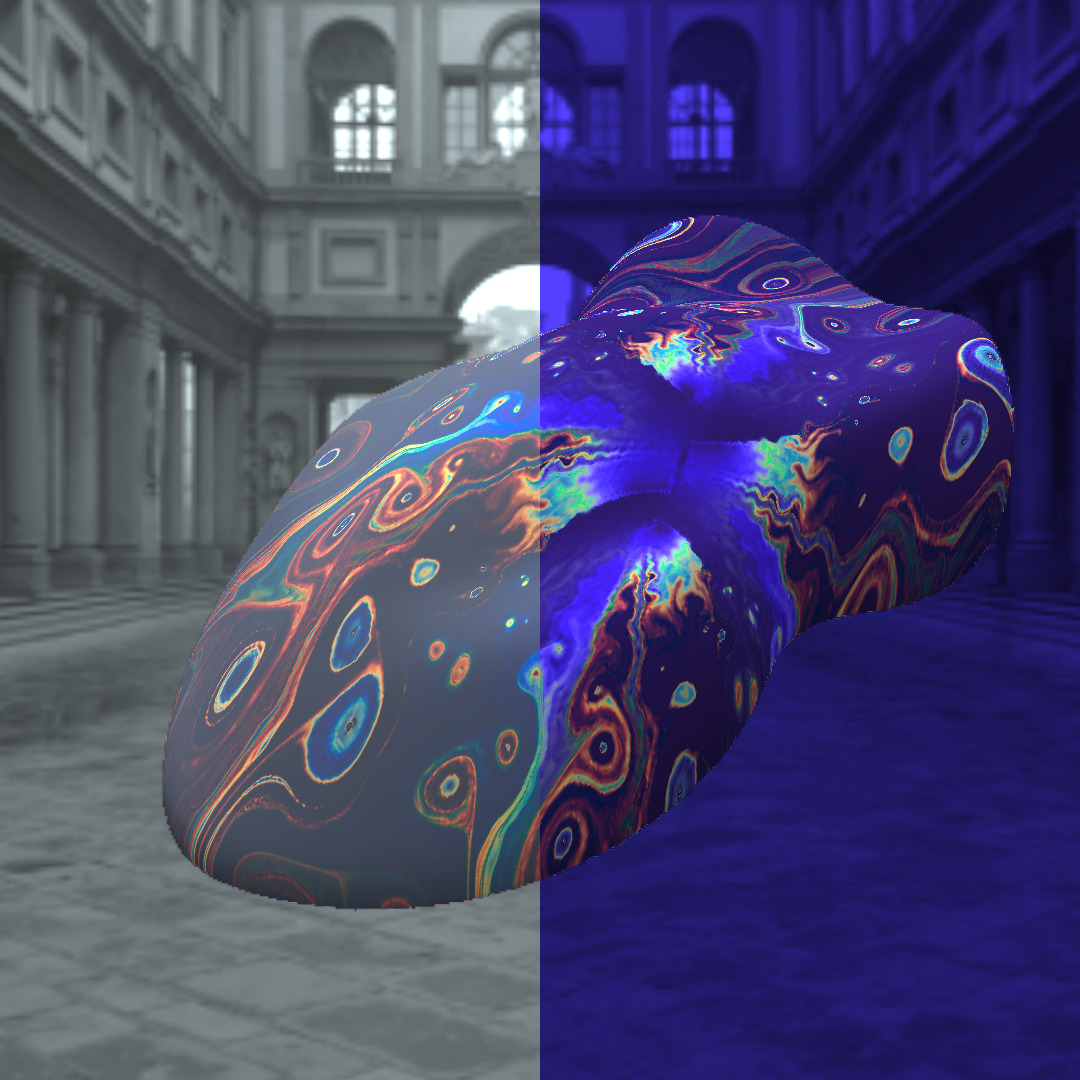}};
			\node[xshift=2pt, inner sep=0pt,anchor=west] at (c.east)   (d) {\includegraphics[width=0.15\linewidth]{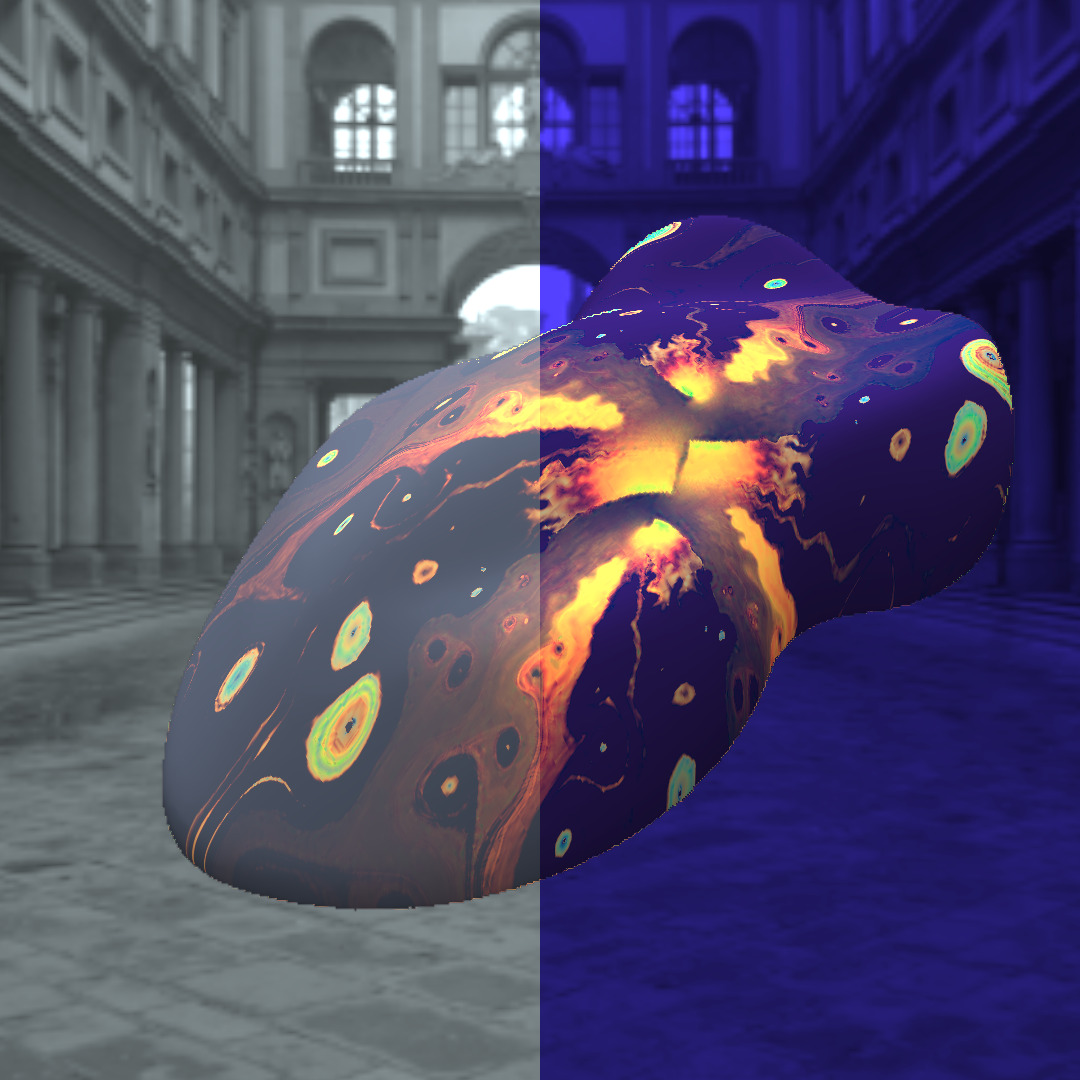}};

			\draw[very thin] (a.north) -- (a.south);
			\draw[very thin] (b.north) -- (b.south);
			\draw[very thin] (c.north) -- (c.south);
			\draw[very thin] (d.north) -- (d.south);

			\draw (a.south west) rectangle (a.north east);
			\draw (b.south west) rectangle (b.north east);
			\draw (c.south west) rectangle (c.north east);
			\draw (d.south west) rectangle (d.north east);
		\end{tikzpicture}
	}
	% \centering
	% \begin{tabular}{@{}c@{\hspace{2pt}}c@{\hspace{2pt}}c@{\hspace{2pt}}c@{\hspace{2pt}}c@{}}
	% \includegraphics[width=0.2\linewidth]{Figures/SpatialVars/var_hsv/fluoTex.jpg} &
	% \includegraphics[width=0.2\linewidth]{Figures/SpatialVars/var_hsv/var_hsv_fluo1.jpg} &
	% \includegraphics[width=0.2\linewidth]{Figures/SpatialVars/var_hsv/var_hsv_fluo2.jpg} &
	% \includegraphics[width=0.2\linewidth]{Figures/SpatialVars/var_hsv/var_hsv_fluo3.jpg} &
	% \includegraphics[width=0.2\linewidth]{Figures/SpatialVars/var_hsv/var_hsv_fluo4.jpg}
	% \end{tabular}
	\vspace{-10pt}
	\caption{\textbf{Spatial variations of fluorescence color.}
	We use the color texture in the leftmost image to control the fluorescence parameters that govern color appearance.
	This is done by first converting the color texture to HSV, and then mapping hue to $\mu_e$, saturation to $\sigma_e$ and value to $\alpha_n$.
	The absorption parameters are set to $(\mu_a,\sigma_a)=(420,100)$ to ensure consistent appearance among illuminants, while the albedo is set to $\pmb{\rho}=(0.14, 0.14, 0.2)$.
	We manually set the mapping in the second image to approximately reproduce the color texture; while in the remaining images, we explore fluorescent color appearance freely.
	\label{fig:var-hsv}}
	\vspace{-10pt}
\end{figure*}

In this section, we explore more complex uses of our fluorescent material model: the edition of measured fluorescent materials, and spatial variations of fluorescence parameters.
The supplemental video demonstrates fluorescent material manipulation in real-time, along with an animated version of Figure~\ref{fig:teaser}. \change{For Figures~\ref{fig:paletteAbsorb}, \ref{fig:paletteEmit} and~\ref{fig:var-alpha} to~\ref{fig:var-hsv}, we used a real-time OpenGL prototype. On a laptop with an Intel Arc 140V integrated GPU, a $1920 \times 1080$ frame with our model takes $8.3$ms. Figures~\ref{fig:teaser} and~\ref{fig:fitInterp} use a CPU path tracer that we compare to a spectral reference following the state-of-the-art method of Mojzik et al.~\cite{Mojzik18}.}

\paragraph{Measured material editing}
In Figure~\ref{fig:fitInterp}, we fit our single-Gaussian model to a pair of measured materials~\cite{Gonzalez00}, then interpolate all fluorescence parameters while keeping the albedo unchanged.
As shown in the reradiation plots (first column), the blending of Gaussian parameters displaces the mass of the first sample towards the second one.
Since the input measured materials fluoresce different colors (orange in the first row, purple in the third row), the resulting interpolated material exhibits yet another color (green in the second row).
As shown in Figure~\ref{fig:fitInterp}, this editing method may be used either in spectral or non-spectral rendering, using full reradiation matrices or reduced matrices respectively.
We further show that we can boost the contribution of the fluorescence component in post-process by increasing the fluorescence strength.

\paragraph{Spatial variations}
In the following we rely on the model of Section~\ref{sec:edit}.
The simplest way to vary fluorescence across a surface is through modulation of the fluorescence strength $\alpha_n$.
This is shown in Figure~\ref{fig:var-alpha} with renderings in two environments, for a fixed $(\mu_a,\sigma_a)$ pair and several parameter choices of $(\mu_e,\sigma_e)$ .
We also vary $\alpha_n$ spatially in Figure~\ref{fig:var-abs}, but this time $(\mu_e,\sigma_e)$ is constant, and we modulate $\mu_a$ spatially using a fixed small absorption spread $\sigma_a$. Altering the range of variations of the $\mu_a$ parameter then results in different behaviors under different illuminants.
Our last example varies the $\mu_e$, $\sigma_e$ and $\alpha_n$  parameters based on an input color map.
More specifically, we have first converted the color map to HSV, then mapped hue to $\mu_e$, saturation to $\sigma_e$ and value to $\alpha_n$.
We show in Figure~\ref{fig:var-hsv} different results obtained with different choices of mapping for $\mu_e$.

%\paragraph{Other usage}

%\begin{itemize}
%    \item
%\todo{What if we can fit the illuminant as a sum of Gaussians? We could use the analytical integral to compute the outgoing radiance for direct illumination and not compute the reduced fluorescence.}
%
%    \item \todo{Mipmapping? }
%\end{itemize}

%\paragraph{Inverse design}
%
%\begin{itemize}
%    \item \todo{\sout{Develop the analytical formula Eq~\ref{eqn:integral_alt_final} to get a direct formula.}}
%
%    \item \todo{Specify the target color for a given illuminant.}
%\end{itemize}

\section{Discussion}
\label{sec:discuss}

We have introduced a new analytical model for fluorescent materials that brings several contributions. 
It is based on a decomposition that not only eases energy conservation, but also grants accurate reproduction of measured fluorescent materials.
We model the normalized fluorescence component with axis-aligned Gaussians, and introduce an analytical bispectral integration over arbitrary sensitivity functions.
This makes our approach particularly well suited to non-spectral rendering since fluorescence parameters may then be edited on the fly with real-time feedback.
Yet our model is also useful in spectral rendering contexts as it opens new avenues for the editing of spectral reradiation matrices.
Observing that a single Gaussian is enough to provide faithful fluorescence appearance, we use it to define a simplified model that enables artist-friendly controls through fluorescence palettes and spatial parameter variations.

\paragraph{Limitations} 
In this work, we have assumed that reflectance and fluorescence are only bound by energy conservation.
From a physical standpoint, the two components might be more tightly coupled, and a more accurate model would need to take such coupling into account.
However, we are not aware of any development in the Optics domain on that topic.
We have also considered Gaussian basis functions of infinite support in our model.
Using Gaussians clamped at the border of the visible range would complicate the derivation of Section~\ref{sec:gauss-int} as we would need to consider explicit integration bounds.
We did not find this to be necessary, as the fitted Gaussian have mimal energy outside of the visible range by construction.

\paragraph{Future work} 
Even though we have demonstrated our approach on diffuse fluorescent materials, it might be adapted to represent angularly varying fluorescence (e.g., as seen in \textit{Troides Magellanus} butterflies). 
However, this will require the availability of a database of measured bispectral BSDFs.
Our model might be useful in this respect to reduce memory footprint or to regularize measurement noise.
We would also like to investigate inverse design approaches on our simplified model, whereby artist provide reradiated colors under various illuminants and model parameters are retrieved. 
This would in particular allow one to explore metameric configurations that arise between fluorescent materials.
Finaly, we have focused our approach on fluorescent surfaces, but we believe it could be readily adapted to the modeling and rendering of fluorescent volumes (e.g., as seen in the white fur of polar bears).

% Bibliography
% \newpage
% \newpage
\pagebreak
\bibliography{main}

\begin{thebibliography}{10}

\bibitem{belcour2017}
{\sc Belcour, L., and Barla, P.}
\newblock {A Practical Extension to Microfacet Theory for the Modeling of
  Varying Iridescence}.
\newblock {\em {ACM Transactions on Graphics} 36}, 4 (July 2017), 65.

\bibitem{Benamira23}
{\sc Benamira, A., and Pattanaik, S.}
\newblock {A Microfacet Model for Specular Fluorescent Surfaces and Fluorescent
  Volume Rendering using Quantum Dots}.
\newblock In {\em Eurographics Symposium on Rendering\/} (2023), The
  Eurographics Association, pp.~11--2313.

\bibitem{Fichet2024}
{\sc Fichet, A., Belcour, L., and Barla, P.}
\newblock {Non-Orthogonal Reduction for Rendering Fluorescent Materials in
  Non-Spectral Engines}.
\newblock {\em {Computer Graphics Forum} 43}, 4 (2024).

\bibitem{Fourneau2024}
{\sc Fourneau, G., Pacanowski, R., and Barla, P.}
\newblock {Interactive Exploration of Vivid Material Iridescence using Bragg
  Mirrors}.
\newblock {\em {Computer Graphics Forum} 43}, 2 (Apr. 2024).

\bibitem{Glassner95}
{\sc Glassner, A.~S.}
\newblock A model for fluorescence and phosphorescence.
\newblock In {\em Photorealistic Rendering Techniques\/} (Berlin, Heidelberg,
  1995), Springer, pp.~60--70.

\bibitem{Gonzalez00}
{\sc Gonzalez, S., and Fairchild, M.}
\newblock Evaluation of bispectral spectrophotometry for accurate colorimetry
  of printing materials.
\newblock In {\em Proceedings of Color Imaging Conference\/} (2000),
  pp.~14--23.

\bibitem{Hua23}
{\sc Hua, Q., Tázlar, V., Fichet, A., and Wilkie, A.}
\newblock Efficient storage and importance sampling for fluorescent
  reflectance.
\newblock {\em Computer Graphics Forum 42}, 1 (2023), 47--59.

\bibitem{Hullin10}
{\sc Hullin, M.~B., Hanika, J., Ajdin, B., Seidel, H.-P., Kautz, J., and
  Lensch, H. P.~A.}
\newblock Acquisition and analysis of bispectral bidirectional reflectance and
  reradiation distribution functions.
\newblock {\em ACM Trans. Graph. (Proc. SIGGRAPH 2010) 29}, 4 (2010),
  97:1--97:7.

\bibitem{Iser2023}
{\sc Iser, T., Lachiver, L., and Wilkie, A.}
\newblock Affordable method for measuring fluorescence using gaussian
  distributions and bounded mese.
\newblock {\em Opt. Express 31}, 15 (Jul 2023), 24347--24362.

\bibitem{Jung18}
{\sc Jung, A., Hanika, J., Marschner, S., and Dachsbacher, C.}
\newblock A simple diffuse fluorescent bbrrdf model.
\newblock In {\em Proceedings of the Eurographics 2018 Workshop on Material
  Appearance Modeling\/} (2018), pp.~15--18.

\bibitem{Mojzik18}
{\sc Mojzík, M., Fichet, A., and Wilkie, A.}
\newblock Handling fluorescence in a uni-directional spectral path tracer.
\newblock {\em Computer Graphics Forum 37}, 4 (2018), 77--94.

\bibitem{Robbins1983}
{\sc Robbins, M.}
\newblock {\em The Collector’s Book of Fluorescent Minerals}.
\newblock Springer New York, NY, 1983.

\bibitem{Travouillon2023}
{\sc Travouillon, K.~J., Cooper, C., Bouzin, J.~T., Umbrello, L.~S., and Lewis,
  S.~W.}
\newblock All-a-glow: spectral characteristics confirm widespread fluorescence
  for mammals.
\newblock {\em Royal Society Open Science 10}, 10 (2023), 230325.

\bibitem{Wilkie2001}
{\sc Wilkie, A., Tobler, R., and Purgathofer, W.}
\newblock Combined rendering of polarization and fluorescence effects.
\newblock pp.~197--204.

\bibitem{Wilkie2006}
{\sc Wilkie, A., Weidlich, A., Larboulette, C., and Purgathofer, W.}
\newblock A reflectance model for diffuse fluorescent surfaces.
\newblock pp.~321--331.

\bibitem{Zenchyzen2024}
{\sc Zenchyzen, B., Acorn, J., Merkosky, K., and Hall, J.}
\newblock Shining a light on uv-fluorescent floral nectar after 50 years.
\newblock {\em Scientific Reports 14\/} (05 2024).

\end{thebibliography}

\end{document}